\documentclass[prb, twocolumn,showpacs,preprintnumbers,amsmath,amssymb,floatfix ]{revtex4}

\usepackage{graphicx}
\usepackage{dcolumn}
\usepackage{bm}

\begin{document}

\input epsf.sty



\title{ Hypothesis of two-dimensional stripe arrangement
and its implications for the superconductivity in high-{$T_c$} cuprates }

\author{B. V. Fine}
\email{fine@mpipks-dresden.mpg.de}

\affiliation{
Max Planck Institute for the Physics of Complex Systems,
Noethnitzer Str. 38, D-01187, Dresden, Germany}


\begin{abstract}
The hypothesis
that holes doped into high-$T_c$ cuprate superconductors 
organize themselves
in two-dimensional (2D) array of diagonal stripes is discussed, and,
on the basis of this hypothesis, 
a new microscopic model  of 
superconductivity is proposed and solved. 
The model describes two kinds of 
hole states 
localized either inside the stripes
or in the antiferromagnetic domains between the stripes.
The characteristic energy difference between these two 
kinds of states 
is identified with the pseudogap.
The onset of superconductivity is caused by the interaction, 
which is assumed to be mediated by the transverse
fluctuations of stripes.
The superconducting (SC) order parameter predicted by the model 
has two components,
whose quantum phases  exhibit a complex dependence on the   
the center-of-mass coordinate.
The model predictions for the tunneling characteristics and for the dependence of
the critical 
temperature ($T_c$)  on the superfluid density show  good quantitative agreement 
with a number of experiments. The model, in particular, 
predicts that the SC peaks  in the tunneling spectra
are asymmetric, only when $\Delta/T_c > 4$, where $\Delta$ is the SC gap.
It is also proposed that, at least in some  
high-$T_c$ cuprates, there  exist
two different superconducting states corresponding to
the same doping concentration and the same critical temperature.
Finally, the  checkerboard pattern in the 
local density of states observed by scanning tunneling microscopy in 
Bi$_2$Sr$_2$CaCu$_2$O$_{8+\delta}$ 
is interpreted  
as coming from the states localized around the centers of stripe elements
forming the 2D superstructure. 
\end{abstract}


\maketitle


\section{Introduction}

At present, the school of thought, which stipulates  the importance of 
local inhomogeneous structures called ``stripes'' for the physics
of high-$T_c$ cuprate superconductors, is
dominated by the view that  stripes form a one-dimensional(1D) array,
where each of them runs parallel to  one of the principal lattice 
directions\cite{ZG,PR,Schulz,Machida,Kato-etal-90,WS,SL,Tranquada-etal-96,Mook-etal-00,CEKO,Kivelson-etal-03}.
This picture, however, entails a number of difficulties 
associated with the stripe geometry.
In particular, in the presence of 
stripes parallel to the principal lattice directions, it is difficult to explain 
why the ``nodal'' quasiparticles having momentum directed along  lattice diagonals (i.e.
at 45 degrees with respect to stripes),
remain least gapped with the onset of both the pseudogap phenomenon and 
superconductivity.

In this work, I propose a microscopic model, 
which  is based on the hypothesis
that, at those doping concentrations, where superconductivity 
is observed in \mbox{high-$T_c$} cuprates, 
the holes organize themselves into a two-dimensional (2D) array of 
diagonal stripes. The 2D stripe superstructure does not incur the geometrical
difficulties associated with 1D stripe arrays.
This superstructure has been mentioned in the
literature\cite{ZG,Kato-etal-90,SG,Mook-etal-00,Tranquada-etal-99,Kivelson-etal-03}
(sometimes under the name of ``grid'' or ``checkerboard''),
but a number of experimental and theoretical arguments have been put forward
against its existence.
However, as I discuss later (in Section~\ref{arguments}),
the 2D stripe scenario has not been yet ruled out entirely.
At the same time, this scenario has never been analyzed
persistently enough, in part,
because no theoretical model has been put forward, which would relate the 2D
stripe superstructure to superconductivity.

The model proposed in this work 
reconciles the 2D stripe geometry
with superconductivity by demonstrating, that the superconductivity can be carried
by states localized in the 2D stripe background. 
The interaction, which, in the model, leads to the 
superconducting (SC) transition, is, presumably, mediated by the transverse
fluctuations of stripes.

The picture  emerging in the framework of the 2D
diagonal stripe hypothesis also
offers a very simple interpretation of 
the pseudogap phenomenon\cite{Marshall-etal-96,Loeser-etal-96,Ding-etal-96,DHS},
including its role in the superconductivity of cuprates. 
Other experimental facts to be interpreted in this work are:
(i) Quasiparticle coherence in $k$-space, which emerges only below 
the SC transition;
(ii) The asymmetry in the tunneling density of states;
(iii) Linear density of states in the vicinity of the chemical potential;
(iii) The checkerboard pattern in the local density of states in
Bi$_2$Sr$_2$CaCu$_2$O$_{8+\delta}$ (Bi-2212) observed by 
scanning tunneling microscopy 
(STM)\cite{Hoffman-etal-02,Howald-etal-03,Hoffman-etal-02A,Vershinin-etal-04}
(iv) Low superfluid density and the universal dependence thereof
on the critical temperature.

The SC order parameter obtained in this work
has complex two-component structure, 
which can not be described as either
$s$-wave or $d$-wave or the combination of two. 
The distinctive unconventional feature of this order parameter is
the non-trivial symmetry with respect to  spatial translations, which includes
the sign change of at least one of the two components.

Reviewing the relevant literature, it should be noted that
the general idea of superconductivity carried by localized states 
has been discussed in the past in
the context of various physical systems including  high-$T_c$ 
cuprates\cite{KK,BS,ML,Beschoten-etal-96,Sadovskii}. At a different level, this work
also contains parallels with several
theoretical proposals\cite{MRB,AA,Ashkenazi}, which involve 
two-component scenarios for high-$T_c$ cuprates.
Since the  2D stripe superstructure can be viewed as a collection of nanodomains
having different electronic properties, the present work can be linked 
to a more general class of ideas stipulating some kind of phase separation
in cuprates (see, e.g., Ref.\cite{GS}). If only the charge ordering associated with 
the 2D stripe superstructure is considered, then it exhibits certain similarities with 
Wigner crystals advocated in some of more recent 
proposals\cite{Chen-etal-02,HK}.

The rest of the paper is organized as follows:
In Section~\ref{2D} the hypothesis of 2D diagonal stripe configuration
is formulated. 
In Section~\ref{arguments}, the arguments in favor and against 
the existence of such a configuration are discussed. 
Possible dynamic properties of this
configuration are briefly analysed in Section~\ref{dynamic}. 
In Section~\ref{potential},
the properties of hole excitations in the presence
of the 2D stripe configuration are described, and the pseudogap phenomenon 
is identified. 
In Section~\ref{model}, a model describing 
the hole excitations in the stripe background is formulated and solved
in the mean-field approximation.  
In Section~\ref{discussion}, the realistic features of the model
and the resulting phase diagram are discussed.
The model predictions are compared with
experiments in Section~\ref{experimental}. 

This paper is quite long, in part, because 
some of the model predictions tested in Section~\ref{experimental}
require detailed calculations.  In the first reading, 
one can, therefore, review all the figures in the theoretical sections,
read Section~\ref{hamiltonian}, and then proceed with reading 
Section~\ref{experimental}.

\section{Two dimensional configuration of diagonal stripes}
\label{2D}

In this section, I introduce the basic assumption of the
present work, namely, the two-dimensional configuration of energetically deep
and spatially narrow stripes.

\begin{figure} \setlength{\unitlength}{0.1cm}
\begin{picture}(50, 205) 
{ 
\put(-23, -5)
{ \epsfxsize= 3.2in
\epsfbox{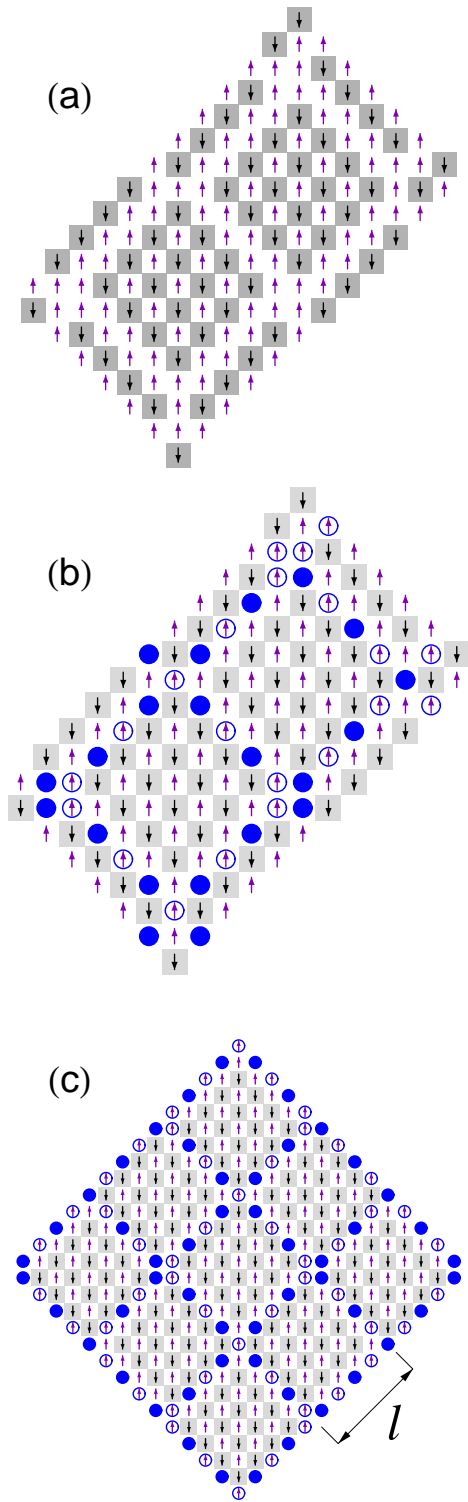} 
}}
\end{picture} 
\caption{(color online) 2D configuration of diagonal stripes: 
(a) Stripes in spin structure isolating
two AF domains with opposite sign of AF vector. (b) The same as (a) but
with holes (filled circles) occupying every second site along the spin 
stripes. Spins on the remaining stripe sites are circled to guide the eye.
(c) The same as (b) but with more stripe supercells shown.} 
\label{fig1} 
\end{figure}

I assume that, at sufficiently high doping concentrations, high-$T_c$ cuprates
find it energetically favorable to organize the spins of copper atoms into the 
background of antiferromagnetic (AF) domains as shown in Fig.~\ref{fig1}(a). 
Such a 
background creates an effective potential for the holes with 
minimum at the boundaries between those domains\cite{WE}.
Indeed, if one considers only the nearest neighbor exchange 
interaction between spins, then placing a hole in the middle
of an AF domain would cost energy $4J$, whereas 
at the domain boundary it costs $0J$. (Here $J$ is the nearest neighbor 
exchange coupling  constant.)  
The hole kinetic energy at the boundary is also lower  than  inside an AF domain,
because in the latter case, the hole
cannot hop
to the neighboring sites without increasing the exchange
energy of the system, while, in the former case, it can. 
I further assume that the gain in the exchange and
the kinetic energies outweighs the loss in the Coulomb energy (associated 
with the repulsion between holes). Therefore, holes
fill the domain boundaries and thus form stripes (see  
Figs.~\ref{fig1}(b) and \ref{fig1}(c)).
It is not important for the present work whether stripes are
centered on copper atoms, as shown in Figs.~\ref{fig1}(b,c),
or on oxygen atoms (i.e. on the ``bonds'' between copper sites). 


The AF domains formed between the stripes fall in two groups, 
which can be distinguished by the value of 
an AF index $\eta$ ($+1$ or $-1$) 
representing the sign of the AF order parameter within
a given domain. 
It is easy to see in Figs.~\ref{fig1}(a-c),
that the AF indices $\eta$ of two neighboring AF domains always 
have opposite signs.

A few mathematical facts about this kind of superlattice:

The spin periodicity of such a structure along each of the two
diagonal directions is $2 l$,
where $l$ is the side length of a single AF domain.
 
\begin{figure} \setlength{\unitlength}{0.1cm}

\begin{picture}(50, 78) 
{ 
\put(-13, 0)
{ \epsfxsize= 3in
\epsfbox{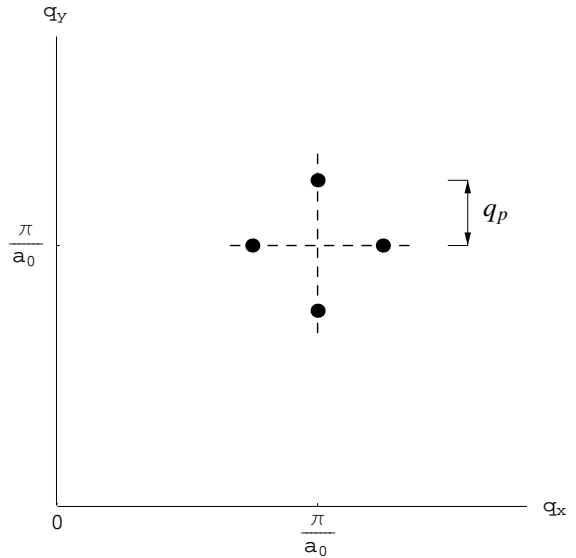} 
}}
\end{picture} 
\caption{Four-fold splitting of magnetic inelastic neutron scattering
peak.} 
\label{fig2} 
\end{figure}

For such a structure, 
the main splitting of the magnetic inelastic neutron scattering 
(INS) peak 
around the AF  wave-vector $({\pi \over a_0},{\pi \over a_0})$ 
is expected to be four-fold as shown in Fig.~\ref{fig2}. 
(Here  $a_0$ is the lattice period.)
The characteristic wave vector $q_p$ of this splitting is
frequently parametrized as
\begin{equation}
q_p = {2 \pi \over a_0} \ \delta,
\label{qp}
\end{equation}
where, in the present case,
\begin{equation}
\delta = {a_0 \over l \sqrt{2}}.
\label{delta}
\end{equation}
Theoretically, there should also exist other peaks corresponding to the
higher order Fourier harmonics of the stripe superstructure. 


Neglecting the intersections between stripes, the fraction of lattice sites
lying on the stripes is 
\begin{equation}
f = {a_0 \sqrt{2} \over l } = 2 \delta.
\label{f}
\end{equation}
The assumption of energetically deep stripes made earlier
implies that most holes are located inside the stripes.
In this case, the fraction of stripe sites occupied by holes is
\begin{equation}
c = {x_d \over f} = {x_d \over 2\delta},
\label{c}
\end{equation}
where $x_d$ is the dimensionless doping concentration.

Yamada et al.\cite{Yamada-etal-98} have discovered experimentally that in underdoped 
La$_{2-x}$Sr$_x$CuO$_4$ (LSCO) 
\begin{equation}
\delta \approx x_d,
\label{deltaxd}
\end{equation}
which means that $c \approx {1 \over 2}$, or the hole content
of diagonal stripes can be set empirically at 1 hole per 
two stripe sites (the same as 1 hole per length $2 a_0 \sqrt{2}$). 

Below, I will use term ``stripe element'' 
to refer to the part of a stripe, which constitutes a side
of one AF domain.
On the basis of empirical relation (\ref{deltaxd}),
I can estimate the number of lattice sites in one stripe element as:
\begin {equation}
{ l \over a_0 \sqrt{2}} = {1 \over 2 x_d}.
\label{la0}
\end{equation}
Doping concentration $x_d = 1/8$, taken as an example,
then corresponds to  $l = 4 a_0 \sqrt{2}$.

\section{Arguments in favor and against the existence of the
2D diagonal stripe structure}
\label{arguments}

If stripes exist, then the primary evidence for 
their geometrical properties comes from
the four-fold splitting of the magnetic INS peak 
(usually called the $(\pi,\pi)$ peak), 
which was described in the previous Section
and shown in Fig.~\ref{fig2}. This splitting was observed in
the underdoped compounds of LSCO\cite{Yamada-etal-98} and 
YBa$_2$Cu$_3$O$_{6+x}$ (YBCO)\cite{Mook-etal-98,Arai-etal-99,DMHD} 
families of high-$T_c$
superconductors.

A straightforward interpretation of this
peak pattern is that the antiferromagnetic spin structure 
\mbox{cos$({\pi \over a_0} x )\ $cos$({\pi \over a_0} y )$} is
modulated along both diagonal directions by function
\mbox{cos$[{\pi \delta \over a_0} (x +y) ]\ $cos$[{\pi \delta \over a_0} (x -y)  ]$}. 
(Coordinates $x$ and $y$ correspond, respectively, to the horizontal and vertical
axes in Fig.1(a-c).)
The modulation function can then be expanded
as the sum ${1\over 4} \left[ e^{i \ {2 \pi \delta x \over a_0}} +
e^{-i \ {2 \pi \delta x \over a_0}} +e^{i \ {2 \pi \delta y \over a_0}} 
+ e^{-i \ {2 \pi \delta y \over a_0}} \right]$ --- hence the four-fold
splitting of the main peak.

However, since the early indications of the stripe nature 
of the $(\pi,\pi)$ peak 
splitting\cite{Tranquada-etal-96}, 
most of the ``stripe community'' has opted for an
interpretation of experiments in terms of the superposition of two
two-peak splittings. The two kinds of splitting were described as 
coming from  two kinds of spatially separated
domains --- each representing a one-dimensional array of stripes
running along one of the principal lattice directions.

There, indeed, exist a number of theories\cite{ZG,PR,Schulz,Machida,Kato-etal-90,WS,SL} 
suggesting that stripes 
extending along the principal lattice directions are more favorable energetically
than diagonal stripes. (The 2D diagonal stripe superstructure
has been explicitly considered, e.g., in Refs.\cite{ZG,Kato-etal-90,SG}.) 
This is, however, a delicate energetic balance, 
which should be sensitive to numerous factors, not all of which are 
taken into account by the above mentioned theories.
For example, the 
interaction with the crystal lattice and the
long-range Coulomb interaction are, typically, neglected 
in the  numerical studies, even though the energy associated with each
of these two interactions can change the outcome of the competition
between different stripe configurations.
Since the chances of bringing the stripe
energetics under the full control of  first principles calculations are 
quite slim, the choice between different stripe configurations
(including the absence of stripes) should be made, eventually, on the basis
of experiments.

On the experimental side, one can find three main observations suggesting 
the one-dimensional nature of the stripe pattern\cite{Kivelson-etal-03}. 

First, in other materials structurally similar to \mbox{high-$T_c$} 
superconducting cuprates, 
such as, e.g.,
insulating La$_{1.95}$Sr$_{0.05}$CuO$_4$\cite{Wakimoto-etal-00} and 
nickelate La$_{2-x}$Sr$_x$NiO$_{4+y}$\cite{CCC,Li-etal-03},
there exists direct experimental evidence of the 1D nature of stripe modulations.

Second, the two-dimensional  spin structure modulation of the  form
\mbox{cos$[{\pi \delta \over a_0} (x +y) ]\ $cos$[{\pi \delta \over a_0} (x -y)  ]$}
should induce an effective potential for electric charge roughly of the 
form 
$\left\{\hbox{cos}[{\pi \delta \over a_0} (x +y) ]\ 
\hbox{cos}[{\pi \delta \over a_0} (x -y)  ]\right\}^2$.
If the spin modulation is weak, then 
it follows from the Landau theory of the second order phase 
transitions\cite{ZKE,Tranquada-etal-99},
that the peak structure for 
the charge inelastic neutron scattering corresponding 
to the above potential should  exhibit four main peaks around 
$(0,0)$.
The orientation of these peaks should be rotated by 45 degrees with
respect to the splitting of the magnetic peak, and characterized by the 
separation ${ 2 \pi \over a_0} \delta  \sqrt{2}$ from 
$(0,0)$.
The argument against the diagonal 2D spin stripes is that the above charge
peaks have not been observed, while, on one occasion, 
the charge peaks expected
for the 1D superstructure were observed\cite{Tranquada-etal-99}.

Finally, the third and, perhaps, the most direct experimental observation
suggesting the one-dimensionality of the stripe superstructure 
comes from Ref.\cite{Mook-etal-00} 
(also supported by Ref.\cite{Stock-etal-04}), 
where it was 
found that partial detwinning of a  YBCO sample leads to a very
strong asymmetry of the $(\pi,\pi)$ peak splitting. This asymmetry is 
in quantitative agreement
with the expectation that each of the twin domains in YBCO 
has only one kind of
one-dimensional stripe pattern, and, therefore, partial 
detwinning should lead to a significant redistribution
of intensities between the four peaks.

The  arguments  counter-balancing the above experimental observations 
can be the following:

The first observation, although important, is only indirect
and thus cannot substitute for direct observations.

The interpretation of the second observation 
is based on the assumption that the spin structure
is weakly modulated by function 
\mbox{cos$[{\pi \delta \over a_0} (x +y) ]\ $cos$[{\pi \delta \over a_0} (x -y)  ]$}.
As a result the charge modulation is also assumed to be small, and, therefore, 
the Landau theory is applied. 
However, if the modulation of the the AF structure is strong,
and the stripes are, indeed,
deep and narrow as  assumed in Section~\ref{2D}, 
then  the four charge peaks proposed as an indicator of
the 2D nature of stripe pattern
would only be  a part of a more complicated peak structure, 
and not necessarily the most pronounced one. 
For example, for a more realistic charge profile modulated as
\mbox{cos$^8[{\pi \delta \over a_0} (x +y) ]\ $
      sin$^2[{\pi \delta \over a_0} (x -y)  ] \ +$
      sin$^2[{\pi \delta \over a_0} (x +y) ]\ $
      cos$^8[{\pi \delta \over a_0} (x -y)  ] \ +$},
the positions of the strongest satellite peaks coincide with those
expected for the 1D picture. 

The effect of the 45 degree rotation of some modulation
peaks can, nevertheless, be relevant in another context. 
It will be shown in Section~\ref{B-states}, that this
effect can lead to the checkerboard pattern observed by STM in the local
density of states of Bi2212\cite{Hoffman-etal-02,Howald-etal-03,Hoffman-etal-02A}.

Concerning the third argument against the 2D diagonal stripe picture, 
the experimental data, as they stand, 
appear quite convincing. Yet, this kind of evidence has been
limited so far only to YBCO. The significant internal anisotropy
of the lattice structure in YBCO 
can, in principle, induce
an anisotropic INS response even of a truly two-dimensional
stripe superstructure. 
Thus the final resolution of the dilemma between
2D structure with anisotropic properties and a 1D structure
cannot be relied on this experiment alone. 
[Later note: 
A very recent INS study\cite{Hinkov-etal-04} 
has shown that the splitting of the $(\pi,\pi)$ peak
measured on {\it detwinned} YBCO crystals is anisotropic
but clearly two-dimensional.]

\section{Dynamic properties of the stripe structure}
\label{dynamic}

Now I address the question, to what extent the stripe picture described
in Section~\ref{2D}  can possess dynamic properties.

There are two possibilities for the time dependence of that structure.

The first possibility is that the stripe boundaries of AF domains can fluctuate 
and then drift away.
One circumstance that favors the fluctuations of stripes  
with or without average drift is that,
in the approximation of the nearest
neighbor exchange and without holes placed inside the stripes, 
the spins located
on the diagonal boundaries of AF domains  have zero exchange energy 
and thus are 
free to flip (see Fig.\ref{fig1}(a)). 
This would not the be case  if the 
AF domain boundaries were oriented along the principal lattice directions.

With holes inside the stripes, it is even 
easier for the stripes to fluctuate locally, but
it is more difficult to drift on average. 
The Coulomb repulsion between different stripe elements,
the topology of the 2D stripe superstructure and pinning on impurities
and structural disorder should inhibit
the average drift of the stripe pattern.

Thus, in the following, I will assume that,
on the timescales relevant to the physics
of superconductivity,  the stripe superstructure
does not drift, even though individual stripe elements
may exhibit strong transverse fluctuations.

The second possibility for the time dependence of the stripe superstructure
is that the AF order parameter of a given AF domain
can fluctuate and then, perhaps, exhibit ``rotational diffusion''. 
The relative spin orientation of neighboring AF
domains is fixed  not by direct 
exchange interaction but by the overall energy
balance of the entire structure, which does not allow holes 
to leave the stripes. 
Spins of a single AF domain cannot simply flip by 180 degrees, 
because such a flip would dissolve the domain boundaries, which, in turn, would 
contradict to the assumption that the high energy balance
favors the stripe structure. It can, however, happen that the AF
order parameter of a given domain fluctuates slightly away from
the the anti-alignment orientation with respect to the neighboring 
domains, and then either this fluctuation is damped back to
the initial orientation, or, on the contrary, the neighboring 
supercells adjust the orientation of their spins, and, in this way, 
the average orientation of  the spin order diffuses 
away for relatively 
large regions of the sample. 
The range of the spin orientations swept by the 
rotational diffusion will, eventually,
depend on the relative strength of 
the local anisotropy interaction with respect to the
the spin fluctuations.

As a consequence of the above kind of rotational diffusion, 
the elastic response from the modulated  AF spin structure 
should be either unobservable or strongly suppressed. 
At the same time, I  assume that
this diffusion is  slow enough
and thus can be neglected in a theoretical model of superconductivity.
(To date,
the elastic response of presumed stripe superstructure 
has been detected from the superconducting samples of
Nd-doped\cite{Tranquada-etal-96,Tranquada-etal-97}, 
Zn-doped\cite{Hirota-etal-98,Kimura-etal-99,Hirota-01} and
``pure'' LSCO\cite{Kimura-etal-99} and from 
La$_{1.875}$Ba$_{0.125-x}$Sr$_x$CuO$_4$\cite{Fujita-etal-02}.)

\section{Potential profile and the two kinds of hole states}
\label{potential}

\begin{figure}[!tbp] \setlength{\unitlength}{0.1cm}
\begin{picture}(50, 145) 
{ 
\put(-13, -15)
{ \epsfxsize= 3in
\epsfbox{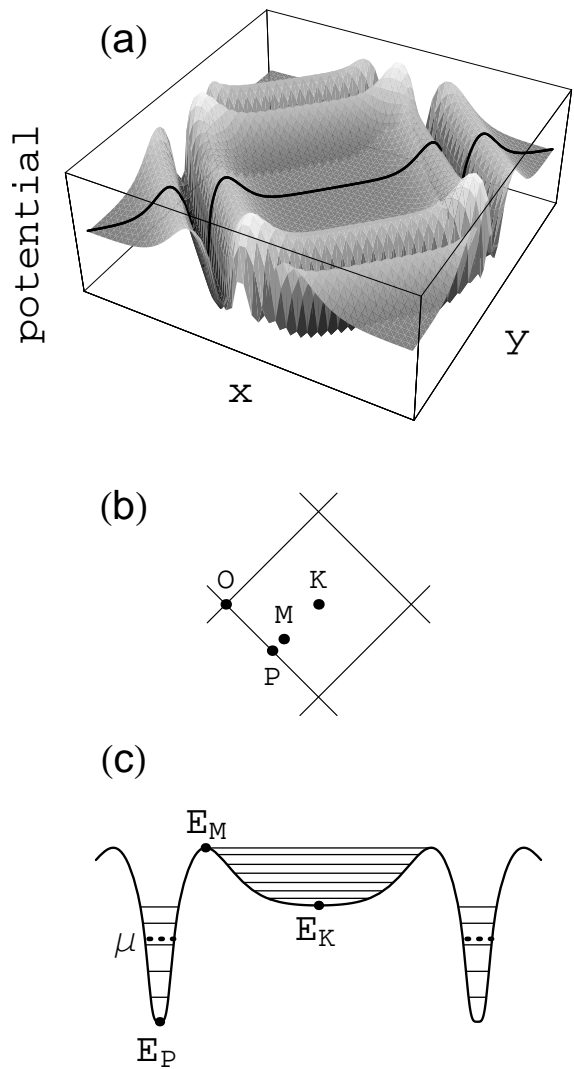} 
}}
\end{picture} 
\caption{(a) Sketch of potential landscape within one stripe supercell. 
(b) Scheme of a stripe supercell. (c) Sketch of potential landscape
along the thick line drawn across plot (a). Energies $E_K$, $E_M$ and $E_P$
correspond to points $K$, $M$ and $P$ marked in figure (b). Solid 
horizontal lines indicate quantum levels inside the respective potential wells.
Dashed line indicates the position of chemical potential $\mu$.} 
\label{fig3} 
\end{figure}

The potential profile for the hole 
excitations in the background of 2D stripe superstructure 
is sketched in Fig.~\ref{fig3}. This profile consists of
two main components: the network of narrow potential wells
running along the stripes and the (repulsive) Coulomb potential 
created around them. 


It is clear from the above picture, that there are two kinds of 
hole states: those localized mainly inside the stripes 
--- I will call them b-states --- and those localized mainly 
in the shallow potential wells 
inside the AF domains  --- I will call them 
a-states.  

It is, particularly easy to discuss the situation,
when (A) the in-stripe potential wells are deep enough,
so that, at low temperatures,   almost all holes stay inside the stripes,
and (B) the AF domains are large enough, so that the energy levels
of both a-states and b-states are spaced closely enough, as shown in 
Fig.~\ref{fig3}(c).
In this case,
the chemical potential $\mu$ of holes 
is 
approximately equal to the highest energy of occupied b-state at
zero temperature. This energy should be below $E_K$ (the minimal
energy  of  a-states indicated in Fig.~\ref{fig3}(c)), which is 
a formal restatement of the
assumption that the stripes are deep.

Such a situation may well correspond
to strongly underdoped (but superconducting) cuprates,
in which case, the difference $E_K - \mu$ can  be identified with the 
pseudogap  defined  as  the leading-edge midpoint in angle-resolved
photoemission spectroscopy (ARPES)\cite{DHS}. The phenomenology of the pseudogap
can then be interpreted as follows.
At energies between
$\mu$ and $E_K - \mu$, the density of states is low, because only b-states contribute
to this energy range. 
Moreover, since b-states are  extended along
the diagonal stripes, the momentum of photoelectrons emitted from
b-states should be preferentially oriented along the 
lattice diagonals. 
Above the pseudogap, the density of states gradually increases
due to the contribution from a-states.
The characteristic energy scale of this increase can be estimated 
as $0.1-0.2$~eV (a typical 
height of the Coulomb potential barrier
$E_M - E_K$ shown in Fig.~\ref{fig3}(c)). 

If the assumptions (A) and (B) hold at higher doping concentrations,
then one can envision some critical concentration 
(perhaps, not far from the optimal doping), at which the pseudogap
measured as $E_K - \mu$ becomes equal to zero. However,  the other measure
of the pseudogap, the gradual increase of the density of states
above $E_K = \mu$, will not disappear as long as stripes remain stable and,
therefore, generate locally inhomogeneous Coulomb potential landscape. 

\

One should note, however, that the validity of assumption (B) above is, 
particularly questionable
at the doping concentrations, corresponding to the physically interesting
sizes of AF domains of the order $l = 4 a_0 \sqrt{2} \approx 23$~\AA 
(for $a_0 = 4$~\AA). If, for an estimate of the level spacing of both
a- and b-states, one takes the spacing between
the lowest levels 
of a free particle having effective mass $m_e^{\ast} = 5 m_e$ 
in a box of size $l$, then one obtains 40~meV. (Here $m_e$  is the bare electron mass.)
The  number 40~meV is
of the order of the experimental values of the pseudogap and also notably larger
than a typical critical temperature ($\approx 7$~meV). 

If the level spacing is, indeed, as large as estimated above,
then it is likely, that 
there are no states (or very few due to disorder) 
in the energy range between
the highest occupied b-state and the lowest a-state.
In such a case, the position of the chemical potential within the 
above energy window becomes uncertain. 
However, the pseudogap
can still be defined as $\varepsilon_a - \varepsilon_b$, where $\varepsilon_a$ is  the  
lowest energy of a-states, 
and $\varepsilon_b$ the energy of a b-state  closest  to the chemical potential.

Formally speaking, the latter definition extends to 
a somewhat counterintuitive situation characterized by inequality 
$\varepsilon_a < \varepsilon_b$.
While, in the underdoped cuprates the strong expectation is 
that $\varepsilon_a > \varepsilon_b$ for both large and small values
of the level spacing,
the analysis of experiments in Section~\ref{superfluid} will 
suggest that inequality $\varepsilon_a < \varepsilon_b$ may 
characterize high-$T_c$ cuprates having doping concentrations above
$x_{dC} \approx 0.19$.
The condition $\varepsilon_a < \varepsilon_b$ 
contradicts to the assumption (A) made above.
However, as long as there are only a few a-states with
energies smaller than $\varepsilon_b$, 
a significant fraction of holes will still be localized inside the stripes
and generate the Coulomb potential required for the validity of 
the sketch shown in Figs.~\ref{fig3}(a,c).

One should also note that the condition $\varepsilon_a < \varepsilon_b$ is 
detrimental to stripe stability, because it implies that holes penetrate
inside the AF domains.
However, the emergence or disappearance of stripes is not just the 
subject of one-particle considerations.
This process is governed by the balance of global energy, 
which, among others, includes the contributions from lattice
strain and  quantum AF fluctuations. 
It is, therefore, not necessary,
and, in fact, unlikely, that stripes become unstable precisely when 
$\varepsilon_a = \varepsilon_b$.  Furthermore, 
in the   present work,
the condition $\varepsilon_a < \varepsilon_b$ will only imply one filled a-state
per AF domain, which the stripe superstructure may, indeed, sustain.

\

Now, I motivate the form of the  model Hamiltonian, 
which will be introduced in the next Section.

Describing b-states, I assume that they are fermionic states
carrying charge $e$ and belonging to one stripe element.
It will only be important for the model  that b-states
can form pairs with total spin 0. Whether or not they carry
spin 1/2 is not of primary importance.

An a-state is the state of one hole injected into 
a finite AF domain. It is not important
for the model to know exactly the orbital and the spin wave functions
of a-states. It is, however, important to note that 
(i) the spin wave function of an a-state 
should be fixed by the AF background, which is assumed to be static
on the time scale
of interest, and (ii)   the AF order parameter has opposite sign for
two neighboring AF domains.
Therefore, if two orbitally equivalent a-states from neighboring
AF domains form a pair, then the total spin of that pair is equal to zero.

It should be mentioned here, that the analogs of a- and b-states have been
identified in the numerical study of spin polarons in the stripe background\cite{WE}. 

As discussed in Section~\ref{dynamic}, the diagonal orientation of stripes
predisposes them to strong transverse fluctuations.  These fluctuations
should, in turn, strongly interact with both a- and b-states.  
The effect of this interaction is then two-fold.
On the one hand, the stripe fluctuations
couple to holes both elastically and inelastically, and thus suppress 
the hole transport across the stripe superstructure. 
On the other hand, they can efficiently mediate the interaction
between different hole states.

The first effect justifies the following ``center-of-mass'' 
selection rule:
{\it The model Hamiltonian  can have transition elements only 
between quantum states having the same center-of-mass coordinate.}

This selection rule, first of all, eliminates the direct hopping terms between
a- and b-states belonging to different AF domains or stripe elements.
Since this is quite a radical assumption, here I list several additional 
factors, which contribute to the suppression of hopping in addition
to the stripe fluctuations. These factors are:
(i) mismatch of AF backgrounds between two neighboring AF domains;
(ii) Coulomb potential barriers between neighboring AF domains and 
between intersecting stripe elements;  (iii) disorder in the stripe 
superstructure (which is not present in Fig.~\ref{fig3} but should be 
present in a real system). For the subsequent treatment, it is not important,
whether a- and b-states are rigorously localized.
All the above factors together should
only ensure that the hopping terms are small in comparison with the interaction term
discussed below.

In the absence of hopping, every a-state and every b-state 
are to be characterized by ``on-site'' energies
$\varepsilon_a$ and $\varepsilon_b$ respectively. 

I assume that
diagonal interactions involving a- and b-states,
and also non-diagonal interactions between a-states inside the same AF domain
and between b-states inside the same
stripe element
can be satisfactorily taken into account by
the renormalization of energies $\epsilon_a$ and $\epsilon_b$.

Considering the alternatives for non-diagonal interaction terms, I limit
the model choices only to the terms of the 
fourth order  with respect to the fermionic creation and
annihilation operators. Most of these terms  fail to
qualify under the ``center-of-mass'' selection rule.
Among the  few terms, which qualify, the only one, 
which will be included in the model, 
corresponds to the transition of two holes occupying two a-states
{\it on the opposite sides} of a given stripe element into  two b-states
inside that stripe element and {\it vice versa} (see Fig.(\ref{fig4}(b)).
One can check that the center of mass of two ``initial'' a-states 
coincides with the center of the stripe element between them,
and thus coincides with the center of mass of two``final'' b-states.

The above transition can be efficiently  mediated by 
the fluctuating stripe element itself. The relevant mechanism 
would involve two steps. {\it Step~1:} A hole hops between
two adjacent a- and b-states. Since the two states have different centers of mass,
the transition between them should be accompanied by a virtual excitation
of the transverse oscillation mode of the stripe element ``housing'' the b-state.
{\it Step~2:} The above oscillation mode is absorbed 
in the course of the symmetric transition of a second hole,
which involves an a-state from the other side of the same stripe element
and, therefore, restores the center-of-mass position. 
In order to appreciate this
mechanism, one can look at Fig.~\ref{fig3}, and imagine 
that, in the course of a transverse fluctuation, 
one of the two potential wells, which represent stripes,
shifts towards the center of the figure.  
As a result of this shift, 
the wave function of an a-state
in the center of the figure strongly overlaps with the wave function of 
a b-state inside the shifted stripe. 
Such a strong overlap constitutes a precondition for a large value of 
the interaction term selected for the model.

The above term, which can be schematically described as  
``aa$\leftrightarrow$bb,"
is sufficient to achieve the primary goal of the present work, which  is
to find at least one plausible channel
for the superconductivity in the presence of the 2D arrangement
of diagonal stripes. 
There exist, however, a few other fourth order non-diagonal terms 
having form
``ab$\leftrightarrow$ab", ``aa$\leftrightarrow$aa," or
``bb$\leftrightarrow$bb," which qualify
under the center-of-mass selection rule. 
At the moment, I rank these  terms as less important, because
either they involve states, which are too far separated from each other,
or they imply no charge flow between different components of the stripe
superstructure.
Nevertheless, 
the effect of including additional non-diagonal (and also diagonal) terms 
in the Hamiltonian would merit further study.

\section{Superconductivity model}
\label{model}

\subsection{Hamiltonian}
\label{hamiltonian}

In this Section, I introduce a model which is minimally sufficient to capture 
all the elements of the qualitative description given 
in Section~\ref{potential}.

\begin{figure}[!tbp] \setlength{\unitlength}{0.1cm}
\begin{picture}(50, 140) 
{ 
\put(-13, 0)
{ \epsfxsize= 3in
\epsfbox{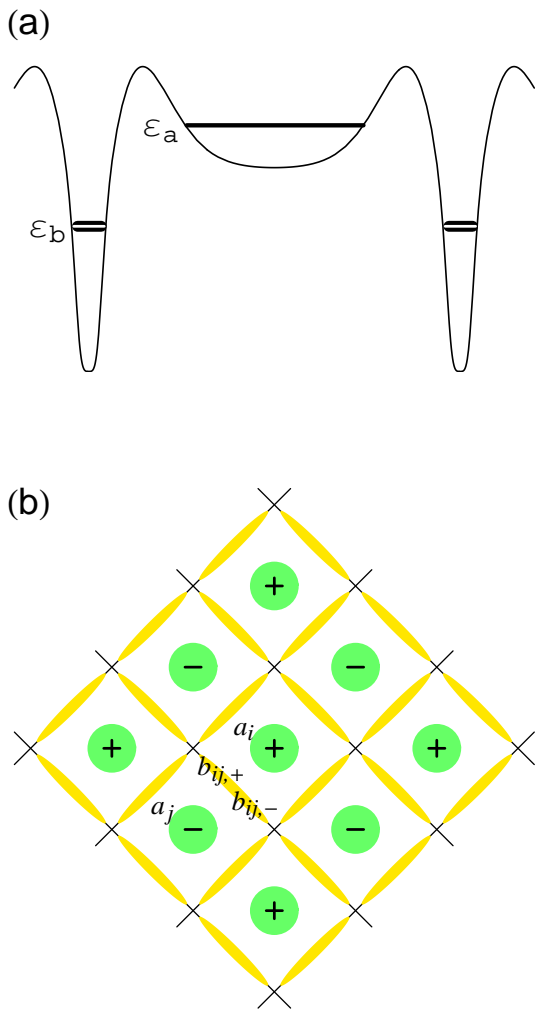} 
}}
\end{picture} 
\caption{(color online) (a) Model quantum states: 
one a-state with energy  $\varepsilon_a$ 
inside every AF domain, and two degenerate states with energy
$\varepsilon_b$ inside every stripe element.
This picture is to be compared with Fig.~\ref{fig1}(c). 
(b) Two-dimensional scheme of a-states and b-states. Each circle represents the center of 
an a-state, while each ellipsoid extended along the stripe boundaries represents the 
location of two  b-states.
Signs ``$+$'' or ``$-$'' inside the circles indicate the sign of AF index $\eta_i$.
Also shown: two operators of a-states
and two operators of b-states in their respective spatial domains. The 
transition between these two pairs of
a-states and b-states represents a typical term  in the
interaction part of Hamiltonian (\ref{H}).} 
\label{fig4} 
\end{figure}

The model is limited to the two-dimensional superlattice of 
a- and b-states, which is shown in Fig.~\ref{fig4}. 
This superlattice is
divided into diamond-shaped supercells, 
like the one shown in Fig.~\ref{fig3}(b).
The supercells are labelled by single index $i$ (or $j$).
Each of them is characterized by the AF
index $\eta_i$ (defined in Section~\ref{2D} but now with subscript $i$).
I will use the terms ``even supercell'' and ``odd supercell''
to refer to the supercells having $\eta_i =1$ and $\eta_i = -1$ 
respectively. The total
(macroscopic) number of supercells in the system is
denoted by variable $N$. I will further assume that
the system has rectangular form of dimensions
$L_x$ and $L_y$ along the $x$- and  the $y$-directions respectively.


In the present model, there exists only one a-state inside each supercell
described by 
hole annihilation and creation operators $a_i$ and $a_i^+$ respectively.
Each a-state is characterized
by the  on-site energy $\varepsilon_a$. 
In order to implement the observation made 
in the previous Section that orbitally equivalent a-states from the neighboring supercells
have opposite spins, the spins of a-states alternate together with the AF index
$\eta_i$, i.e. the spin of $i$th a-state is equal to ${1 \over 2} \eta_i$.  
Operators $a_i$ do not need an additional spin index, because,
in this model, the spin of an a-state is not an independent quantum number
but instead fixed by the lattice index $i$.

The next assumption is that, inside each stripe element separating  the $i$th
and the $j$th neighboring supercells,  there exist
only two b-states both characterized by the same ``on-site'' energy 
$\varepsilon_b$ but having different spins (+1/2 or -1/2). 
The hole  annihilation and creation operators for these states
are $b_{ij,\sigma}$ and $b_{ij,\sigma}^+$ respectively. 
Here index $\sigma$ 
represents the spin of a b-state and assumes values $+$ and $-$. 
(The assumption of a well defined spin is made just for the concreteness of the model.
In reality, it can, indeed, be spin but also any other kind of quantum number.)

In total, this model  contains one a-state and four 
b-states per one supercell (two b-states per stripe element
times two stripe elements per supercell).

The on-site energies 
$\varepsilon_a$ (the same for all a-states) and
$\varepsilon_b$ (the same for all b-states) are both measured 
from the chemical potential.  More detailed assumptions about their values  
will be made later.

Finally, the model Hamiltonian is: 
\begin{eqnarray}
{\cal H} &=&\varepsilon_a \sum_i a_i^+ a_i \ + \ 
\varepsilon_b \sum_{i, j(i), \sigma}^{\eta_i= 1} b_{ij,\sigma}^+ b_{ij,\sigma} 
\nonumber \\ 
&& + \ g \sum_{i, j(i)}^{\eta_i= 1} ( b_{ij,+}^+ b_{ij,-}^+ a_i a_j + \hbox{h. c.}),
\label{H}
\end{eqnarray}
where $g$ is the interaction constant.
Here and below the notation $j(i)$ in the sum subscript  
implies that the sum over index
$j$ extends only over the nearest neighbors of the $i$th supercell.
Expression $\eta_i= 1$ in the sum superscript means that the summation over
index $i$  includes only even supercells 
(marked by pluses  in Fig.\ref{fig4}(b)).

Hamiltonian (\ref{H}) can be described as an exotic 2-band model, 
where the non-interacting states are
localized but still have the same on-site energy, 
and the interaction includes only inter-band coupling.
In such a model, the variational SC ground state 
exists independently of the sign
of the coupling constant $g$. In the calculations, I will, therefore, 
assume that $g >0$.

Two b-states per stripe element and one a-state per AF domain,
represent the minimal configuration required for implementing the
interaction term in the Hamiltonian~(\ref{H}). At the same time,
it should be noted here that the  ratio ``one hole per $2 a_0 \sqrt{2}$''
extracted in Section~\ref{2D} from the scaling of INS data
also corresponds to 2 holes per stripe element in 
the most interesting case of  $l \approx 4 a_0 \sqrt{2}$. Furthermore,
if the level spacing inside AF domains is, indeed, as large 
as estimated in Section~\ref{potential}, then keeping only one a-state
per AF domain also constitutes a meaningful approximation
for the description of low-energy properties of the system.

\subsection{Classification of the model regimes}
\label{classification}

The present work is mostly limited to the analytical results
describing the following regimes:

\

Case IA: \ $\varepsilon_a \geq 0$, $\varepsilon_b = 0$;

\

Case IB: \ $\varepsilon_a \leq 0$, $\varepsilon_b = 0$;

\

Case IIA: \ $\varepsilon_a = 0$, $\varepsilon_b \geq 0$;

\

Case IIB: \ $\varepsilon_a = 0$, $\varepsilon_b \leq 0$;

\

Critical case: \ $\varepsilon_a = \varepsilon_b = 0$.

\

\begin{figure} \setlength{\unitlength}{0.1cm}

\begin{picture}(50, 55) 
{ 
\put(-20, 0)
{ \epsfxsize= 3.5in
\epsfbox{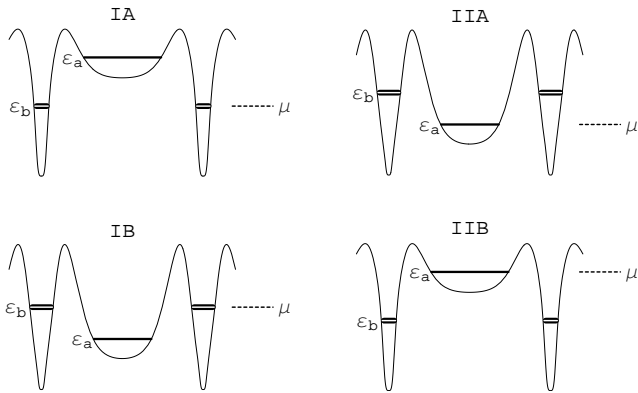} 
}}
\end{picture} 
\caption{Sketches representing Cases IA, IB, IIA and IIB discussed in the text.
} 
\label{fig-cases} 
\end{figure}

 The diagrams of energy levels representing 
Cases IA, IB, IIA and IIB are sketched
in Fig.~\ref{fig-cases}.
The reasons for distinguishing those special
regimes from the general case  ($\varepsilon_a \neq 0$, $\varepsilon_b \neq 0$) 
are the following:
(i) As shown later in subsection~\ref{chemical}, for the fixed difference
$\varepsilon_a - \varepsilon_b$, the situations, when either
$\varepsilon_a$ or $\varepsilon_b$ coincides with the chemical potential
(i.e. equal to zero) correspond to sharp minima in the
ground state energy. 
(ii)  The condition  $\varepsilon_a=0$ or $\varepsilon_b=0$
leads to a significant simplification of the model calculations.


Since most observables 
characterizing Cases IA and IB are identical,
these two Cases will be referred to as Case~I, whenever the difference
between IA and IB is not important. Similarly,  ``Case II''
will refer simultaneously to  Cases IIA and IIB.

\subsection{Mean-field solution}
\label{canonical}

In this Subsection I shall proceed with finding the variational
ground state using the method of Bogoliubov transformation.
The variational procedure will consist of
(i) making Bogoliubov transformation for b-states in real space; 
(ii) truncating
the Hamiltonian (\ref{H}) by leaving only the 
diagonal terms with respect to 
the new Bogoliubov quasiparticles and then averaging those terms;
(iii) making the Fourier transform of a-states;  
(iv) introducing the Bogoliubov transformation of a-states in 
$k$-space; and, finally, (v) minimizing the energy with respect to
the both transformations. 
Although straightforward, the above procedure is somewhat involved.
For this reason, in Appendix~\ref{non-canonical},
I also present an approximate version of the same mean-field
solution. The content of Appendix~\ref{non-canonical} 
also reveals a number of interesting facts about the
robustness of the full solution.

The first step of the variational scheme consists 
of the following Bogoliubov transformation:
\begin{eqnarray}
b_{ij+} = s B_{ij+} \  + \ w e^{i \varphi_{ij}} B^+_{ij-};
\label{bB+2}
\\
b_{ij-} = s B_{ij-} \ - \ w e^{i \varphi_{ij}} B^+_{ij+},
\label{bB-2}
\end{eqnarray}
where $B_{ij,\sigma}$ are the annihilation operators of
the Bogoliubov quasiparticles,  $s$ and $w$ are the (real)
transformation coefficients satisfying the normalization
constraint 
\begin{equation}
s^2 + w^2 = 1,
\label{sw}
\end{equation} 
and $\varphi_{ij}$ are the transformation phases chosen to be the same
for all translationally equivalent stripe elements.

There exist four translationally non-equivalent types of stripe elements.
Each type corresponds to one of the four possible orientations of 
vector ${\mathbf{r}}_j - {\mathbf{r}}_i$,
where ${\mathbf{r}}_i$ and ${\mathbf{r}}_j$ are 
the positions
of the centers of two neighboring supercells.
One of these two supercells is always even, while the other one
is always odd.
I will use the convention assigning  ${\mathbf{r}}_i$ to an even
supercell, and  ${\mathbf{r}}_j$ to an odd one. The four 
possible realizations of vector
${\mathbf{r}}_j - {\mathbf{r}}_i$  are:
\begin{eqnarray}
{\mathbf{R}}_1 = {l \over \sqrt{2}} (1, 1) ;
\label{R1}
\\
{\mathbf{R}}_2 = {l \over \sqrt{2}} (-1, 1) ;
\label{R2}
\\
{\mathbf{R}}_3 = {l \over \sqrt{2}} (-1, -1) ;
\label{R3}
\\
{\mathbf{R}}_4 = {l \over \sqrt{2}} (1, -1).
\label{R4}
\end{eqnarray} 

Phases $\varphi_{ij}$ can now be presented as
\begin{equation}
\varphi_{ij} = \varphi({\mathbf{r}}_j - {\mathbf{r}}_i).
\label{phir}
\end{equation}
They can have, at most, four different values
$\varphi_1 = \varphi({\mathbf{R}}_1)$, $\varphi_2 = \varphi({\mathbf{R}}_2)$, 
$\varphi_3 = \varphi({\mathbf{R}}_3)$ 
and $\varphi_4 = \varphi({\mathbf{R}}_4)$. 
I label these four phases by index $\alpha$ 
and refer to them using the notation
\begin{equation}
\varphi_{\alpha} = \varphi({\mathbf{R}}_{\alpha}).
\label{phialfa}
\end{equation}
The physical explanation, why the four phases $\varphi_{\alpha}$ 
should be tracked in 
the variational solution, is given in the end of Appendix~\ref{non-canonical}

Substituting transformation (\ref{bB+2}, \ref{bB-2}) 
into the Hamiltonian (\ref{H}), and then averaging the result with respect to  
B-operators, I obtain
\begin{eqnarray}
\nonumber
{\cal H}_a = 4 \ \varepsilon_b \ N \left[ s^2 n_B + w^2 (1-n_B) \right]
+ \varepsilon_a \sum_i a_i^+ a_i
\\
+ g s w (1- 2 n_B) \sum_{\alpha}
\left[ e^{-i \varphi_{\alpha}} \ 
\sum_i^{\eta_i = 1}  a_i a_{j(i,{\mathbf{R}}_{\alpha})}
+ \hbox{h. c.} \right],
\label{Hint}
\end{eqnarray}
where 
\begin{equation}
n_B = {1 \over \hbox{exp}\left({\varepsilon_B \over T}\right) +1}.
\label{nB}
\end{equation}
Here, $n_B$ and $\varepsilon_B $ are, respectively, the occupation number 
and the energy of a B-quasiparticles; and $T$ is the temperature measured
in energy units.
Index $j(i,{\mathbf{R}}_{\alpha})$  in Eq.(\ref{Hint}) labels
the nearest neighbor of $i$th supercell such that
$ {\mathbf{r}}_j - {\mathbf{r}}_i = {\mathbf{R}}_{\alpha} $.

At this point, it is convenient to replace supercell
indices $i$ and $j$ in Eq.(\ref{Hint}) by the set of the radius-vectors
and also to separate explicitly the summations over even and 
odd supercells. This gives
\begin{eqnarray}
\label{Hint1}
{\cal H}_a = 4 \varepsilon_b N \left[ s^2 n_B + w^2 (1-n_B) \right]
\ \ \ \ \ \ \ \ \ \ \ \ \ \ \ \ \ \ \ \ \ \ \ \ \ \ \ \ \ \
\\
\nonumber
+ \varepsilon_a 
\sum_{{\mathbf{r}}_{\hbox{e}}} 
a^+({\mathbf{r}}_{\hbox{e}})  a({\mathbf{r}}_{\hbox{e}})
+ \varepsilon_a 
\sum_{{\mathbf{r}}_{\hbox{o}}} 
a^+({\mathbf{r}}_{\hbox{o}})  a({\mathbf{r}}_{\hbox{o}})
\ \ \ \ \ \ \ \ \ \ \ \ \ \ \ \ \ \ \  
\\
+ g s w (1- 2 n_B) \sum_{\alpha}
\left[ e^{-i \varphi_{\alpha}} 
\sum_{{\mathbf{r}}_{\hbox{e}}}  
a({\mathbf{r}}_{\hbox{e}}) 
a({\mathbf{r}}_{\hbox{e}} + {\mathbf{R}}_{\alpha})
+ \hbox{h.c.} \right],
\nonumber
\end{eqnarray}
where even supercells are characterized by the discrete set of 
radius-vectors $\{ {\mathbf{r}}_{\hbox{e}}  \}$
and odd supercells by the complementary discrete set 
$\{ {\mathbf{r}}_{\hbox{o}}  \}$. Note: any vector of the form
${\mathbf{r}}_{\hbox{e}} + {\mathbf{R}}_{\alpha}$
belongs to the ``odd'' subset.

Now I introduce the Fourier transform {\it separately} for even and odd 
supercells:
\begin{eqnarray}
a_{\hbox{e}}({\mathbf{k}}) = \sqrt{{2 \over N}} 
\sum_{{\mathbf{r}}_{\hbox{e}}} 
a({\mathbf{r}}_{\hbox{e}})  \ 
e^{- i {\mathbf{k}} {\mathbf{r}}_{\hbox{e}}}
\label{aek}
\\
a_{\hbox{o}}({\mathbf{k}}) = \sqrt{{2 \over N}} 
\sum_{{\mathbf{r}}_{\hbox{o}}} 
a({\mathbf{r}}_{\hbox{o}})  \ 
e^{- i {\mathbf{k}} {\mathbf{r}}_{\hbox{o}}}.
\label{aok}
\end{eqnarray}
The two Fourier transforms, although involve different parts of 
real space, still performed with the same set of $k$-vectors,  
because the even and the odd subsets have 
the same periodicity. 
The projections $k_x$  and $k_y$ of the $k$-vectors, 
change in discrete steps 
${2 \pi \over L_x}$ and ${2 \pi \over L_y}$ respectively.
They fall in the interval 
$-{\pi \over d} \leq k_x, k_y \leq {\pi \over d}$, where 
$d$ is
the period of the sublattice of even (or odd) supercells equal to $l \sqrt{2}$. 
The total number of $k$-vectors is 
\begin{equation}
N_k = L_x L_y/(d^2) = N/2.
\label{Nk}
\end{equation}

After transformation (\ref{aek},\ref{aok}), the Hamiltonian~(\ref{Hint1}) can be written as
\begin{eqnarray}
\label{Hint2}
{\cal H}_a &=& 4 \varepsilon_b N \left[ s^2 n_B + w^2 (1-n_B) \right]
\\
\nonumber
&&
+ \varepsilon_a \  
\sum_{{\mathbf{k}}} 
a^+_{\hbox{e}}({\mathbf{k}})  a_{\hbox{e}}({\mathbf{k}})
+ \varepsilon_a 
\sum_{{\mathbf{k}}} 
a^+_{\hbox{o}}({\mathbf{k}})  a_{\hbox{o}}({\mathbf{k}})
\\
&& + \ g s w (1- 2 n_B) 
\nonumber
\\
&&
\ \ \  
\times
\sum_{{\mathbf{k}}}
\left[ a_{\hbox{e}}({\mathbf{k}})   a_{\hbox{o}}(- {\mathbf{k}})
V({\mathbf{k}})
+ \hbox{h.c.} \right],
\nonumber
\end{eqnarray}
where 
\begin{eqnarray}
\nonumber
V({\mathbf{k}}) & \equiv   \sum_{\alpha} 
e^{- i \varphi_{\alpha} - i {\mathbf{k}} {\mathbf{R}}_{\alpha}} 
\ \ \ \ \ \ \ \ \ \ \ \ \ \ \ \ \ \ \ \ \ \ \ \ \ \ \ \ 
\\
\nonumber
&=
2 \ \hbox{exp} \left[ -i \ {\varphi_1 + \varphi_3 \over 2} \right]
\hbox{cos} \left[ {\mathbf{k}} {\mathbf{R}}_1 + 
{\varphi_1 - \varphi_3 \over 2} \right] 
\\
&
+
2 \ \hbox{exp} \left[ -i \ {\varphi_2 + \varphi_4 \over 2} \right]
\hbox{cos} \left[ {\mathbf{k}} {\mathbf{R}}_2 + 
{\varphi_2 - \varphi_4 \over 2} \right]. 
\label{V}
\end{eqnarray}
Given the form of the interaction term in the Hamiltonian (\ref{Hint2}), 
the choice of canonical 
transformation for a-states is clear:
\begin{eqnarray}
a_{\hbox{e}}({\mathbf{k}}) &=& u({\mathbf{k}}) A_{\hbox{e}}({\mathbf{k}}) 
+ v({\mathbf{k}}) e^{i \phi_a({\mathbf{k}})} 
A^+_{\hbox{o}}(-{\mathbf{k}}) , 
\label{aeA}
\\
a_{\hbox{o}}(-{\mathbf{k}}) &=& u({\mathbf{k}}) 
A_{\hbox{o}}(-{\mathbf{k}}) 
- v({\mathbf{k}}) e^{i \phi_a({\mathbf{k}})} 
A^+_{\hbox{e}}({\mathbf{k}}),  
\label{aoA}
\end{eqnarray}
where $A_{\hbox{e}}({\mathbf{k}})$ 
and $A_{\hbox{o}}({\mathbf{k}})$ are annihilation operators of the new Bogoliubov quasiparticles;
$\phi_a({\mathbf{k}})$ is the phase of this transformation; and
$u({\mathbf{k}})$ and $v({\mathbf{k}})$ are the real
numbers 
obeying the following
normalization condition:
\begin{equation}
u^2({\mathbf{k}}) + v^2({\mathbf{k}}) = 1 .
\label{uvknorm}
\end{equation}

An important conceptual detail to be noted here is that 
transformation (\ref{aeA}, \ref{aoA}) will eventually
lead a coherent one-particle dispersion of A-quasiparticles 
in $k$-space. 
This  kind of $k$-space coherence emerges only 
in the SC phase and appears to be ``protected''
by the Fermi statistics 
(see the end of Appendix~\ref{non-canonical}).

Substitution of transformation (\ref{aeA}, \ref{aoA}) into the 
Hamiltonian (\ref{Hint1}) results in the following expression 
for the energy of the system:
\begin{eqnarray}
\label{E}
E &=& 4 \varepsilon_b N \left[ s^2 n_B + w^2 (1-n_B) \right]
\\
\nonumber
&&
+ 2 \varepsilon_a 
\sum_{{\mathbf{k}}} 
\left\{ 
u^2({\mathbf{k}})  n_A ({\mathbf{k}}) +  v^2({\mathbf{k}}) [1 - n_A ({\mathbf{k}})] 
\right\}
\\
\nonumber
&&
+ \ 2 g s w  (1- 2 n_B) 
\\
\times & \sum_{{\mathbf{k}}} &
u({\mathbf{k}})   v({\mathbf{k}}) (2n_A ({\mathbf{k}}) -1 )
|V({\mathbf{k}})| 
\hbox{cos} [\phi_V({\mathbf{k}}) + \phi_a({\mathbf{k}}) ],
\nonumber
\end{eqnarray}
where $|V({\mathbf{k}})|$ and $\phi_V({\mathbf{k}})$ are the absolute value 
and the phase of the complex-valued function (\ref{V}), and 
\begin{equation}
n_A({\mathbf{k}}) = {1 \over \hbox{exp}\left({\varepsilon_A({\mathbf{k}}) \over T}\right) +1}.
\label{nAk}
\end{equation}
Here $n_A({\mathbf{k}})$ and $\varepsilon_A ({\mathbf{k}})$ are, respectively,
the occupation number and the
energy of a Bogoliubov quasiparticle created by operator
$A^+_{\hbox{e}}({\mathbf{k}})$ or $A^+_{\hbox{o}}({\mathbf{k}})$.

It is immediately obvious that the minimization of the above expression
requires the last term to have maximally negative value.
All the sign conventions used below will be such that
\mbox{$g s w (1- 2 n_B) u({\mathbf{k}})   v({\mathbf{k}}) (2n_A ({\mathbf{k}}) -1 ) < 0$}.
Therefore, the maximally negative value  will be reached,
when
\begin{equation}
\hbox{cos} [\phi_V({\mathbf{k}}) + \phi_a({\mathbf{k}}) ] = 1,
\label{cosphia}
\end{equation} 
which means that (up to an integer number  of $2\pi$'s)
\begin{equation}
\phi_a({\mathbf{k}}) = - \phi_V({\mathbf{k}}).
\label{phia}
\end{equation}

For the reasons discussed in subsection~\ref{classification}, 
the rest of the calculations will be mostly limited to 
Cases IA, IB, IIA and IIB.
Formulas for the critical case will be summarized in subsection~\ref{critical}. 
For the general case, only the condensation energy
will be obtained (in subsection~\ref{chemical}). 

\

{\bf Case IA:} $\varepsilon_b =0, \varepsilon_a \geq 0 $.

A natural sign convention in this case is:  
$\varepsilon_A ({\mathbf{k}})>0$ and $\varepsilon_B <0$, i.e., at $T=0$,
$n_A({\mathbf{k}}) = 0$ and $n_B = 1$. 

Given the above convention, the minimization of energy~(\ref{E}) gives
\begin{eqnarray}
u({\mathbf{k}}) &=& \sqrt{
{1 \over 2} +  {1 \over 2} 
\sqrt{{
       1 
       \over 
       1 + {{\cal T}^2({\mathbf{k}}) \over {\cal Q}^2({\mathbf{k}})}
     }}
} ;
\label{ukI}
\\
v({\mathbf{k}}) &=& \sqrt{
{1 \over 2} -  {1 \over 2} 
\sqrt{{
       1 
       \over 
       1 + {{\cal T}^2({\mathbf{k}}) \over {\cal Q}^2({\mathbf{k}})}
     }}
} ;
\label{vkI}
\\
s &=& {1 \over \sqrt{2}};
\label{skI}
\\
w &=& - {1 \over \sqrt{2}},
\label{wkI}
\end{eqnarray}
where
\begin{eqnarray}
{\cal Q}({\mathbf{k}}) &=& 2 \varepsilon_a (2 n_A({\mathbf{k}}) - 1);
\label{Qk}
\\
{\cal T}({\mathbf{k}}) &=&  g (1 - 2 n_B) (2 n_A({\mathbf{k}}) - 1) |V({\mathbf{k}})|;
\label{Tk}
\end{eqnarray}

By varying $E$ with respect to $n_A({\mathbf{k}})$ and $n_B$,
one can now obtain the quasiparticle energies:
\begin{equation}
\varepsilon_A({\mathbf{k}}) = \sqrt{\varepsilon_a^2 + 
{1 \over 4} g^2 (2 n_B -1)^2 |V({\mathbf{k}})|^2
},
\label{epsAkI}
\end{equation}
\begin{equation}
\varepsilon_B = - {g^2 (2 n_B -1) \over 8N}
\sum_{{\mathbf{k}}} {(1 - 2 n_A({\mathbf{k}}) |V({\mathbf{k}})|^2
\over \varepsilon_A({\mathbf{k}})},
\label{epsBI}
\end{equation}
and then express the total energy of the system as
\begin{equation}
E = - \sum_{{\mathbf{k}}} [(1 - 2 n_A({\mathbf{k}}) )
\varepsilon_A({\mathbf{k}})
- \varepsilon_a ] .
\label{EI}
\end{equation}
Here and everywhere, the summation over ${\mathbf{k}}$ can
be replaced by integration according to the following
rule
\begin{equation}
{1 \over N} \sum_{{\mathbf{k}}} \rightarrow {d^2 \over 8 \pi^2} 
\int_{-{\pi \over d}}^{{\pi \over d}} dk_x 
\int_{-{\pi \over d}}^{{\pi \over d}} dk_y
\label{sum-integral}
\end{equation}

Energy (\ref{EI}) is the function of $|V({\mathbf{k}})|$ 
(via $\varepsilon_A({\mathbf{k}})$ and 
$n_A({\mathbf{k}}) \equiv n_A[\varepsilon_A({\mathbf{k}})]$ \ ). 
In turn, $|V({\mathbf{k}})|$  is the function of four phases 
(\ref{phialfa}) (via Eq.(\ref{V})). Therefore, energy (\ref{EI}) should further be minimized
with respect to the values of those phases.

As shown in Appendix~\ref{phasemin}, such a minimization
imposes only one constraint:
\begin{equation}
{\varphi_2 + \varphi_4 -\varphi_1 - \varphi_3 \over 2} =
{\pi \over 2} + \pi n,
\label{phases}
\end{equation}
where $n$ is an integer number.
Among three other independent combinations of phases 
$\varphi_1$, $\varphi_2$, $\varphi_3$ and  $\varphi_4$,
one should remain free as a consequence of the global gauge invariance,
while two remaining combinations should, in principle, be fixed, but
not in the framework of the present model.

\begin{figure} \setlength{\unitlength}{0.1cm}

\begin{picture}(50, 199) 
{ 
\put(-13, -7)
{ \epsfxsize= 2.6in
\epsfbox{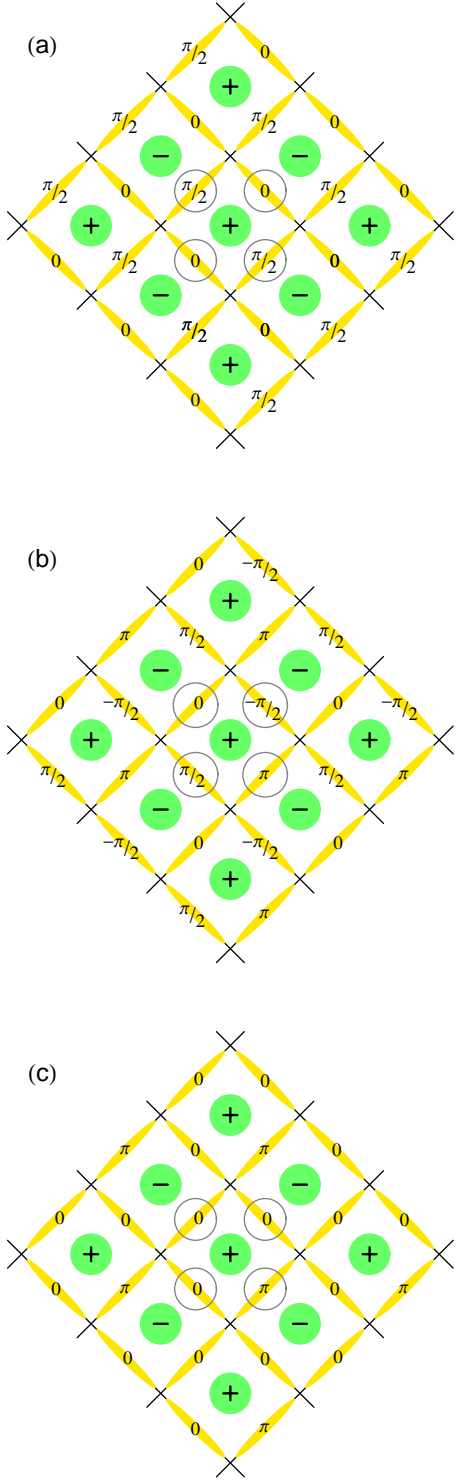} 
}}
\end{picture} 
\caption{(color online) Three examples of particularly symmetric patterns of phases
$\varphi_{ij}$ consistent with the variational SC solution. The values
of the phases are indicated on the top of the corresponding stripe elements.
Each pattern is obtained by the periodic translation of four circled
phases  denoted in the text as $\varphi_{\alpha}$. These phases are constrained by
Eq.(\ref{phases}). 
The expressions for $V({\mathbf{k}})$ corresponding to the
phase patterns (a), (b) and (c) are given, in respective order,  
by Eqs.(\ref{Vcc}), (\ref{Vss}) and (\ref{Vcs}).
} 
\label{fig-phases} 
\end{figure}

The main thermodynamic and transport properties
of the model are independent of the choice
of phases $\varphi_1$, $\varphi_2$, $\varphi_3$ and  $\varphi_4$
as long as this choice is consistent with Eq.(\ref{phases}).
This situation
is somewhat similar to that of superfluid $^3$He, where the interaction
causing the superfluid transition does not fix the values of all variables
characterizing the order parameter\cite{Leggett-75}.
In $^3$He, the remaining freedom is eliminated by magnetic dipolar 
interaction between nuclei, and by other small interactions.
In the present case, the same role can be played, e.g., by pair hopping
between a-states or b-states belonging to different supercells.
In this work, the issue of the ``phase freedom'' is not resolved.
It is, however, possible to speculate that the additional  terms
will lead to sufficiently symmetric selection of phases,  such that
\begin{equation}
V({\mathbf{k}}) = 2 \left\{
\hbox{cos} \left[ {\mathbf{k}} {\mathbf{R}}_1  \right]  -
i \  \hbox{cos} \left[ {\mathbf{k}} {\mathbf{R}}_2  \right] 
\right\},
\label{Vcc}
\end{equation}
or 
\begin{equation}
V({\mathbf{k}}) = 2 \left\{
\hbox{sin} \left[ {\mathbf{k}} {\mathbf{R}}_1  \right]  -
i \ \hbox{sin} \left[ {\mathbf{k}} {\mathbf{R}}_2  \right] 
\right\},
\label{Vss}
\end{equation}
or
\begin{equation}
V({\mathbf{k}}) = 2 \left\{
\hbox{cos} \left[ {\mathbf{k}} {\mathbf{R}}_1  \right]  -
i \ \hbox{sin} \left[ {\mathbf{k}} {\mathbf{R}}_2  \right] 
\right\},
\label{Vcs}
\end{equation}
The first choice corresponds to
$\varphi_1 = \varphi_3 = 0$, 
$\varphi_2 = \varphi_4 = \pi/2$;  the second one to
 $\varphi_1 = -\pi/2$, $\varphi_2 = 0$, 
$\varphi_3 =  \pi/2$, and $\varphi_4 = \pi$;
and the third one to
$\varphi_1 = 0$, $\varphi_2 = 0$, 
$\varphi_3 =  0$, and $\varphi_4 = \pi$;
The resulting  
patterns of phases are shown in Fig.~(\ref{fig-phases}).


Having specified the parameters of transformations
(\ref{bB+2}, \ref{bB-2}, \ref{aeA}, \ref{aoA}), one can 
calculate the temperature dependence
of various thermodynamic quantities.
This requires the numerical solution of the system
of equations (\ref{nB}, \ref{nAk}, \ref{epsAkI}, \ref{epsBI}),
which is not done in the present work. Without the full numerical solution
only the zero-temperature characteristics and the SC transition temperature 
$T_c$ can be evaluated. 
The evaluation of $T_c$ is based on
a manipulation described in Appendix~\ref{TcEq}, which gives the following 
simple 
equation:
\begin{equation}
T_c = { g^2 \left[\hbox{exp}\left({\varepsilon_a \over T_c}\right) - 1\right] \over 
8 \varepsilon_a \left[\hbox{exp}\left({\varepsilon_a \over T_c}\right) + 1\right]}.
\label{Tceq}
\end{equation}

When $ g \leq \varepsilon_a$, 
the approximate solution of Eq.(\ref{Tceq}) is 
\begin{equation}
T_c \cong { g^2  \over 8 \varepsilon_a }.
\label{Tc}
\end{equation}
Another simple limit is $\varepsilon_a =  0$,
in which case,  Eq. (\ref{Tceq}) yields 
\begin{equation}
T_c = { g  \over 4 }.
\label{Tc1}
\end{equation}
In general, however, Eq.(\ref{Tceq}) has to be solved numerically.

Since
the operators of both real holes and real electrons do not commute with
A- and B-operators defined by Eqs.(\ref{aA}-\ref{bB-}),  
the tunneling studies of both 
A- and B-quasiparticles (via a contact with  normal metal) 
should show the density of states on the both sides of the chemical potential,
i.e. at $\varepsilon = \pm \varepsilon_A$ and $\varepsilon = \pm \varepsilon_B$. 
Moreover, as long as $\varepsilon_b = 0$,
tunneling into B-states should result in 
the density of states symmetric with respect to the 
chemical potential. 
As far as A-states are concerned, then 
tunneling into them should show asymmetric density of states. 
This asymmetry is characterized by the ratio 
\begin{equation}
{ D[\varepsilon_A({\mathbf{k}})] \over D[-\varepsilon_A({\mathbf{k}})] }  =
{ u^2({\mathbf{k}}) \over v^2({\mathbf{k}}) } = 
{ \varepsilon_A({\mathbf{k}}) + \varepsilon_a \over \varepsilon_A({\mathbf{k}}) - \varepsilon_a}.
\label{DDI}
\end{equation}

\begin{figure} \setlength{\unitlength}{0.1cm}

\begin{picture}(50, 142) 
{ 
\put(-20, -3)
{ \epsfxsize= 3in
\epsfbox{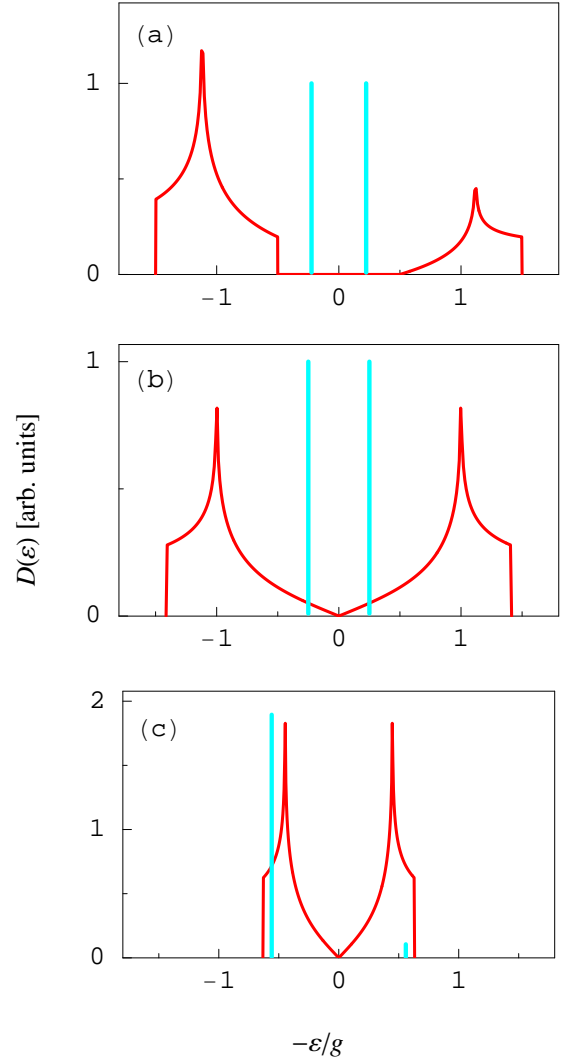} 
}}
\end{picture} 
\caption{(color online) Examples of the tunneling density of states at $T=0$: 
(a) Case IA ($\varepsilon_a = 0.5 g$, $\varepsilon_b = 0$);
(b) critical case ($\varepsilon_a = 0$, $\varepsilon_b = 0$); 
(c) Case IIA ($\varepsilon_a = 0$, $\varepsilon_b = 0.5 g$).  
In each case, the spectra
contain two vertical delta-peaks representing B-states and
located at $\pm \varepsilon_B$ given by 
(a) Eq.(\ref{epsAkI}); (b,c) Eq.(\ref{epsAkII})
The continuous part  in each spectrum represents A-states. 
It is calculated from: (a) Eq.(\ref{epsAkI}); (b,c) Eq.(\ref{epsAkII}).
In all three cases, the spectra of A-states have Van-Hove singularities
located at $\pm \varepsilon_{A0}$ and the   termination points located at 
$\pm \varepsilon_{A1}$. The spectrum of A-states in figure (a)
also has a gap between $\varepsilon_a$ and $- \varepsilon_a$. 
The asymmetry of the spectra is obtained from: (a) Eq.(\ref{DDI});
(c) Eq.(\ref{DDII}). 
Note: the positive direction
of the horizontal axis corresponds  to negative hole energies.
(This reflects a convention of tunneling spectroscopy.)
} 
\label{fig-tunnel} 
\end{figure}

The zero-temperature 
tunneling spectra of A- and B-quasiparticles are shown in 
Fig.~\ref{fig-tunnel}(a). 
The spectrum of B-quasiparticles consists
of two delta peaks located at $\pm \varepsilon_B$.
A-quasiparticles have a continuous spectrum, which is 
fully gapped with minimal
energy $\varepsilon_a$.
It was obtained
by first calculating the density of states
following from Eq.(\ref{epsAkI}) as a function of positive energies
$\varepsilon_A$ and then dividing
the weight between positive and negative tunneling energies according to
formula (\ref{DDI}). 


\begin{figure} \setlength{\unitlength}{0.1cm}

\begin{picture}(50, 130) 
{ 
\put(-30, 0)
{ \epsfxsize= 4in
\epsfbox{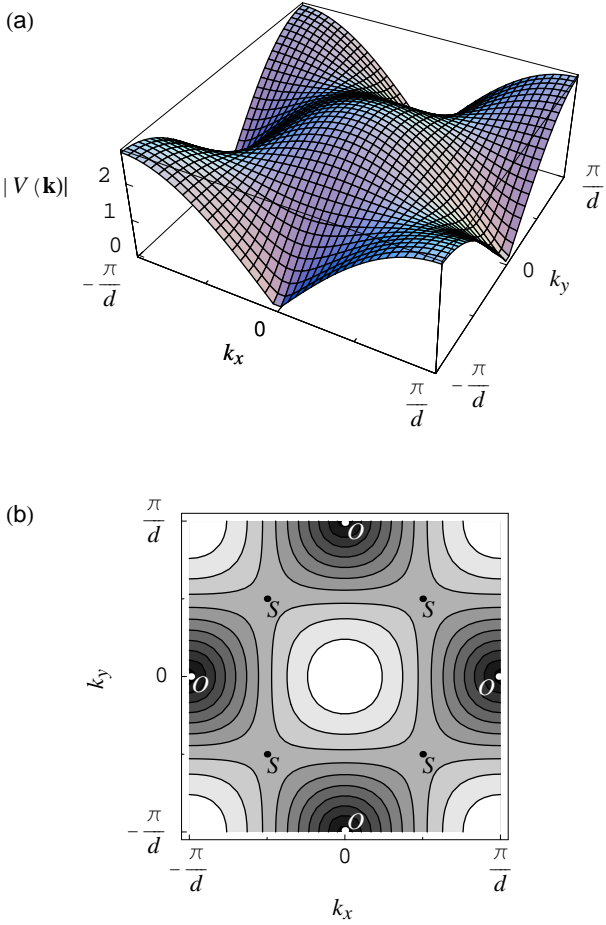} 
}}
\end{picture} 
\caption{ (a) Three-dimensional plot and (b) contour plot of
$|V({\mathbf{k}})|$  corresponding to Eq.(\ref{Vcc}).
Each plot shows the first Brillouin zone of the stripe superstructure.
In the contour plot, the saddle points are indicated with ``$S$'',
and zeros with ``$O$''. 
} 
\label{fig-Vk2} 
\end{figure}

In addition to the gap and the asymmetry, 
two other important features of the spectrum of A-quasiparticles are:
the Van-Hove singularity and the sharp termination
point at a higher energy.
These two features correspond, respectively, to the saddle points
and to the maxima of $|V({\mathbf{k}})|$.
Function $|V({\mathbf{k}})|$
obtained from Eq.(\ref{Vcc}) is shown in Fig.~\ref{fig-Vk2}. 
It has four saddle points at 
${\mathbf{k}}_{\hbox{s}} = {\pi \over 2 d} (\pm 1, \pm 1)$.
With another choice of phases consistent with Eq.(\ref{phases}), 
the $k$-space position of the saddle
points may change but not the value of 
$|V({\mathbf{k}}_{\hbox{s}})| = 2$.
Therefore, according to Eq.(\ref{epsAkI}), 
the density of A-states exhibits a Van Hove singularity at
\begin{equation}
\varepsilon_{A0} = \sqrt{
\varepsilon_a^2 +   g^2 (2 n_B -1)^2 
}.
\label{epsA0I}
\end{equation}
The maximum energy of A-states, which corresponds to 
the tunneling spectrum termination point, 
can be found by substituting the maximum value of $|V({\mathbf{k}})|$ equal to $2\sqrt{2}$
into Eq.(\ref{epsAkI}), which gives 
\begin{equation}
\varepsilon_{A1} = \sqrt{
\varepsilon_a^2 +   2 g^2 (2 n_B -1)^2 
}.
\label{epsAI1}
\end{equation}


It should be noted here, that the appearance of the tunneling spectrum of A-states
shown in Fig.~\ref{fig-tunnel}(a)
is quite similar to that of the fermion spectrum obtained
by Altman and Auerbach from the plaquette boson-fermion
model\cite{AA}.

\

{\bf Case IB:} $\varepsilon_b =0$, $\varepsilon_a  \leq 0$.

In this case, if one keeps the same sign convention as in Case IA,
then $u({\mathbf{k}})$ and 
$v({\mathbf{k}})$ given by 
Eqs.(\ref{ukI}, \ref{vkI}) should switch values.
All formulas for the quasiparticle 
energies and the tunneling spectrum asymmetry   
obtained 
for Case IA  apply without modification to 
the present case.  Equation (\ref{Tceq}) 
for the critical temperature
also applies but  with the trivial substitution
of $|\varepsilon_a|$  instead of 
$\varepsilon_a$.

The only observable difference between
Cases IA and IB is the  opposite asymmetry with respect 
to the chemical potential: in Case IA, the density of A-states is greater
on the hole side, while, in Case IB, on the electron side.

\

{\bf Case IIA:} $\varepsilon_a =0$, $\varepsilon_b  \geq 0$.

The sign convention  in this case is chosen to be opposite to that of Case IA,
namely: 
$\varepsilon_A ({\mathbf{k}})<0$ and $\varepsilon_B >0$, i.e., at $T=0$,
$n_A({\mathbf{k}}) = 1$ and $n_B = 0$.

In this case,
\begin{eqnarray}
u({\mathbf{k}}) &=& {1 \over \sqrt{2}};
\label{ukII}
\\
v({\mathbf{k}}) &=& {1 \over \sqrt{2}};
\label{vkII}
\\
s &=& \sqrt{
{1 \over 2} +  {1 \over 2} 
\sqrt{{
       1 
       \over 
       1 + {{\cal T}^2 \over {\cal Q}^2}
     }}
} ;
\label{skII}
\\
w &=& - \sqrt{
{1 \over 2} -  {1 \over 2} 
\sqrt{{
       1 
       \over 
       1 + {{\cal T}^2 \over {\cal Q}^2}
     }}
} ,
\label{wkII}
\end{eqnarray}
where
\begin{eqnarray}
{\cal Q} &=& 4 N \varepsilon_b (2 n_B - 1);
\label{Q}
\\
{\cal T} &=&  g (1 - 2 n_B) C_a N;
\label{T}
\\
C_a &=& {1 \over N} \sum_{{\mathbf{k}}} 
(2 n_A({\mathbf{k}}) -1) |V({\mathbf{k}})| 
\label{Ca}
\end{eqnarray} 
One can then obtain the quasiparticle energies 
\begin{eqnarray}
\varepsilon_A({\mathbf{k}}) &=&
- {g^2 \ |V({\mathbf{k}})| \ C_a \ (1 - 2 n_B) \over 8 \varepsilon_B},
\label{epsAkII}
\\
\varepsilon_B &=& \sqrt{\varepsilon_b^2 + 
{1 \over 16} \ g^2 \ C_a^2
},
\label{epsBII}
\end{eqnarray}
and the total energy 
\begin{equation}
E = - 2 N [(1 - 2 n_B) 
\varepsilon_B
- \varepsilon_b ] .
\label{EII}
\end{equation}

The choice of phases (\ref{phialfa}) is still 
constrained by condition (\ref{phases}) (see Appendix~\ref{phasemin}). Given this constraint,
the zero temperature value of $C_a$ (obtained numerically) is
\begin{equation}
C_{a0} \equiv {1 \over N} \sum_{{\mathbf{k}}} |V({\mathbf{k}})|   
= 0.958\ldots \ .
\label{Ca0}
\end{equation}

A manipulation analogous to the one described in
Appendix~\ref{TcEq} gives the following equation 
for the critical temperature:  
\begin{equation}
T_c = { g^2 \left[\hbox{exp}\left({\varepsilon_b \over T_c}\right) - 1\right] \over 
8 \varepsilon_b \left[\hbox{exp}\left({\varepsilon_b \over T_c}\right) + 1\right]}.
\label{TceqII}
\end{equation}

The tunneling density of states corresponding to Eqs.(\ref{epsAkII},\ref{epsBII})
is shown in Fig.~\ref{fig-tunnel}(c). 
Contrary to the result for Case~I, the tunneling density of A-states
in Case II is symmetric,
while the density of B-states is asymmetric. 
This asymmetry is characterized by the ratio 
\begin{equation}
{ D(\varepsilon_B) \over D(-\varepsilon_B) }  =
{ s^2 \over w^2 } = 
{ \varepsilon_B + \varepsilon_b \over \varepsilon_B - \varepsilon_b}.
\label{DDII}
\end{equation}

An important feature of Case II, which is absent
in Case~I, is that the energy spectrum of A-quasiparticles is gapless
with the linear density of states around the chemical potential. 
Indeed, $\varepsilon_A({\mathbf{k}})$ given by Eq.(\ref{epsAkII}) touches zero
in an isolated set of non-analytic
points corresponding to
$|V({\mathbf{k}})| = 0$.  
For the specific choice of $V({\mathbf{k}})$ 
given by Eq.(\ref{Vcc}), the zeros of 
$|V({\mathbf{k}})|$ are shown in Fig.~\ref{fig-Vk2}.
They are located at 
${\mathbf{k}}_0 = {\pi \over  d} (\pm 1, 0)$ and 
${\mathbf{k}}_0 = {\pi \over  d} (0, \pm 1)$.)
This feature is a direct consequence 
of the phase relation (\ref{phases}).  
A deviation from that relation would produce
a line of zeros, which implies a non-zero density of states at $\varepsilon_A = 0$.

The density of A-states in Case IIA has 
Van Hove singularity and the termination point
located, respectively, at  
\begin{equation}
\varepsilon_{A0} = 
- {g^2 C_a (1 - 2 n_B) \over 4 \varepsilon_B}
\label{epsA0II}
\end{equation}
and
\begin{equation}
\varepsilon_{A1} = \sqrt{2} \ \varepsilon_{A0}.
\label{epsAII1}
\end{equation}

\

{\bf Case IIB:} $\varepsilon_a =0$, $\varepsilon_b  \leq 0$.

All formulas  for the quasiparticle 
energies and the tunneling spectrum asymmetry   
obtained for Case IIA  apply without modification to 
Case IIB.  Equation (\ref{TceqII}) 
for the critical temperature
only requires the  substitution
of $|\varepsilon_b|$  instead of 
$\varepsilon_b$.
The only difference between
Cases IIA and IIB is the opposite asymmetries in
the tunneling spectra of B-quasiparticles. In Case IIA,
the hole side of the B-quasiparticle spectrum has more weight, 
while in Case IIB, the larger weight is on the electron side.

\subsection{Chemical potential as a variational parameter}
\label{chemical}

In this subsection, I argue 
that the situations
corresponding to  
$\mu = \varepsilon_b$ (Case I)
or $\mu = \varepsilon_a$ (Case II) should be considered as
likely scenarios describing realistic stripe systems.

The constraint on the total number of particles, which is
usually used to fix $\mu$, cannot be straightforwardly applied
to the present model for the following reasons:
(i) The model quantum states,
form a subset of all quantum states of a real ``striped'' system,
and, therefore, the actual total number of particles
cannot be reliably counted. 
(ii) The system can always readjust the periodicity of the stripe
superstructure, which would change the ratio between
the number of model states and the number of holes doped into
CuO$_2$ planes. (iii) The chemical potential can change within
the model pseudogap without affecting the total number of particles
occupying model states (at $T=0$).

It is, therefore, reasonable to treat the chemical potential
as a variational parameter, which is fixed by the minimization
of the total energy of the real system considered as the sum 
of the contribution
from the model states and the contribution from 
environment (unspecified here). 

The model contribution to the total energy
as a function of the chemical potential can be obtained by
solving the model in the general case:   
$\varepsilon_a \neq  0$, $\varepsilon_b \neq  0$.
The description of general case is 
as straightforward as that of Cases I and II. 
However, the minimization routine
produces an integral equation, which couples the values of 
$u({\mathbf{k}})$, $v({\mathbf{k}})$, $s$ and $w$, 
and which has to be solved numerically.

\begin{figure} \setlength{\unitlength}{0.1cm}

\begin{picture}(50, 120) 
{ 
\put(-28, -42)
{ \epsfxsize= 5.0in
\epsfbox{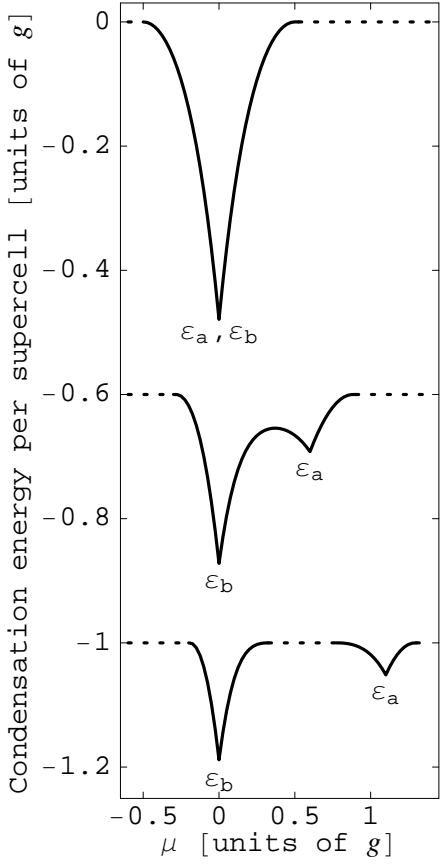} 
}}
\end{picture} 
\caption{Condensation energy (with the negative sign) 
as a function of the chemical potential $\mu$. 
Each of the three curves  was calculated on the basis of Hamiltonian (\ref{H}) 
by fixing  $\varepsilon_a$, $\varepsilon_b$ and $g$ and then varying
$\mu$. In the calculations, $g$ and $\varepsilon_b$ were equal, respectively, 
to 1 and 0
in all three cases, while 
$\varepsilon_a$ admitted the following values 
(top to bottom): 0, 0.6 and 1.1. The solid/dashed lines indicate the 
SC/non-SC ground state. The deeper minima of the lower two curves
correspond to Case I, and the shallower ones to Case II.
The single minimum of the top curve corresponds to the critical case.
The vertical coordinates of
the middle and the bottom curves are shifted by -0.6 and -1 respectively.
} 
\label{fig-cusps} 
\end{figure}

Figure~\ref{fig-cusps} shows three representative
curves for the evolution of the SC ground state energy
as a function of the chemical potential.
Each curve was obtained numerically for the fixed values of 
$\varepsilon_a $, $\varepsilon_b $ and $g$ indicated in the caption.
In order to allow for the variation of $\mu$,   the reference point
for one-particle energies was shifted (in this part only)  
from $\mu=0$ to some arbitrary value
$\varepsilon_b =  0$. 
The SC ground state energy was measured from the energy of the normal state.
The absolute value of thus defined quantity is 
conventionally called condensation energy.


In a generic situation (two lower curves
in Fig.~\ref{fig-cusps}),
the ground state energy has two minima: the deeper one --- corresponding to
Case I, and the shallower one --- corresponding to Case II.
For a fixed value of $g$, 
the SC transition is possible only,
when the value of $\mu$ lies close enough to either
$\varepsilon_a$ or $\varepsilon_b$. Dependently on the ratio 
$|\varepsilon_a - \varepsilon_b|/g$, the  values of $\mu$
compatible with superconductivity fill either
one  finite interval including both 
$\varepsilon_a$ and $\varepsilon_b$ 
or two disconnected finite intervals around 
$\varepsilon_a$ and $\varepsilon_b$.
The top curve in Fig.~\ref{fig-cusps} illustrates 
a non-generic situation, when the two 
minima coincide. The resulting single minimum 
then corresponds to the critical case.

The main purpose of Fig.~\ref{fig-cusps} is to illustrate, that,
not only Cases I and II  correspond to local minima in the 
$\mu$-dependence of the ground state energy,
but also that these minima have the form of cusps. 
This implies that, if the environment contribution to the
total
energy of the model-plus-environment system 
is a smooth function  of $\mu$, then
the total energy should have the cusp minima 
due to the model contribution  
at exactly the same values of $\mu$ as those obtained without
environment. 
If only the model contribution were taken into account, then 
it would follow from Fig.~\ref{fig-cusps}, that
the minimum corresponding to Case I is the global one.
However, the environment contribution
can change the relative values
of energies, corresponding to the two minima and thus shift 
the global minimum to that of Case II.

In principle, it is also possible,
that environment introduces
an extra energy minimum in addition to the two cusps described above,
and, moreover, that additional minimum is the global one.

The resulting possibilities can be summarized as follows: 
Cases I and II of the present model
describe two different SC states corresponding to two different positions
of the chemical potential. In the ``best case'' scenario 
(which I also consider more likely), one of these
two states corresponds to the global energy minimum of the real system,
while the other one represents a metastable state. 
In the ``worst case''
scenario, both SC states are metastable, but the true SC state may still 
be describable by the present model with 
$\varepsilon_a \neq 0$ and $\varepsilon_b \neq 0$.

It is further possible that the position of the chemical potential
near the sample surface  is different from that in the bulk.
Therefore, the  SC state, which is metastable in the bulk
may become stable near the surface and {\it vice versa}. 
In principle, it is also possible that the system phase separates and forms domains
describable either by Case~I or by Case~II.

In Section~\ref{correspondence}, I will try to discriminate
between the SC states corresponding to Cases~I and II 
by making comparison
between the model predictions and the experiments.

\subsection{Anomalous correlation functions}
\label{odlro}

Bogoliubov transformations (\ref{bB+2},\ref{bB-2},\ref{aeA},\ref{aoA})  
imply that, below $T_c$, the following  anomalous
correlation functions have non-zero values:
\begin{equation}
\Psi_a({\mathbf{k}}) = 
\langle a_{\hbox{e}}({\mathbf{k}}) a_{\hbox{o}}(-{\mathbf{k}})
\rangle = u({\mathbf{k}}) v({\mathbf{k}}) e^{\phi_a({\mathbf{k}})}[2 n_A({\mathbf{k}}) -1],
\label{Psiak}
\end{equation}
\begin{equation}
 \Psi_b({\mathbf{r}}_{ij}, {\mathbf{r}}_{pn})   \equiv   
\langle b_{ij,-} b_{pn,+} \rangle  = s w e^{i \varphi_{ij}} (1 - 2 n_B)
\delta({\mathbf{r}}_{ij} - {\mathbf{r}}_{pn}),
\label{Psib1}
\end{equation}
where ${\mathbf{r}}_{ij}$ represents the positions of the centers of stripe
elements, and $\delta(\ldots)$  is defined as  
Kronneker delta on the discrete
superlattice, i.e. it is equal to 1, when its argument is zero,
and 0 otherwise. The two correlation functions (\ref{Psiak})
and (\ref{Psib1}) are the two components
of the SC order parameter corresponding to a- 
and b-states respectively.

In the real space, the first component can be defined as
 \begin{equation}
\Psi_a({\mathbf{r}}_i, {\mathbf{r}}_j) 
\equiv 
\langle a({\mathbf{r}}_i) a({\mathbf{r}}_j) \rangle.
\label{Psiarrdef}
\end{equation}
It has non-zero values only when its two arguments
correspond to the supercells of different kind  (i.e. even and odd).
The formal structure of $\Psi_a({\mathbf{r}}_i, {\mathbf{r}}_j)$
can be expressed as follows:
\begin{equation}
\Psi_a({\mathbf{r}}_{\hbox{e}}, {\mathbf{r}}_{\hbox{o}}) 
= 
{2 \over N} \sum_{\mathbf{k}} 
\Psi_a({\mathbf{k}}) 
e^{i {\mathbf{k}} ({\mathbf{r}}_{\hbox{e}} - {\mathbf{r}}_{\hbox{o}})},
\label{Psiarr}
\end{equation}
\begin{equation}
\Psi_a({\mathbf{r}}_{\hbox{o}}, {\mathbf{r}}_{\hbox{e}})
= - \Psi_a({\mathbf{r}}_{\hbox{e}}, {\mathbf{r}}_{\hbox{o}}),
\label{Psiarr-}
\end{equation}
\begin{equation}
\Psi_a({\mathbf{r}}_{\hbox{e}}, {\mathbf{r}}_{\hbox{e}}^{\prime})
= \Psi_a({\mathbf{r}}_{\hbox{o}}, {\mathbf{r}}_{\hbox{o}}^{\prime}) = 0.
\label{Psiarr0}
\end{equation}
where $\Psi_a({\mathbf{k}}) $  is given by Eq.(\ref{Psiak}).
Note: Eq.(\ref{Psiarr-}) is the consequence of the
fermionic anticommutation rule.

The coherence length of the order parameter 
$\Psi_a( {\mathbf{r}}_i, {\mathbf{r}}_j)$
should  be inversely proportional
to the one characterizing $V({\mathbf{k}})$ in $k$-space. 
The examination of 
Eq.(\ref{V}) reveals that the characteristic scale of $V({\mathbf{k}})$
is $\pi / |{\mathbf{R}}_1| = \pi/l $. Therefore, the coherence length associated with 
$\Psi_a( {\mathbf{r}}_i, {\mathbf{r}}_j)$ 
can be estimated as the supercell size $l$. It is likely, that on a longer
scale $\Psi_a( {\mathbf{r}}_i, {\mathbf{r}}_j)$ exhibits an oscillatory
power law decay with the period of oscillations being of the order of $l$.

The coherence length associated with  $\Psi_b$ is equal to
zero, which means that only
b-states belonging to the same stripe element form coherent
pairs.

Two useful quantities, which will later be required
in the calculation of supercurrent are:
\begin{equation}
\Psi_{a(ij)}  \equiv  \Psi_a^{\hbox{\footnotesize n.n.}}( {\mathbf{r}}_i, {\mathbf{r}}_j)
\equiv \langle a_i a_j \rangle^{\hbox{\footnotesize n.n.}}
\label{Psiaij}
\end{equation}
and
\begin{equation}
\Psi_{b(ij)} \equiv  \Psi_b({\mathbf{r}}_{ij}, {\mathbf{r}}_{ij}) 
\equiv \langle b_{ij,-} b_{ij,+} \rangle ,
\label{Psibij}
\end{equation}
where the superscript ``n.n.'' indicates that indices
$i$ and $j$ represent the nearest neighbors.

In Case~I, the explicit expression for $\Psi_{b(ij)}$ 
can be obtained by substituting
the values of  $s$ and $w$ given by Eqs.(\ref{skI},\ref{wkI}) 
into Eq.(\ref{Psib1}) 
for ${\mathbf{r}}_{ij} = {\mathbf{r}}_{pn}$,
which gives
\begin{equation}
\Psi_{b(ij)}  = 
{1 \over 2} e^{i \varphi_{ij}} (2 n_B - 1).
\label{Psib1I}
\end{equation}
One can then obtain $\Psi_{a(ij)}$ by making use of the fact that
\begin{equation}
\Psi_{b(ij)}^* \Psi_{a(ij)} = {E_{\hbox{\footnotesize int}} \over 4 g N }
\label{Eint}
\end{equation}
where $E_{\hbox{\footnotesize int}}$ is the interaction part
of the energy (\ref{E}) 
(i.e. $E_{\hbox{\footnotesize int}} = 2 g s w (1-2n_B)  \sum_{{\mathbf{k}}} ...$).
After $E_{\hbox{\footnotesize int}}$ is evaluated with the help of Eqs.(\ref{ukI}-\ref{wkI}),
one can use Eqs.(\ref{Psib1I}, \ref{Eint}) to obtain:
\begin{equation}
\Psi_{a(ij)} = {g (1 - 2 n_B) \eta_i e^{i \varphi_{ij}} \over 8 N} 
\sum_{{\mathbf{k}}}  { [1 - 2 n_A({\mathbf{k}})] |V({\mathbf{k}})|^2 
\over \varepsilon_A({\mathbf{k}})}.
\label{Psia1I}
\end{equation}
The role of index $\eta_i$ in Eq.(\ref{Psia1I}) is to
supply factor 1 or $-1$ dependently
on whether the first index of $\Psi_{a(ij)}$ corresponds
to an even or an odd supercell. 

In Case II, the expressions analogous to (\ref{Psia1I},\ref{Psib1I}) are
\begin{equation}
\Psi_{a(ij)}  =  {1 \over 4} \eta_i e^{i \varphi_{ij}} ,
\label{Psia1II}
\end{equation}
\begin{equation}
\Psi_{b(ij)}  = - { g C_a  e^{i \varphi_{ij}} (1 - 2 n_B) \over 8 \varepsilon_B}.
\label{Psib1II}
\end{equation}

The essential elements of the symmetry of the SC order parameter
(\ref{Psib1I},\ref{Psia1I}) or, alternatively, (\ref{Psia1II},\ref{Psib1II}) 
are captured in Fig.~\ref{fig-phases}. This symmetry 
is characterized by the pattern of phases
$\varphi_{ij}$  indicated
on the top of each stripe element, and, in addition,
by the pattern of index $\eta_i$. 

The order parameter 
$\Psi_b({\mathbf{r}}_{ij}, {\mathbf{r}}_{pn})$ has no
dependence on the relative orientation of ${\mathbf{r}}_{ij}$ 
and ${\mathbf{r}}_{pn}$. Therefore, it can be described as having
orientational s-wave symmetry  with additional  
strong phase dependence on the center-of-mass
position of the paired holes.

The symmetry of $\Psi_a({\mathbf{r}}_i, {\mathbf{r}}_j)$
is more different
from conventional analogs. It includes the strong 
dependence on $\varphi_{ij}$, which traces the phase dependence of 
$\Psi_b({\mathbf{r}}_{ij}, {\mathbf{r}}_{pn})$, but, in addition,  
$\Psi_a({\mathbf{r}}_i, {\mathbf{r}}_j)$  also exhibits
a sign change  under the translation by
one period of the stripe superstructure.
This sign change reflects the switching between even/odd and odd/even
order in Eq.(\ref{Psiarr-}).

\subsection{Supercurrent and the penetration depth}
\label{supercurrent}

The superconducting properties of the present model are unusual,
because the superconducting  phase stiffness comes solely from
the interaction term of the Hamiltonian (\ref{H}). This term induces
the fundamental ``internal supercurrent''  associated with the
transfer of particle density from a-states to b-states and 
{\it vice versa}. The translational supercurrent then appears as
a gradient of the internal one.

The operator expression for the internal current from
$i$th a-state to the surrounding b-states can be obtained by considering the 
time derivative
of the particle density operator: 
\begin{eqnarray}
J_{ab(i)} \equiv - {d \over dt} (a_i^+ a_i) = 
- {i \over \hbar} [{\cal H} a_i^+ a_i - a_i^+ a_i {\cal H}]
\nonumber
\\
= - {i g \over \hbar} \sum_{j(i)} 
( b_{ij,+}^+ b_{ij,-}^+ a_i a_j -  \hbox{h.c.}) .
\label{Jabi}
\end{eqnarray}
Here and everywhere in this subsection,
index $i$ corresponds to an even supercell, and index $j$ to an odd one.

The sum in Eq.(\ref{Jabi}) has four terms --- each corresponding 
to the transfer of
the particle density from the $i$th AF domain into  a nearby 
stripe element labelled by the pair of indices $ij$. 
Therefore, the operator of translational current
through the $i$th supercell (to be denoted as ${\mathbf{J}}^t_i$) 
can be obtained by assigning the direction 
to the flow of particle density associated 
with each of the above four terms, i.e.
\begin{equation}
{\mathbf{J}}^t_i
= - {i g \over 2 \hbar} \sum_{j(i)} \ 
\hat{{\mathbf{n}}}_{ij} \  \left( b_{ij,+}^+ b_{ij,-}^+ a_i a_j - \hbox{h.c.} \right) ,
\label{Jti}
\end{equation}
where $\hat{{\mathbf{n}}}_{ij}$ is the unit vector in the direction from the 
$i$th to the $j$th supercell. 

The translational current is created, when
the probability of an a-particle to hop into one of the 
surrounding stripe elements is greater 
in one direction than in the opposite one.
For this reason, the translational current can only be carried by 
a-states. The number of
b-particles hopping on the both sides of a given stripe element
is the same for each quantum transition generated 
by the Hamiltonian (\ref{H}).

Now, I show that the phase, which drives the internal {\it supercurrent}, is
\begin{equation}
\phi_{ab} = - \phi_a({\mathbf{k}}) -  \phi_V({\mathbf{k}}).
\label{phiab}
\end{equation}

In equilibrium, according to Eq.(\ref{phia}), $\phi_{ab} = 0$.
The state with non-zero $\phi_{ab}$
can be obtained by modifying the Bogoliubov transformation
for b-states 
(Eqs. (\ref{bB+2}, \ref{bB-2})) as follows:
\begin{eqnarray}
b_{ij+} = s B_{ij+}  + w e^{i (\varphi_{ij} + \phi_{ab})} B^+_{ij-};
\label{bB+3}
\\
b_{ij-} = s B_{ij-}  - w e^{i (\varphi_{ij} + \phi_{ab})} B^+_{ij+},
\label{bB-3}
\end{eqnarray}
while keeping the transformation (\ref{aeA}, \ref{aoA})
for a-states unchanged.
In this case,  the anomalous correlation functions are, for a-states belonging
to the neighboring supercells,
\begin{equation}
 \langle a_i a_j \rangle   =  \Psi_{a(ij)};
\label{aa}
\end{equation}
and, for b-states belonging to the same stripe element,
\begin{equation}
\langle b_{ij-} b_{ij+} \rangle  =  \Psi_{b(ij)}  e^{i \phi_{ab}}, 
\label{b+b+}
\end{equation}
where $\Psi_{a(ij)}$ and $\Psi_{b(ij)}$ are 
the {\it equilibrium} values of the 
two SC components given, in Case I, by Eqs.(\ref{Psib1I},\ref{Psia1I}), and,
in Case II, by
Eqs.(\ref{Psia1II},\ref{Psib1II}).

The averaging of operator (\ref{Jabi}) with the subsequent substitution 
of Eqs.(\ref{aa},\ref{b+b+})  gives the internal supercurrent:
\begin{equation}
\langle J_{ab(i)} \rangle = - {2 \ g \ \hbox{sin} \phi_{ab} \over \hbar} \ 
\sum_{j(i)} \Psi_{b(ij)}^* \Psi_{a(ij)}.
\label{sJabi}
\end{equation}
Each term in the above sum is a real number given by Eq.(\ref{Eint}).
(Spatially homogeneous internal supercurrent (\ref{sJabi})
can exist, when the particle density oscillates between 
a- and b-states.)

The translational supercurrent emerges, when $\phi_{ab}$ 
becomes position-dependent [to be denoted as $\phi_{ab}({\mathbf{r}}_{ij})$].
In this case, the relevant Bogoliubov transformation for b-states is
\begin{eqnarray}
b_{ij+} = s B_{ij+}  + w \ e^{i [\varphi_{ij} + \phi_{ab}({\mathbf{r}}_{ij})]} \ B^+_{ij-};
\label{bB+4}
\\
b_{ij-} = s B_{ij-}  - w \ e^{i [\varphi_{ij} + \phi_{ab}({\mathbf{r}}_{ij})]} \ B^+_{ij+},
\label{bB-4}
\end{eqnarray}
The transformation for a-states is still given by 
Eqs.(\ref{aeA}, \ref{aoA}).
The averaging of Eq.(\ref{Jti}) under the assumption 
that phases $\phi_{ab}({\mathbf{r}}_{ij})$ are small and have weak 
positional dependence, gives the following expression for 
the translational supercurrent:
\begin{equation}
\langle {\mathbf{J}}^t_i \rangle = - { g l  \over \hbar} 
\Psi_{b(ij)}^* \Psi_{a(ij)}
\nabla \phi_{ab}
\label{sJti}
\end{equation}
Note: according to Eq.(\ref{Eint}), the value of the product 
$\Psi_{b(ij)}^* \Psi_{a(ij)}$
is independent of  the orientation of the stripe element labelled by indices $ij$.

From Eq.(\ref{sJti}), the supercurrent density can be obtained as
\begin{equation}
{\mathbf{j}} = {e \over l z_0}  \langle {\mathbf{J}}^t_i \rangle
= S_{\phi} \nabla \phi_{ab},
\label{j}
\end{equation}
where 
\begin{equation}
S_{\phi} = - { e g \over \hbar z_0}  \Psi_{b(ij)}^* \Psi_{a(ij)} ,
\label{Sphi}
\end{equation}
$z_0$ is the transverse distance per one SC plane, and
$e$ the charge of electron. As follows from
Eq.(\ref{j}), the unconventional feature of the present model
is that the supercurrent is induced not by the phase gradients
of $\Psi_a$ and $\Psi_b$ separately, but by the gradient of
the phase difference between $\Psi_a$ and $\Psi_b$. Keeping up
with convention in the literature, I will refer
to the phase stiffness $S_{\phi}$ as ``superfluid density,''
but the intuitive associations with some kind of real density
would be misleading in this case. 

In Case I, the substitution of Eqs.(\ref{Psib1I}, \ref{Psia1I}) into
Eq.(\ref{Sphi}) gives 
\begin{equation}
S_{\phi} = {e g^2 (1 - 2 n_B)^2 \over 16 N \hbar z_0} 
\sum_{\mathbf{k}} 
{(1 - 2 n_A({\mathbf{k}})) |V({\mathbf{k}})|^2 \over 
\varepsilon_A({\mathbf{k}}) },
\label{SphiI}
\end{equation}
and, in Case II,  the analogous equations (\ref{Psia1II}, \ref{Psib1II}) lead to
\begin{equation}
S_{\phi} = {e g^2 (1 - 2 n_B) C_a^2 \over 32 \hbar z_0 \varepsilon_B }.
\label{SphiII}
\end{equation}

\

Now I calculate the in-plane penetration depth $\lambda$
of magnetic field directed perpendicularly to the SC planes.

The natural expectation is that the gauge-invariant generalization of Eq.(\ref{j})
accounting for the presence of the vector potential of electromagnetic
field ${\mathbf{A}}$
has form  
\begin{equation}
{\mathbf{j}} 
= S_{\phi} \left( \nabla \phi_{ab} - {2 e \over \hbar c} {\mathbf{A}} \right),
\label{jA}
\end{equation}
where  $c$ is the speed of light. 
In the present work, I do not derive Eq.(\ref{jA}) but take it as an additional
postulate.

The equivalent of  the London
limit in the present model is $\lambda  \gg l$. 
In this limit, 
the standard result\cite{LP} for the penetration depth,
which follows from Eq.(\ref{jA}), is
\begin{equation}
\lambda = \sqrt{{\hbar c^2 \over 8 \pi e S_{\phi}}}.
\label{lambdaab}
\end{equation}
For the numbers relevant to high-$T_c$ cuprates,
the value of $S_{\phi}$ is very small, i.e. $\lambda$ is
large, and, therefore,  the London limit
is well fulfilled
(see the estimate in the end of subsection~\ref{critical}).

The vector potential entering Eq.(\ref{jA}) should be interpreted 
as describing
magnetic field averaged over a large number of supercells.
It is, therefore, possible, that, on the scale of $l$,
the true magnetic field fluctuates around the exponentially
decaying penetration profile characterized by $\lambda$.

\

\begin{figure} \setlength{\unitlength}{0.1cm}

\begin{picture}(50, 70) 
{ 
\put(-13, 0)
{ \epsfxsize= 2.7in
\epsfbox{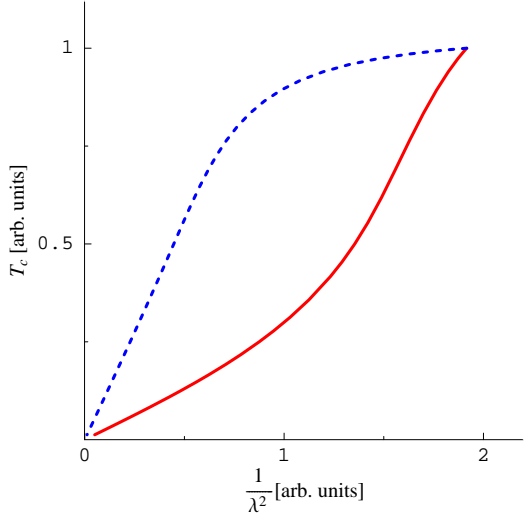} 
}}
\end{picture} 
\caption{(color online) Critical temperature vs. superfluid density 
(represented as $1/\lambda^2$) at $T=0$:
solid line --- Case I;  dashed line --- Case~II. The plot points are calculated as
described in the text.
} 
\label{fig-fish} 
\end{figure}

The relationship between $T_c$ and the zero-temperature
superfluid density (represented as $1/\lambda^2$) is plotted in 
Fig.~\ref{fig-fish}. 
The points in this plot were obtained by 
fixing the value of the interaction constant $g$ and then varying
$\varepsilon_a$ (in Case I) or $\varepsilon_b$ (in Case II) from zero to very large values.
The $T_c$-coordinate of each plot point was obtained by taking a pair
of values $(\varepsilon_a, g)$ or $(\varepsilon_b, g)$ and then
solving numerically Eq.(\ref{Tceq}) or 
(\ref{TceqII}) for Cases I and II, respectively. The 
$1/\lambda^2$-coordinate was obtained for the same values of 
$(\varepsilon_a, g)$ or $(\varepsilon_b, g)$ by calculating
$S_{\phi}$ according to Eq.(\ref{SphiI}) or (\ref{SphiII}).
The theoretical plot of Fig.~\ref{fig-fish} 
is compared with experiments in Section~\ref{superfluid}.


\subsection{Critical case: $\varepsilon_a = \varepsilon_b = 0$}
\label{critical}

In the critical case, 
all the formulas obtained earlier have very
simple form. 
In particular, the SC transition temperature is given by
Eq.(\ref{Tc1}), which is rewritten here as
\begin{equation}
g = 4 T_c.
\label{Tccr}
\end{equation}
Substituting Eq.(\ref{Tccr}) and $\varepsilon_a = 0$  
into Eqs.(\ref{epsAkI},\ref{epsBI},\ref{EI}, \ref{epsA0I},\ref{SphiI})
and using Eq.(\ref{Ca0}) when necessary, 
the zero-temperature values of several key quantities can be expressed as
follows:
\begin{equation}
\varepsilon_{A0} = 4 T_c;
\label{epsA0cr}
\end{equation}
\begin{equation}
\varepsilon_B =  C_{a0} T_c;
\label{epsBcr}
\end{equation}
\begin{equation}
E_{\hbox{\sc \scriptsize gs}} =  - 2 C_{a0} T_c N;
\label{EGScr}
\end{equation}
and, finally, 
\begin{equation}
\lambda = { \hbar c \over 2 e } \sqrt{{z_0 \over \pi C_{a0} T_c}}.
\label{lambdacr}
\end{equation}
The tunneling spectrum for the critical case is shown in Fig.~\ref{fig-tunnel}(b).

I now assume that
one model layer represents one copper-oxide layer of a real compound,
which allows me to express 
the condensation energy per one in-plane copper atom as
\begin{equation}
U_0 = 
{|E_{\hbox{\sc \scriptsize gs}}| a_0^2 \over N l^2 } =  
{2 C_{a0} T_c a_0^2 \over l^2 }
\label{U0cr}
\end{equation}

Substituting $T_c = 90$~K, $l = 4 a_0 \sqrt{2} \approx 23$\AA \ and 
$z_0 = 6$\AA \  into Eqs.(\ref{epsA0cr},\ref{epsBcr},\ref{lambdacr},\ref{U0cr})
and recalling that $C_{a0} = 0.958...$,
I obtain: $\varepsilon_{A0} = 31$~meV,
$\varepsilon_B =7.4$meV, $U_0 = 5.4$~K and $\lambda = 417$~nm.

\section{Discussion}
\label{discussion}

\subsection{Realistic features of the model}
\label{realistic}

In this subsection I list the features, which I expect
will survive the adaptation of the above model to the the
properties of real materials, if the 2D diagonal stripe
hypothesis turns out to be correct.

The basic underlying feature of this model --- the 
potential background, which localizes both a- and b-states, --- should
survive the generalizations. In particular, the actual shape of the
AF domains can be quite distorted and thus noticeably
different from the perfect diamond shape
drawn in Fig.~\ref{fig1}. It is only important that the stripes divide
the plane into finite AF domains with the alternating sign
of the AF order parameter. 
In the model, the disorder in the shape of  AF domains
can be accommodated through the disorder
in the values of $\varepsilon_a$, $\varepsilon_b$ and $g$.

Other supposedly realistic features include the coherence length
of the order of  $l$ and
the decrease of the critical temperature 
with the increase of $|\varepsilon_a - \varepsilon_b|$,
which is associated with the pseudogap. 

Finally, the very unconventional translational symmetry
of the SC order parameter  described in Section~\ref{odlro}
should also survive generalizations.

\subsection{Phase diagram}
\label{phase}

\begin{figure} \setlength{\unitlength}{0.1cm}

\begin{picture}(50, 56) 
{ 
\put(-21, 0)
{ \epsfxsize= 3.5in
\epsfbox{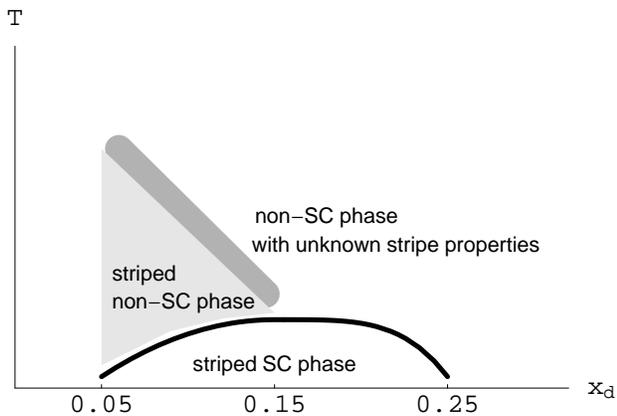} 
}}
\end{picture} 
\caption{Cuprate phase diagram proposed in the text.} 
\label{fig5} 
\end{figure}

In this part, I give the description of the phase diagram of
high-$T_c$ cuprates, which is based, in part, on
the SC model of Section~\ref{model}, and, in another part, 
on a few facts borrowed
from the next Section, where the model predictions are 
compared with experiments. 
The phase diagram itself is shown in Fig.~\ref{fig5}.


I start at doping concentration $x_d \approx 0.06$. From the viewpoint
of the hypothesis adopted in this work, this is the lowest concentration at
which the 2D stripe pattern stabilizes by virtue of some unknown 
energy balance. At this concentration, according 
to the description given in Sections~\ref{2D}, \ref{potential}
and \ref{model}: 
(i)the length of the stripe
supercell can be estimated as $l=1/(2x_d) \approx 8$ lattice sites;
(ii) $\varepsilon_a > \varepsilon_b$
(iii) the value of the pseudogap $\varepsilon_a - \varepsilon_b$  is maximal;  
and (iv) the superconducting transition 
temperature is minimal. 

At higher doping concentrations, the size of the AF domains
decreases, eventually saturating at $l \approx 4 a_0 \sqrt{2}$.
Simultaneously, the pseudogap $\varepsilon_a - \varepsilon_b$ 
also decreases. The rationale for the latter assertion is that, 
at some threshold  doping concentration, stripes should disappear, 
which implies that holes should be expelled from them.
According to Eq.(\ref{Tceq}) or (\ref{TceqII}), the decrease of the pseudogap
is accompanied by the increase of  
the critical temperature.

The interaction strength $g$  should also change with the increase of doping.
It is not obvious {\it a priori} whether
it increases or decreases, but, at least,
it is likely that the value of $g$ never
approaches zero, while the value of
$\varepsilon_a - \varepsilon_b$ 
either reaches zero at some critical doping or becomes
very small. This suggests that 
the relative change of 
$\varepsilon_a - \varepsilon_b$ across the SC doping range has
stronger effect on the observable quantities 
than the relative change of $g$.

If the value of $g$ were independent of the doping concentration, 
then the maximum of $T_c$ would be achieved at a critical doping
corresponding to  $\varepsilon_a - \varepsilon_b = 0$. However,
because of the presumed dependence of $g$ on the doping level, the
optimal doping concentration, $x_{d0}$,
corresponding to the actual maximum of $T_c$ can 
be slightly shifted with respect to the critical one. 
The comparison with  experiments in Section~\ref{superfluid} indicates that 
the above notion of the critical doping coincides 
with the ``critical doping'' 
$x_{dC} \approx 0.19$ identified
experimentally in Refs.~\cite{Bernhard-etal-01,Loram-etal-01,Tallon-etal-03}.
As expected,  $x_{d0} \approx 0.16$ is not much different from $x_{dC}$.
The fact that $x_{d0} < x_{dC}$ suggests that $g$ decreases
as the doping level increases.

Below the critical doping, the inequality $\varepsilon_a > \varepsilon_b$ implies
that
the real materials 
can be describable by the model either as Case~IA or as Case~IIB. 
(See Section~\ref{correspondence} for the discussion of this issue).

Above the critical doping, it is, in principle, 
possible that
either $\varepsilon_a - \varepsilon_b$ stays equal to zero,
while only $g$ changes, or $\varepsilon_a - \varepsilon_b$
becomes negative, which means that 
the absolute value of the ``inverted'' pseudogap
starts growing again and thus additionally suppresses 
the critical temperature. The analysis of experiments
in Section~\ref{superfluid} favors the inverted pseudogap scenario
and, given the model choice between Cases IB and IIA,
clearly points to Case~IB.
As discussed in Section~\ref{potential}, 
the stripe superstructure may, for a while, remain stable
even after  $\varepsilon_b$ becomes greater than 
$\varepsilon_a$.
It is further possible that the SC transition can contribute 
to the stabilization of
stripes by lowering 
the total energy of the stripe phase. 

In the context of the present proposal, stripes should exist in the SC phases
of both underdoped and overdoped cuprates. It is, however, unclear, whether,
in overdoped cuprates, stripes  can be stabilized
without the SC transition. Negative answer to this question
would imply that above the SC transition, overdoped cuprates
enter a
stripeless normal state. 
Otherwise, the normal state of overdoped cuprates may still exhibit 
some kind of stripe order. 
In turn, if  overdoped cuprates
enter stripeless phase simultaneously with the SC transition, 
then this important aspect cannot be captured 
by the model of Section~\ref{model}.
In such a case, the model scenario for the overdoped
situations  becomes doubtful.

The conceptual difference between the above description and the popular idea
of competing orders
is that, in the present proposal, the two orders are not competing, 
but, on the contrary, cooperating: the stripe order 
is crucial for the existence of the SC
transition, while  the SC transition can also help stabilizing the stripe
order.

\section{Comparison with experiments}
\label{experimental}

\subsection{Qualitative aspects}

The primary concern in the context of the present proposal is that no evidence
of stripes has been observed so far in most optimally doped and overdoped
materials.  It should be pointed out, however, that this proposal
stipulates that strong transverse fluctuations of stripes mediate
superconductivity.  If true, this would imply that any attempt to observe
stripes by  pinning charges or freezing spins, 
would suppress their transverse fluctuations,
and thus suppress the SC transition or, at least, significantly reduce the 
critical temperature.  
Such an inverse relation between the amplitude
of the stripe fluctuations and $T_c$ can explain
why the stripes are best observable in cuprates having lower $T_c$,
such as underdoped\cite{Yamada-etal-98,Kimura-etal-99}, 
Nd-doped\cite{Tranquada-etal-97} and 
Zn-doped\cite{Hirota-etal-98,Kimura-etal-99,Hirota-01} LSCO.
The transverse fluctuations of stripes should be strongly coupled to the lattice.
Therefore, some kind of isotope effect should also be present in such a system.

The proper treatment of the single particle 
excitations in the non-superconducting (normal) 2D stripe phase 
is not developed in this work. 
It is, however, difficult not to see
that the description of the normal state pseudogap given in Section~\ref{potential}, 
while following only from the basic facts about the 2D stripe geometry,
bears a strong resemblance of the experimental facts\cite{DHS}.
In particular, the disappearance of the pseudogap  in the diagonal 
crystal directions can be naturally explained by the presence of holes inside the
diagonal stripes. In the model framework, the absolute value of
the pseudogap $|\varepsilon_a - \varepsilon_b|$ constitutes the primary factor suppressing
the SC transition (see Eqs.(\ref{Tceq},\ref{TceqII})).

\

In the SC state, the model predicts such distinctive properties as
(i) the suppression of $T_c$ with the growth of the
pseudogap $|\varepsilon_a - \varepsilon_b|$
(see Eqs.(\ref{Tceq},\ref{TceqII}));
(ii) the emergence of quasiparticles having coherent dispersion 
in $k$-space only at $T<T_c$
(see the remark following Eq.(\ref{uvknorm})); 
(iii) the asymmetry of the tunneling density of states 
(Fig.~\ref{fig-tunnel}a); (iv) linear
density of states around the chemical potential (Fig.~\ref{fig-tunnel}b,c); 
and (v) low superfluid density (Section~\ref{supercurrent}).
In the following subsections (\ref{tunneling}-\ref{superfluid}), 
I show that some of the quantitative 
model predictions made without adjustable parameters
also agree with experiments.

\subsection{Tunneling characteristics}
\label{tunneling}

In this subsection I compare the theoretical
tunneling spectra at $T=0$ with experimental tunneling spectra at 
$T \ll T_c$. Therefore, the discussion will imply 
the zero- temperature values  of all relevant quantities.

The model predicts two kinds of contributions to the tunneling density 
of states corresponding to A- and B- Bogoliubov quasiparticles (see Fig.~\ref{fig-tunnel}). 
If the Van Hove singularities at 
$\pm \varepsilon_{A0}$   and the 
delta-function peaks at $\pm \varepsilon_B$ exist at all, 
they should be identifiable in experimental data.

The difficulty now is that the tunneling 
spectra of high-$T_c$ cuprates have, in
general, only one prominent feature, namely, two SC peaks at the opposite
values of the bias.
The energies of the SC peaks are, usually, denoted as $\pm \Delta$,
and $\Delta$ is referred to as the SC gap.  
In the superconductivity model of Bardeen, Cooper
and Schrieffer (BCS), $\Delta/T_c = 1.76 $. In high-$T_c$ cuprates,
the reported values of  $\Delta/T_c$ 
show significant variations but, typically, 
fall in the range between 1.5 and 7, with 4 being 
a representative value. 
In the present model, the  ratios   
$|\varepsilon_{A0}|/T_c$ and $|\varepsilon_B|/T_c$  can also vary broadly.
Their representative values are 4 and 1, respectively
(see Section~\ref{critical}).
Having observed that $\Delta/T_c \sim |\varepsilon_{A0}|/T_c$, 
I identify the experimental SC peak with 
the Van Hove singularity in the density of A-states. This identification
will later prove to be consistent with a 
number of other qualitative and quantitative facts. 
In the following, the variables $|\varepsilon_{A0}|$  
and $\Delta$ will refer, respectively, to the theoretical and experimental
values of the same quantity.

The identification of B-states
is more problematic. The evidence for their existence
is non-trivial, but largely indirect. It is based on the
STM observations of the checkerboard patters 
in the local density of states (LDOS) 
of Bi-2212\cite{Hoffman-etal-02,Howald-etal-03,Hoffman-etal-02A,Vershinin-etal-04} 
(to be discussed in subsection~\ref{B-states}).
Another possibly related observation is that of asymmetric resonance peaks 
in YBCO by Derro~{\it et al.} The asymmetry and the energy range of
those peaks agree well with the expectations for the delta-peaks 
due to B-states (see Fig.~\ref{fig-tunnel}(c)).  

There are two possible explanations why, in general,
B-states are more difficult to observe experimentally.

The first explanation is that, in a 
real system, the on-site energies $\varepsilon_b$ can be distributed. 
As a result, the spectrum of B-states
may become broad and featureless.  

The second explanation is that the matrix elements for
tunneling into b-states (and hence B-states) 
can be much smaller than those for a-states.
This, in turn, can be related to the fact that b-states are  
localized in the narrow regions inside the stripes,
while a-states spread over the AF domains and thus have a broader
``interface'' with the environment. Alternatively,
it might happen that b-states have exotic quantum numbers,
in which case tunneling into them can be suppressed at all.

In the rest of this subsection,  I assume that B-states are mostly unobservable,
and, unless specified otherwise, the tests of the model will amount 
to the comparison between the density of A-states and the experimental 
tunneling spectra.

\

I limit the choices to the 
special Cases~I and II defined in Section~\ref{classification}. 
Therefore, the model calculation of the density of A-states
only requires the knowledge of  two parameters:
$g$ and $\varepsilon_a$ in Case I, or  $g$ and $\varepsilon_b$ in Case II. 
Using  $T_c$ and $\Delta$ as input parameters,
I can   
both  discriminate between Cases I and II,  and  determine the 
values of $g$, $\varepsilon_a$ and $\varepsilon_b$. 

The inequality $\Delta/T_c > 4 $ can appear only in the
framework of Case I, in which case, $|\varepsilon_a|$ and 
$g$ should be obtained numerically from Eqs.(\ref{Tceq},\ref{epsA0I}) 
(with $n_B=1$ and $\varepsilon_{A0} = \Delta$).
The opposite inequality, $\Delta/ T_c < 4$, can only correspond to Case II,
i.e.  $|\varepsilon_b|$ and 
$g$ have to be obtained from Eqs.(\ref{epsBII},\ref{TceqII},\ref{epsA0II})
(with $n_B=0$, $C_a=C_{a0}$ and $\varepsilon_{A0} = -\Delta$). The situation
$\Delta/T_c = 4 $ corresponds to the critical case described in Section~\ref{critical}.

After the model Hamiltonian is specified,  the following
tunneling characteristics  
can be predicted without adjustable parameters: 
(i) the asymmetry in the density of A-states or the absence thereof,
(ii) the maximum energy of A-states $\varepsilon_{A1}$, and also
(iii) the expected positions $\pm \varepsilon_B$ 
of the delta-peaks representing
B-states.

The selection of Case I implies a strong qualitative prediction of 
asymmetric SC peaks.  However, the knowledge of $\Delta$ and $T_c$ alone
cannot help to discriminate between Cases IA and IB, which correspond to the opposite
``polarities''  of the SC peak  asymmetry . 
However, if Case I is identified in underdoped
cuprates, then the strong expectation is that  $\varepsilon_a > \varepsilon_b$,
which implies Case~IA (larger SC peak at the negative bias). 
The identification of Case~II implies 
that the SC peaks are symmetric. 
(The inequality $\varepsilon_a > \varepsilon_b$ would then
favor Case~IIB for underdoped cuprates.)

\begin{figure} \setlength{\unitlength}{0.1cm}

\begin{picture}(50, 143) 
{ 
\put(-17, 73){ \epsfxsize= 3in \epsfbox{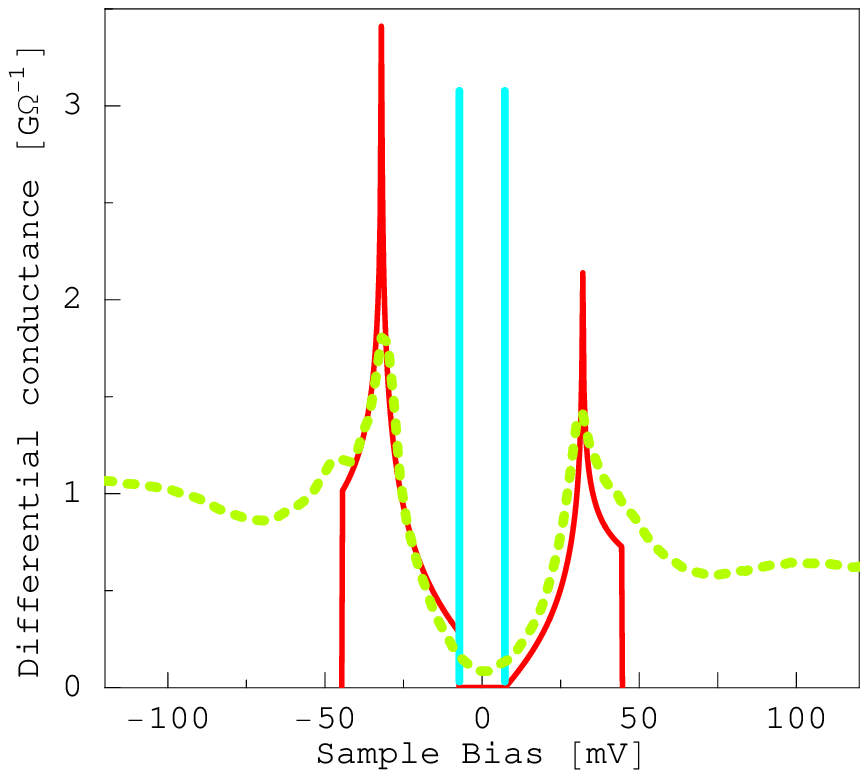} }
\put(-17, 0){ \epsfxsize= 3in \epsfbox{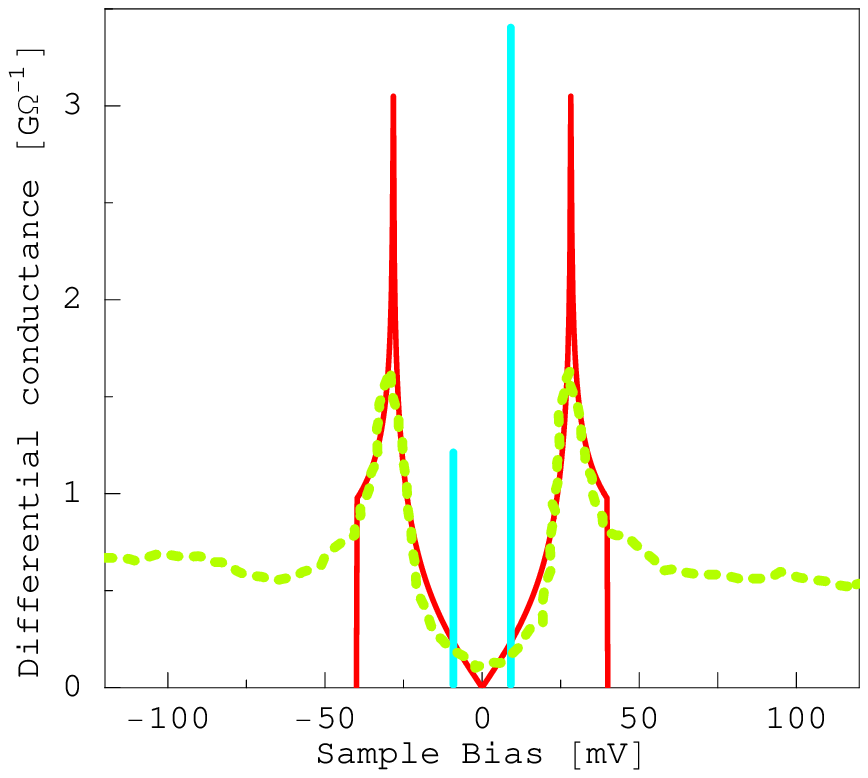} }
\put(-1,133){(a)}
\put(-1,60){(b)}
}
\end{picture} 
\caption{(color online) Comparison between zero-temperature model calculations  and 
low-temperature STM spectra of Bi-2212. 
Solid lines represent the theoretical 
density of states 
in the same way as in Fig.~\ref{fig-tunnel}. Dashed lines represent 
experimental spectra extracted from the following
references: (a) Fig. 2 of Ref.\cite{Pan-etal-00} 
(regular part of Bi-2212  surface); (b) Fig. 7 of Ref.\cite{RF}. 
} 
\label{fig-tunnel1} 
\end{figure}

In
Fig.~\ref{fig-tunnel1}, 
the model calculations are compared with two particularly well
resolved STM spectra of Bi-2212.  The experimental spectrum in
Fig.~\ref{fig-tunnel1}(a) was extracted from Fig.~2 of Ref.~\cite{Pan-etal-00}.
It is representative of ``regular'' parts of the sample surface (i.e.  vortex
free and impurity free). In this case, the experimentally determined numbers
$\Delta = 32$~meV, $T_c = 87$~K and $\Delta/T_c = 4.3$ imply
Case~I with $|\varepsilon_a| = 7.3$~meV and $g= 31$~meV.
The theoretical spectrum presented in Fig.~\ref{fig-tunnel1}(a) corresponds
to $\varepsilon_a > 0$, i.e. to Case IA. 

The experimental spectrum in
Fig.~\ref{fig-tunnel1}(b) was extracted from Fig.~7 of Ref.~\cite{RF}.
The relevant experimental numbers  
$\Delta = 28$~meV, $T_c = 92.3$~K and $\Delta/T_c = 3.5$ 
imply Case II with $|\varepsilon_b| = 4.3$~meV and $g= 32$~meV.
The asymmetry of the theoretical 
delta-peaks in Fig.~\ref{fig-tunnel1}(b)  corresponds 
to $\varepsilon_b < 0$, i.e. to Case~IIB. The vertical scale of 
the theoretical plots in Fig.~\ref{fig-tunnel1} was chosen to fit best
the experimental data.


It thus appears, that not only the model gives the correct prediction
of the presence or  the absence of the SC peak asymmetry, but also
it predicts the degree of asymmetry quantitatively. One should
also note  that the termination points
of the theoretical spectra  have 
experimental counterparts in the form of
the shoulder-like features located approximately at 
energies $\pm \varepsilon_{A1}$ predicted theoretically. 
Finally, the  delta-peaks shown in Fig.~\ref{fig-tunnel1}(a)
are located at $\pm \varepsilon_B = \pm 7.3$~meV. 
These are precisely the energies, 
at which, in Ref.~\cite{Pan-etal-00}, 
the vortex cores have shown anomalous {\it symmetric} ``humps'' 
absent in the regular SC regions.
A subsequent study\cite{Hoffman-etal-02} 
has revealed that 

The local density of states (LDOS)
associated with the above humps was later shown to exhibit a checkerboard
pattern inside the vortex cores\cite{Hoffman-etal-02}. 
In subsection~\ref{B-states}, I will 
show that this 
pattern is precisely what one should expect from 
LDOS associated with B-states.

\

Now I discuss to what extent the model-based rule
\begin{equation}
\begin{array}{rcl}
\hbox{``{\it $\ \Delta/T_c > 4 \  \Leftrightarrow  \ $ asymmetric SC peaks;}}
\\ 
\hbox{{\it $\ \ \Delta/T_c \leq 4 \  \Leftrightarrow  \ $ symmetric SC peaks}\ ''}
\end{array}
\label{rule}
\end{equation}
is supported by other STM or point-contact 
superconductor---insulator---normal-metal 
(S-I-N) experiments. 
The values of $\Delta$ used below are obtained as half of the difference
between the energies of the SC peak maxima.   

\

{\it Supporting evidence:}

Clearly asymmetric SC peaks corresponding to $\Delta/T_c > 4$
have been reported for bi-layer compounds

\noindent Bi-2212 
in Refs.\cite{Koltun-etal-91,Koltun-etal-91-comment,DeWilde-etal-98,Renner-etal-98,Renner-etal-98A,Pan-etal-00,Misra-etal-02,McElroy-etal-04}, 

\noindent HgBa$_2$CaCu$_2$O$_{6+\delta}$ (Hg-1212) in 
Ref.\cite{Wei-etal-98}; 

\noindent and tri-layer compounds

\noindent Bi$_2$Sr$_2$Ca$_2$Cu$_3$O$_{10+\delta}$ (Bi-2223) in 
Ref.\cite{Chen-etal-94}, 

\noindent HgBa$_2$Ca$_2$Cu$_3$O$_{8+\delta}$ (Hg-1223) in 
Ref.\cite{Jeong-etal-94,Wei-etal-98}.

\noindent The values of $\Delta/T_c $ extracted from the above references
cover the range between 4.1 and 6.9~.

 Symmetric SC peaks corresponding to $\Delta/T_c \leq 4$
have been reported for single layer compounds 

\noindent HgBa$_2$CuO$_{4+\delta}$ (Hg-1201) in 
Ref.\cite{Chen-etal-94,Wei-etal-98},

\noindent Tl$_2$Ba$_2$CuO$_6$ (Tl-2201) in 
Ref.\cite{Ozyuzer-etal-98}; 

\noindent and bi-layer compounds 

\noindent Bi-2212 in 
Refs.\cite{Koltun-etal-91,RF}, also Pb-doped Bi-2212 in Ref.\cite{Huang-etal-89A},

\noindent YBCO in 
Ref.\cite{Maggio-Aprile-etal-95},

\noindent Tl$_2$Ba$_2$CaCu$_2$O$_x$ (Tl-2212) in 
Ref.\cite{Huang-etal-89}.

\noindent The values of $\Delta/T_c $ extracted from the above references
cover the range between 1.7 and 3.9.

\

{\it Contradicting evidence:} 

Symmetric  SC peaks corresponding 
to $\Delta/T_c > 4$
have been reported 
for 

\noindent Bi-2212 in Refs.\cite{MA,RF,Renner-etal-98,Renner-etal-98A}.

Asymmetric SC peaks  corresponding to $\Delta/T_c \leq 4$
have been reported for 

\noindent LSCO in Ref.\cite{MOA}, 

\noindent Tl-2201 in Ref.\cite{Ozyuzer-etal-98}

\noindent Tl-2212  in Ref.\cite{Huang-etal-89}. 

\noindent However, in the case of LSCO, the difference between the peak heights was 
certainly within the limits of  experimental uncertainty. 
In the case of Tl-2201 and Tl-2212, 
the overall impression from
the cited references is that  the SC peaks are largely symmetric.
(Most of the spectra reported in the same references and measured on similar junctions
pass as symmetric and contribute to the ``supporting evidence.'')

\

Unlike the two spectra shown in Fig.~\ref{fig-tunnel1}, most of 
the measured tunneling spectra 
have more rounded SC peaks, which may be the consequence of limited 
experimental resolution. Since
a significant 
broadening of a SC peak  also shifts
the position of its maximum, it is possible
that, a measured symmetric spectrum indicates the ratio $\Delta/T_c $
greater than 4, while 
the true ratio $\Delta/T_c $ is slightly smaller than 4. 
(One such an example is given in Ref.\cite{RF}.) 
The resolution-limited broadening
of the SC peaks can, therefore, be responsible for, at least a part
of the ``contradicting evidence'' in Bi-2212.

Taken as a whole, the above review of experimental data
clearly supports
the rule (\ref{rule}).
Furthermore, this  rule (\ref{rule}) seems to unify 
the experimental data, which, otherwise, may appear contradicting 
to each other. (I will return to this issue in Section~\ref{correspondence}.)

Another interesting fact is that, in the references cited above,
the SC peak asymmetry of the bi-layer compounds is opposite to that
of the tri-layer
compounds:  the bi-layer compounds have the higher SC peak 
mostly at the negative
bias (as in Fig.~\ref{fig-tunnel1}(a)), while the tri-layer 
compounds have the higher peak at the positive bias.

Rule~(\ref{rule}) can be compared with a more simple prediction
made by Altman and Auerbach\cite{AA}, that the asymmetry of the kind shown
in Fig.~\ref{fig-tunnel1}(a) is inherent in all high-$T_c$ cuprates.
The lack of the asymmetry in some of the tunneling spectra 
and the opposite asymmetry of the tri-layer compounds would contradict
to the above prediction.

\

Finally, one additional clear prediction of the model is that  
the inequality $\Delta/T_c > 4$
implies that, as $T$ approaches $T_c$,
the energy of the SC peaks approaches the finite value $|\varepsilon_a|$
(see Eq.(\ref{epsA0I})). The inequality  $\Delta/ T_c < 4$ implies 
the zero energy of the SC peaks at $T=T_c$ (see Eq.(\ref{epsA0II})).
The above prediction is difficult to test,
because the SC peaks tend to be totally ``washed out'' in the 
vicinity of $T_c$. Nevertheless, one can observe that
the first part of this prediction is consistent 
with the trend in the tunneling
data from 
Refs.\cite{Renner-etal-98,Miyakawa-etal-98,SW,Krasnov-02,Yamada-etal-03}.
The second part is more difficult to test, but
it also appears to be consistent with the results reported in 
Refs.\cite{Vedeneev-etal-94,Krasnov-02,Krasnov-02A}, though the results from
Refs.\cite{SW,Yamada-etal-03} leave either ambiguous or the opposite
impression.

\subsection{B-states and the checkerboard patterns observed by STM}
\label{B-states}

The only experimental evidence, which, at the moment, I can  identify
with B-states is the checkerboard modulation of the LDOS observed by STM in
the vortex cores\cite{Hoffman-etal-02}.  
The analogous modulations observed in the normal
state of Bi-2212\cite{Vershinin-etal-04} can  be attributed to  b-states.
(For the alternative interpretations of  these (and other) checkerboard patterns   
see  
Refs.\cite{ZMB,PVS,Chen-etal-02,Hoffman-etal-02A,Howald-etal-03,WL,Alexandrov-03,Kivelson-etal-03,Chen-etal-04}).

\begin{figure} \setlength{\unitlength}{0.1cm}

\begin{picture}(50, 65) 
{ 
\put(-13, 0)
{ \epsfxsize= 2.5in
\epsfbox{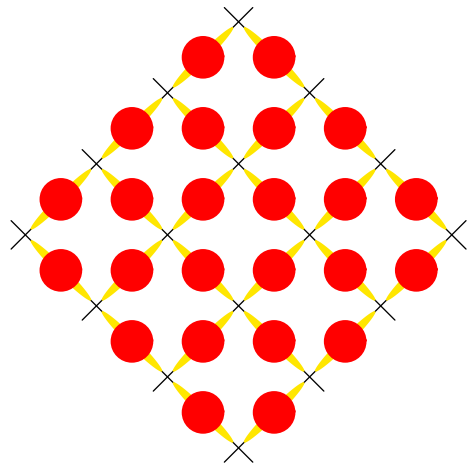} 
}}
\end{picture} 
\caption{(color online) Checkerboard pattern formed by the centers of stripe elements.
Circles indicate the regions, where the densities
of B-states in the SC phase or b-states in the normal phase are expected to be
particularly high.
} 
\label{fig-checker} 
\end{figure}

The checkerboard modulations  revealed by the above experiments have 
periodicity of approximately 4 lattice periods along the principal lattice directions.
From the viewpoint of 
the present proposal,  this periodicity is due 
to the response from B-states in the SC state
or b-states in the normal state. 
These model states are localized around the centers of the
stripe elements, which, as shown in Fig.~\ref{fig-checker}, form a pattern with the required
orientation of the ``checkerboard''. 


As a corroborating evidence, I can mention that,
in the vortex core experiments\cite{Pan-etal-00, Hoffman-etal-02}, 
the LDOS modulation was most pronounced at energies  
approximately equal to $\pm 7$~meV, in almost exact  correspondence with 
$\pm \varepsilon_B$ obtained in Section~\ref{tunneling} just from
the knowledge of $\Delta$ and $T_c$ for the spectrum  presented in Fig.~\ref{fig-tunnel1}(a). 
In the normal state experiment\cite{Vershinin-etal-04}, 
the typical range of energies, where the modulations were most pronounced,
was also consistent with the possible energies of b-states.

In regular SC regions of 
nearly optimally doped B-2212, a different kind of
LDOS modulations showing strong energy dependence
has also been observed\cite{Howald-etal-03,Hoffman-etal-02A}.
From the view point of the present proposal, 
this energy dependence can be caused by two factors:
(i) the crossover from the pattern corresponding to B-states
to the pattern corresponding to A-states; and (ii) the
defect-induced interference of 
A-quasiparticles. 
The first factor can be appreciated after one observes
that the spatial patterns of A-states and B-states are characterized by different sets of
wave vectors. The pattern of A-states has diagonal periods   
$l$ coinciding with
that of the underlying stripe superstructure, while the periods  of 
B-states (shown in Fig~\ref{fig-checker}) are equal to $l/\sqrt{2}$
and oriented 
along the principal lattice directions.
Therefore, as the energy probed by STM increases, the pattern representative of B-states
gradually transforms into the pattern representative of A-states, hence
the energy dependence of the characteristic wave vectors.
The description of the second (interference) factor would require quantitative
analysis extending beyond the scope of the present work. 
(The idea, that a different kind of the quasiparticle interference can entirely
explain the energy-dependent modulation patterns, was  advocated  
in Refs.\cite{Hoffman-etal-02A,WL}.

Since, in 
the 2D diagonal stripe picture, the experimentally observed
checkerboard periodicity of $4a_0$ implies 
the true underlying period $l = 4 a_0 \sqrt{2}$ along the
diagonal directions, a direct test for the existence of the 
diagonal superstructure can consist of  reconstructing the 
position of hypothetical diagonal stripes
from the knowledge of the 
checkerboard pattern at lower energies, 
and then checking whether the
LDOS modulation at higher energies has more
pronounced correlations between the (approximately) equivalent
positions belonging to different supercells.

Finally, the in-stripe hole content
corresponding to the present interpretation can be estimated by  
substituting  $x_d \approx 0.16$ and $f = a_0\sqrt{2}/l = 1/4$, 
into Eq.(\ref{c}), which gives 
$c = x_d/f= 0.6$. This number is only slightly greater than 0.5 extracted in
Section~\ref{2D}  from the experimentally observed INS peak
splitting in underdoped LSCO
compounds.

\subsection{Superfluid density}
\label{superfluid}

In the model framework,
the calculation of  the critical temperature $T_c$ and 
the superfluid density, $S_{\phi}$,
requires the knowledge of three numbers: $|\varepsilon_a - \varepsilon_b|$,
$g$ and the prefactor of $S_{\phi}$.
In general, one should expect that
both $|\varepsilon_a - \varepsilon_b|$ and $g$ 
change as functions of doping concentration. However, since 
$|\varepsilon_a - \varepsilon_b|$, presumably, approaches zero
not far from the optimal doping, the relative effect of this change 
on the observable quantities 
should be stronger than the effect of the change of $g$.
Therefore, for a given family of high-$T_c$ cuprates, 
one can obtain an approximate relation between $S_{\phi}$
and $T_c$ by fixing the value of $g$ and then varying
$\varepsilon_a$ in Case~I or $\varepsilon_b$ in Case~II. 
The two theoretical
curves shown in Fig.~\ref{fig-fish} were obtained  precisely in this way.

\begin{figure} \setlength{\unitlength}{0.1cm}

\begin{picture}(50, 205) 
{ 
\put(-24, 0) { \epsfxsize= 1.9in \epsfbox{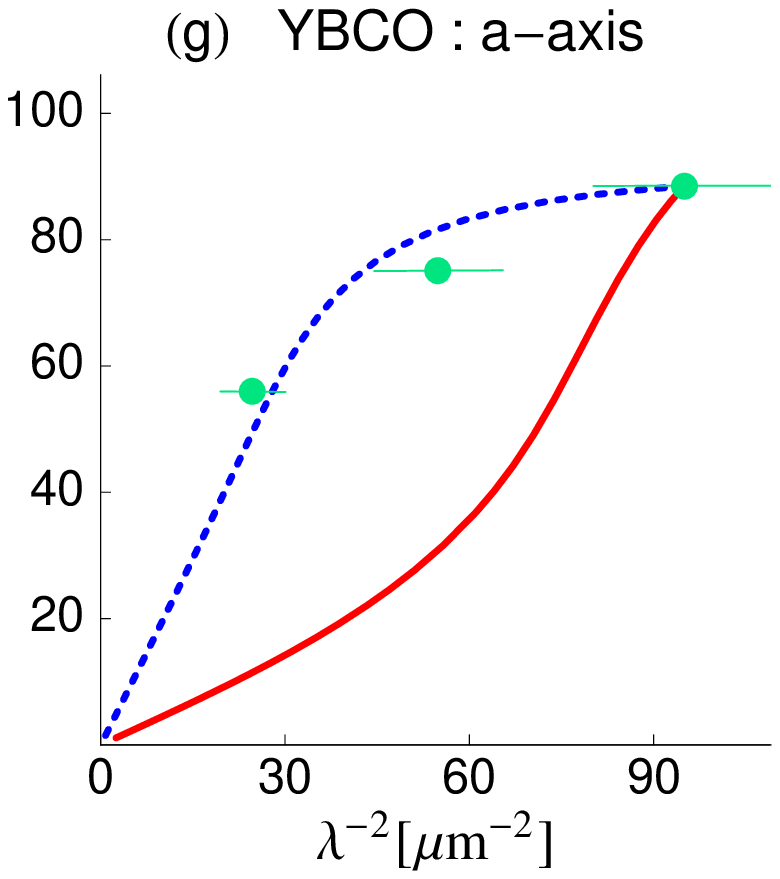}  }
\put(20, 0) { \epsfxsize= 1.9in \epsfbox{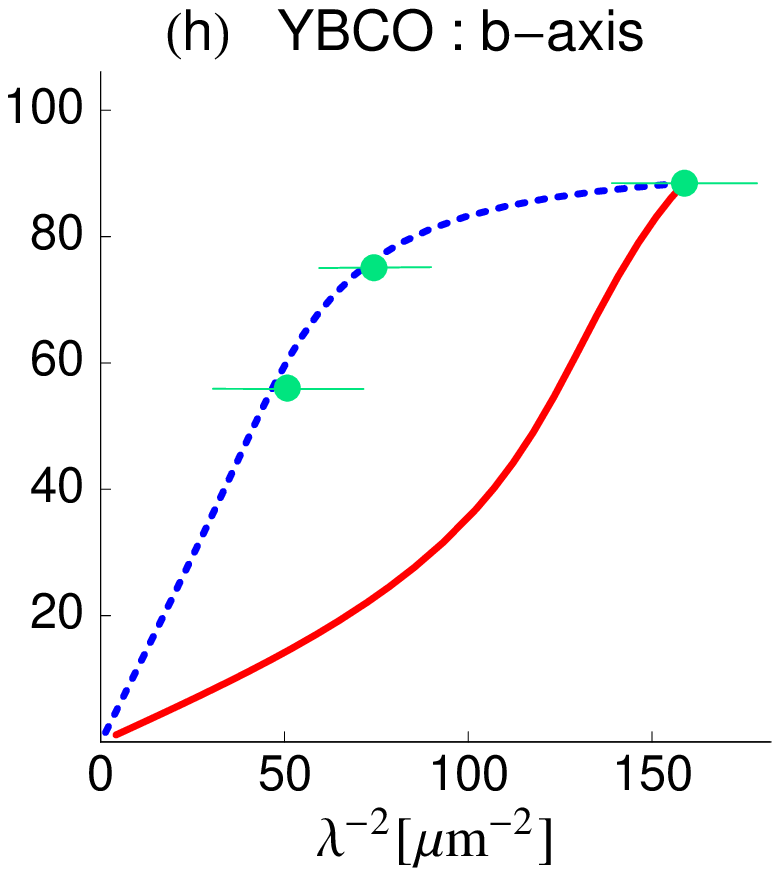}  }
\put(-24, 52) { \epsfxsize= 1.9in \epsfbox{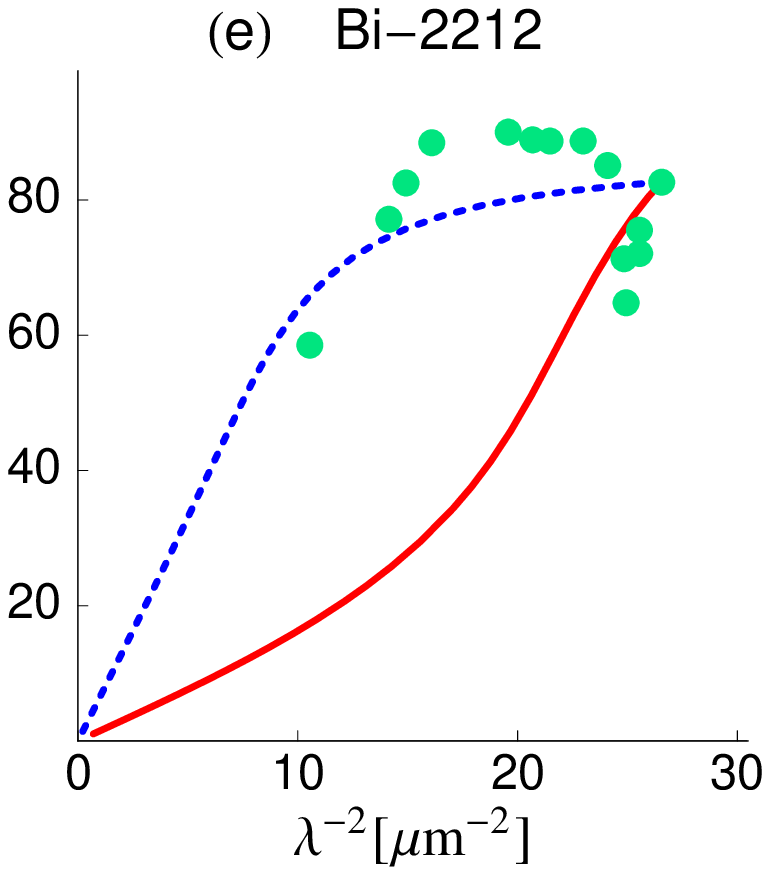}  }
\put(20, 52) { \epsfxsize= 1.9in \epsfbox{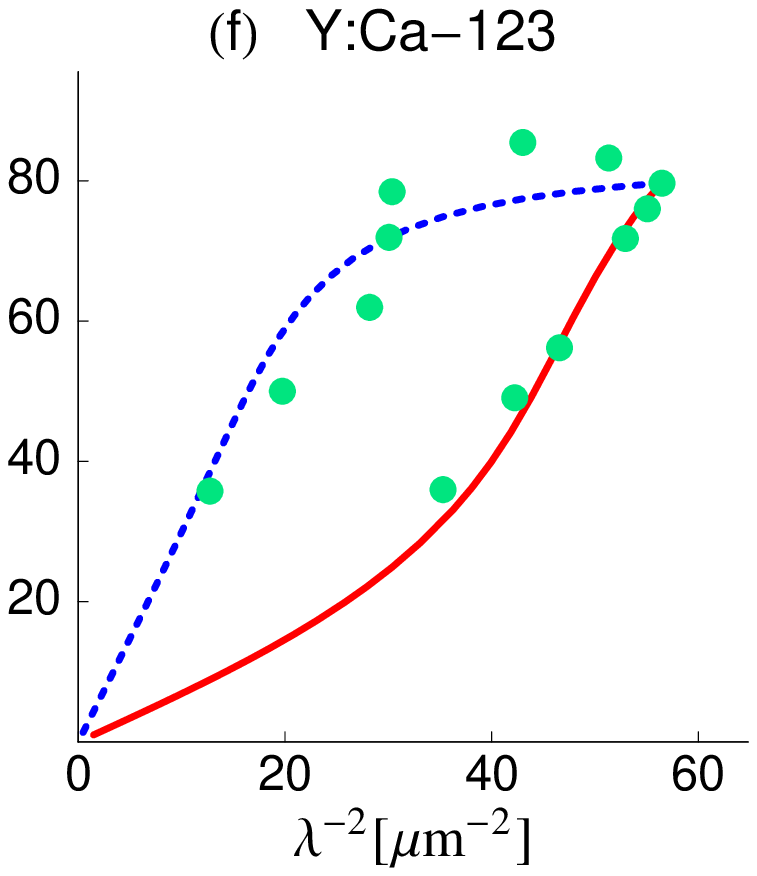}  }
\put(-24, 104) { \epsfxsize= 1.9in \epsfbox{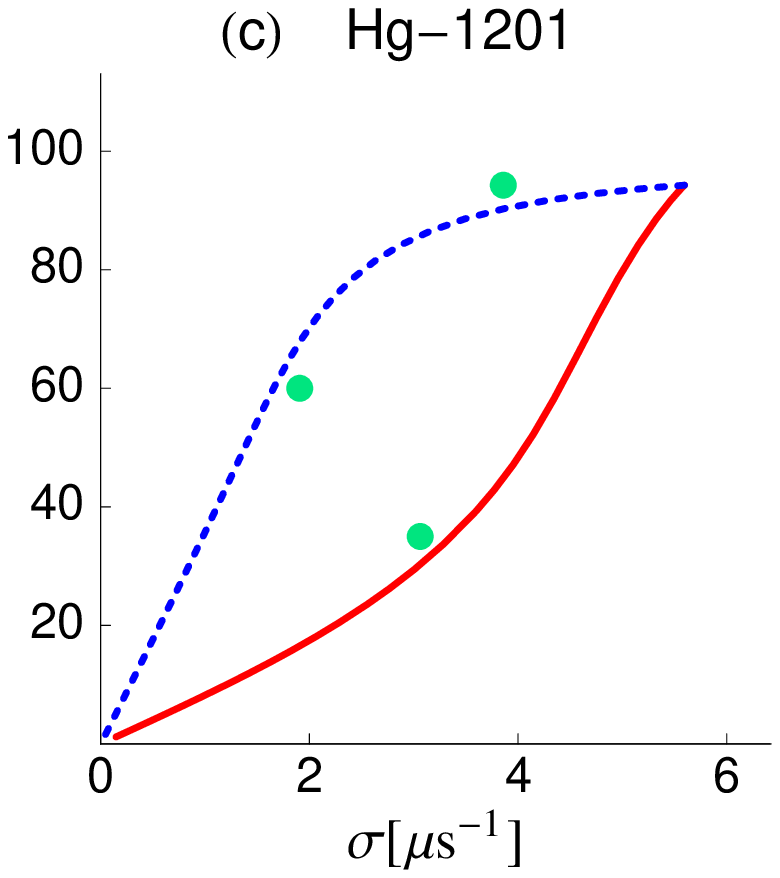}  }
\put(20, 104) { \epsfxsize= 1.9in \epsfbox{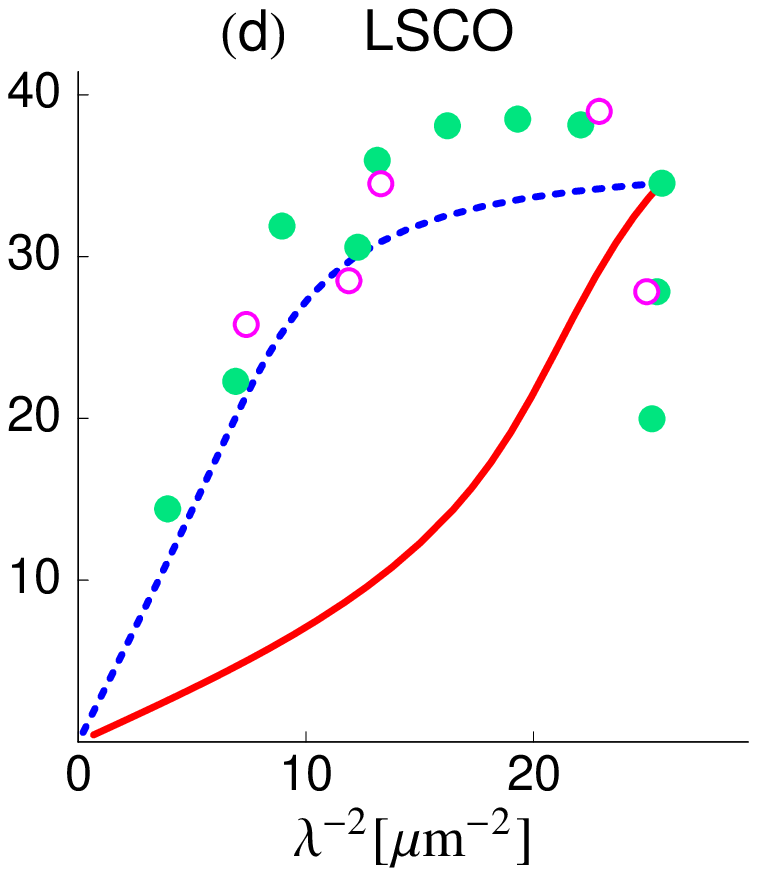} }
\put(-24, 156) { \epsfxsize= 1.9in \epsfbox{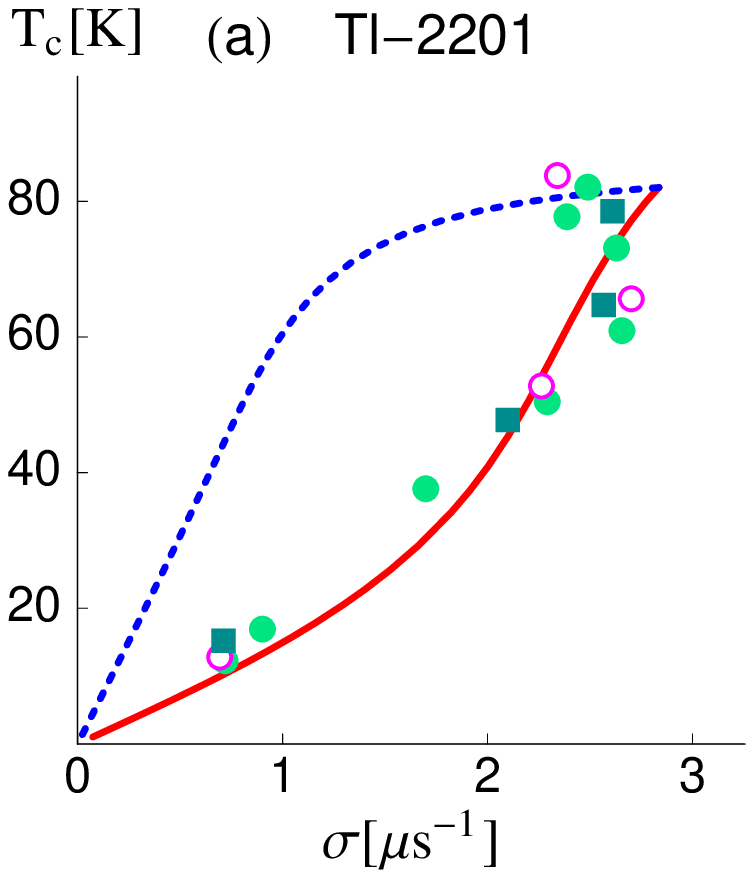}  }
\put(20, 156) { \epsfxsize= 1.9in \epsfbox{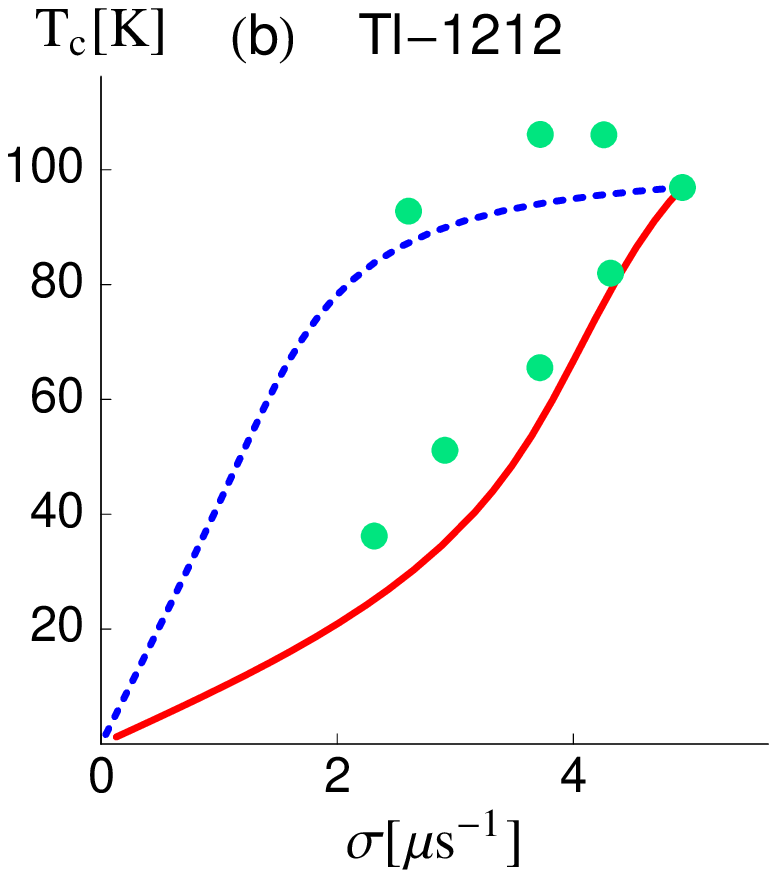}  }  
}
\end{picture} 
\caption[]{(color online) Critical temperature vs. superfluid density 
(presented as $\sigma$ or $\lambda^{-2}$ ) at $T\cong 0$.
The theoretical plots (solid and dashed lines) are obtained by simple
rescaling of the plot presented in Fig.~\ref{fig-fish}.
The experimental points are extracted from the following references:
(a) filled circles - Ref.\cite{Bernhard-etal-95}, open circles - 
Ref.\cite{Niedermayer-etal-93}, squares - Ref.\cite{Uemura-etal-93};
(b) Ref.\cite{Bernhard-etal-01};
(c) Ref.\cite{Panagopoulos-etal-99};
(d) filled circles - Ref.\cite{Tallon-etal-03}, 
open circles - Ref.\cite{Uemura-etal-89};
(e,f) Ref.\cite{Tallon-etal-03};
(g,h) Ref.\cite{Pereg-Barnea-etal-03}.
The doping ranges corresponding to the experimental points are shown
in Fig.~\ref{fig-sfexpCombined}.
} 
\label{fig-sfexp} 
\end{figure}

\begin{figure} \setlength{\unitlength}{0.1cm}

\begin{picture}(50, 77) 
{ 
\put(-20, 0) { \epsfxsize= 3in \epsfbox{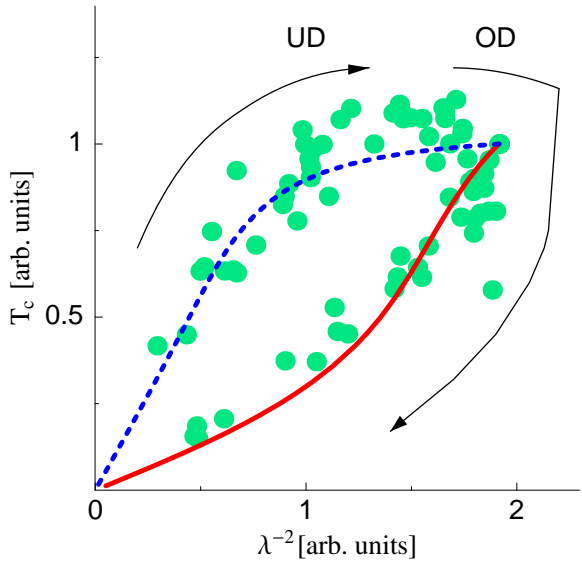}  } 
}
\end{picture} 
\caption{ (color online) All the data points 
from Fig.~\ref{fig-sfexp} rescaled in such a way that the theoretical plots
coincide with each other. The arrows indicate the direction, in which the 
doping concentrations increase. ``UD'' and ``OD'' indicate
underdoped and overdoped samples, respectively.
} 
\label{fig-sfexpCombined} 
\end{figure}

In this subsection, I test the model relationship between  $S_{\phi}$
and $T_c$ by superimposing the (rescaled) theoretical plot of Fig.~\ref{fig-fish} 
on the experimental results
for Tl-2201\cite{Uemura-etal-93,Niedermayer-etal-93,Bernhard-etal-95},
Tl$_{0.5-y}$Pb$_{0.5 + y}$Sr$_2$Ca$_{1-x}$Y$_x$Cu$_2$O$_7$ \mbox{(Tl-1212)}~\cite{Bernhard-etal-01},
Hg-1201~\cite{Panagopoulos-etal-99},
LSCO\cite{Uemura-etal-89,Tallon-etal-03}, 
Bi-2212\cite{Tallon-etal-03},
Ca-doped YBCO (Y:Ca-123)~\cite{Tallon-etal-03} and
YBCO\cite{Pereg-Barnea-etal-03}, which report either relaxation rate $\sigma$ measured
by muon spin rotation ($\mu$SR) technique, or the inverse square of the
penetration depth $\lambda$ extracted from the $\mu$SR
data, field-dependent thermodynamic measurements, or 
electron spin resonance (ESR) studies. Both  $\sigma$ and $\lambda^{-2}$
should be proportional to $S_{\phi}$.
The result is shown in Figs.~\ref{fig-sfexp} and \ref{fig-sfexpCombined}. 


In each of Figs.~\ref{fig-sfexp}(b,d,e,f), the theoretical plot is rescaled
in such a way that the critical case point (the one, where the 
the dashed and the solid 
curves intersect) coincides  with the experimental critical
point.
The latter is defined as the point,
where the derivative of the experimental  $T_c$ vs. $S_{\phi}$ dependence
undergoes an apparent discontinuous change. 
In Refs.\cite{Bernhard-etal-01,Loram-etal-01,Tallon-etal-03}, 
the same point was found to correspond to the  
so-called ``critical doping concentration''
approximately equal to 0.19.  
In  Figs.~\ref{fig-sfexp}(g,h) the theoretical critical
point simply matched the experimental point of the maximal superfluid density.
(Here I ignored the issue of anisotropy and treated
a-axis and b-axis data as independent data sets.)
In Figs.~\ref{fig-sfexp}(a,c), the rescaling relied on 
the overall best fit to the experimental 
data rather then on matching a specific point. 

The comparison with experiments in Fig.~\ref{fig-sfexp} amounts to
the following (crude) quantitative test of the model: After the interaction
constant $g$ and the prefactor of $S_{\phi}$ are fixed by establishing the
absolute scale for the theoretical critical point,
the values of {\it two} independent numbers $S_{\phi}$ and $T_c$
for non-critical points are obtained by varying only {\it one}
theoretical parameter: $\varepsilon_a$ in Case~I or $\varepsilon_b$ in Case~II.

As evident from Figs.~\ref{fig-sfexp} and \ref{fig-sfexpCombined}, 
the theoretical ``fish-like'' plot clearly captures the main features of the 
experimental data, namely: the existence of two different regimes, with a critical
point  of the maximal superfluid density separating them. 
The quality of the quantitative agreement in Figs.~\ref{fig-sfexp}(a,f) 
is particularly  surprising, 
given the crudeness of the test and the fact that the data extend
into the overdoped region, where the model assumptions appear less reliable 
{\it a priori}. 
 
At the point corresponding to the critical case,
one can also estimate the absolute value of the penetration depth from Eq.(\ref{lambdacr}).
This estimate, which only requires the knowledge of the critical 
temperature and the transverse
distance per one CuO$_2$ plane, was made in Section~\ref{critical}
with the numbers close to those of YBCO or Bi-2212. 
The number obtained ($\lambda = 417$~nm)
is  about 3-4 times greater than the numbers typically cited for 
YBCO (see, e.g., Ref.\cite{Pereg-Barnea-etal-03}) 
and about 2 times greater than the numbers cited for Bi-2212 
(see, e.g., Ref.\cite{Tallon-etal-03}). This comparison is representative
of the general trend: the theoretical formula (\ref{lambdacr}) overestimates
the penetration depth by about factor of three.

For a simple estimate, which involves only the fundamental constants and
two well-known material parameters ($T_c$ and $z_0$), the factor-of-three
agreement with the experimental numbers is quite reasonable. 
One should also be conscious of the possibility
(mentioned in Section~\ref{supercurrent}) that the profile of magnetic field 
may strongly fluctuate within the penetration depth layer.  At the same time, the 
absolute values of the penetration depth are typically extracted from the experimental data
on the basis of theoretical formulas, which do not take into account such a possibility.

\subsection{Correspondence between the model regimes and the doping concentrations}
\label{correspondence}

In this subsection, I attempt to establish the correspondence between
model Cases IA, IB, IIA and IIB and the doping concentrations of high-$T_c$ cuprates.
On the basis of the content of 
Sections~\ref{potential}, \ref{tunneling} and \ref{superfluid},
the following four criteria discriminating between 
the four model Cases can be proposed:

{\it Criterion 1}: The inequality $\Delta/T_c > 4$ indicates Case~I, while
$\Delta/T_c < 4$ indicates Case II.

{\it Criterion 2}: When the theoretical ``fish''-plot
from Fig.~\ref{fig-fish}  is superimposed on the experimental dependence of $T_c$ on the 
superfluid density, the proximity of a data point
to the  solid line indicates Case I, while the proximity to
the dashed line indicates Case II.

{\it Criterion 3}: The asymmetry in the tunneling density of states 
characterized by a larger SC peak at negative voltages indicates Case IA.
The opposite asymmetry indicates Case IB.

{\it Criterion 4}: When the asymmetry in the tunneling density of states is not
accessible, I will rely on the postulate,
that, in underdoped cuprates, $\varepsilon_a > \varepsilon_b$,
which favors Case IA over IB, and IIB over IIA.

{\it Criterion 2}, when applied to Fig.~\ref{fig-sfexpCombined}, 
suggests a very simple picture: 
The cuprates are describable
by Case II  at subcritical doping concentrations 
$x_d < x_{dC} \approx 0.19$, and by Case I at the supercritical concentrations 
$x_d > x_{dC} $. {\it Criterion 4} then further narrows
the choice to Case IIB for $x_d < x_{dC}$. 
This identification implies that, 
at low doping, $\mu = \varepsilon_a > \varepsilon_b$,
and then, as the doping concentration increases, 
the difference $\varepsilon_a - \varepsilon_b$ 
decreases until, at the critical doping,  it becomes equal to zero.
The identification of  Case~I 
for $x_d > x_{dC}$ then suggests that $|\varepsilon_a - \varepsilon_b|$
starts increasing again (but now with $\mu = \varepsilon_b$). 
Since  the derivative of $\varepsilon_a - \varepsilon_b$
as a function of doping is unlikely to change sign exactly at $x_d = x_{dC}$,  
I conclude that the model  pseudogap
$\varepsilon_a - \varepsilon_b$ {\it changes sign} as $x_d$ passes $x_{dC}$.
This means that, at the supercritical doping concentrations, 
$\varepsilon_a < \varepsilon_b$, which implies
Case IB. Thus the assignment following from the above discussion is:
\begin{equation}
\begin{array}{rcl}
x_d < x_{dC} & \Rightarrow & \hbox{Case IIB}
\\
x_d > x_{dC} & \Rightarrow & \hbox{Case IB}.
\end{array}
\label{assignment}
\end{equation}

The clear systematics of the superfluid density data
should now be contrasted with a less systematic picture
emerging from the tunneling data.

Most of the point-contact and STM tunneling spectra discussed in 
subsection~\ref{tunneling}, as well as break junction and interlayer
tunneling spectra, 
are collected at
doping concentrations $x_d \leq x_{dC}$.
In this doping range, the tunneling data for 
Tl-2201~\cite{Ozyuzer-etal-98}, 
Tl$_2$Ba$_2$CaCu$_2$O$_{8+\delta}$ (Tl-2212)~\cite{Huang-etal-89},
Hg-1201~\cite{Chen-etal-94,Wei-etal-98},
LSCO~\cite{MOA},
YBCO~\cite{Ponomarev-etal-95,Ponomarev-etal-95A,Maggio-Aprile-etal-95}
(and also YbBa$_2$Cu$_3$O$_{7-x}$~\cite{Ponomarev-etal-95,Ponomarev-etal-95A})
support inequality $\Delta/T_c \leq 4$, which, according to  {\it Criterion 1},
suggests Case~II in agreement with the assignment~(\ref{assignment}).

The tunneling studies of Bi-2212 do not reveal a coherent picture
either in terms of the ratio $\Delta/T_c$ or in terms of the SC peak asymmetry.  
Most of more recent \mbox{Bi-2212} tunneling data for 
$x_d \leq x_{dC}$~\cite{MA,DeWilde-etal-98,Miyakawa-etal-98,Renner-etal-98,Renner-etal-98A,Pan-etal-00,SW,Zasadzinski-etal-01,Misra-etal-02,Krasnov-02}
show $\Delta/T_c > 4$, and whenever the asymmetry is evident in
the data, it mostly points to Case~IA --- 
in clear contradiction with the assignment
(\ref{assignment}). At the same time, many other (and some of the same) 
tunneling 
studies\cite{Huang-etal-89A,Koltun-etal-91,BBG,Vedeneev-etal-94,RF,Krasnov-02} 
find in the same doping range 
the gap values corresponding $\Delta/T_c \leq 4$ and thus, according to {\it Criterion 1}
support the assignment~(\ref{assignment}). 

The remarkable fact is that, at least on two 
occasions\cite{Koltun-etal-91,Koltun-etal-91-comment,Krasnov-02},
the tunneling spectra of Bi-2212  presented in the same paper
and measured on samples with nearly equal critical temperatures
have shown two different values of $\Delta/T_c$ --- one significantly greater
than 4, and the other one smaller than 4. 

The tunneling phenomenology
of Bi-2212 can be explained by the existence of
two different SC states, which, for $x_d < x_{dC}$, correspond either
to Case IA or to Case IIB. As argued in Section~\ref{chemical}, both SC states
are characterized by sharp minima of the total energy of the system and by
the same critical temperature. One of them can, e.g., constitute
a stable bulk state, while the other one a stable or metastable surface state.

The asymmetric STM spectrum of Hg-1212 reported in Ref.\cite{Wei-etal-98}
clearly suggest Case IA, which cannot be placed within the assignment 
(\ref{assignment}).

{\it Criteria 1} and {\it 3}, when applied to the asymmetric STM spectra
for the tri-layer compounds Bi-2223\cite{Chen-etal-94A} and Hg-1223\cite{Wei-etal-98}
suggest Case~IB, and thus could be compatible to the assignment~(\ref{assignment})
provided that the doping concentration in those samples exceeds 
the (unknown) critical concentration
for the corresponding families of cuprates.
However, the recent interlayer tunneling results for 
\mbox{Bi-2223}\cite{Yamada-etal-03} indicate
that, as the doping concentration increases, the ratio $\Delta/T_c$ decreases
from values larger than 4 to values smaller than 4, which, in combination with the 
STM data\cite{Chen-etal-94A} rather suggests the assignment
opposite to (\ref{assignment}), namely: Case IB for $x_d \leq x_{dC}$
and Case IIB for $x_d > x_{dC}$.

\

Summary of the findings of this subsection: Assignment~(\ref{assignment})
is consistent with the superfluid density data and/or the tunneling data reported
for Tl-2201, Tl-2212, Tl-1212, Hg-1201, LSCO and YBCO. For Bi-2212,
the same  assignment  is partially supported by experiments, while the overall
phenomenology rather suggests the occurrence of two different
SC states at the same doping concentration. The limited tunneling data on
Hg-1212, Hg-1223 and Bi-2223 appear to contradict to the assignment~(\ref{assignment})
with varying degrees of certainty. Assignment~(\ref{assignment}), if true,
implies that the pseudogap 
changes sign at the critical doping concentration.

\subsection{Symmetry of the SC order parameter}
\label{symmetry}

The prediction, which distinguishes the present proposal from many others, 
is the non-trivial translational symmetry of the SC order parameter.
This property has not been yet investigated experimentally.  
One straightforward experimental test would be to construct a nanoscale probe, 
which is sensitive to the SC phase
difference between two points separated by one period of the 
expected stripe superstructure.
Slightly prior to the appearance of the present work,
the position-dependent sign change of the SC order parameter has
also been proposed by Ashkenazi\cite{Ashkenazi} in the context of a different
stripe-related model. 

An important question is whether the SC order parameter introduced in this work 
is consistent with the phase sensitive experiments,  which are
usually interpreted as the evidence for
$d_{x^2-y^2}$ symmetry of a spatially uniform SC state (see e.g. Ref.\cite{AGL}).
Here I am primarily concerned with the corner SQUID experiments\cite{Wollman-etal-93} 
and with the tricrystal junction experiments\cite{Tsuei-etal-94,Tsuei-etal-04}.

A definitive discussion of these experiments 
cannot be given at this time 
because of the following two uncertainties: 

(i) The continuous family of possible SC solutions obtained in Section~\ref{canonical} 
has not been narrowed to a single one.   

(ii) The boundary conditions necessary for 
the discussion of the phase sensitive 
experiments have not been specified.  

\begin{figure} \setlength{\unitlength}{0.1cm}

\begin{picture}(50, 72) 
{ 
\put(-33, 0) { \epsfxsize= 3.6in \epsfbox{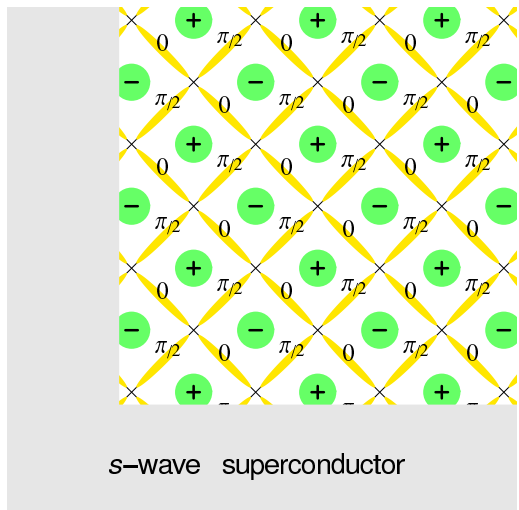}  } 
}
\end{picture} 
\caption{(color online) Possible geometry of a corner SQUID experiment.
} 
\label{fig-corner} 
\end{figure}

In view of the above uncertainties, I limit further discussion
to presenting just one  of several possible interpretations
of the well-known $\pi$ phase shift observed in the corner SQUID experiments\cite{Wollman-etal-93}. 
This interpretation  is illustrated in Fig.~\ref{fig-corner}. It  is based on the same 
SC phase pattern as the one shown   in Fig.~\ref{fig-phases}(a). 
The examination of Fig.~\ref{fig-corner} reveals that the 
phases ``0'' and ``$\pi/2$'', which characterize both $\Psi_a$ and $\Psi_b$,  
form the same pattern along both of the SQUID interfaces,
which means that this aspect of the phase symmetry is unlikely 
to contribute to the relative phase shift between the two interfaces. 
However, in addition, the $\Psi_a$-component also has the
position dependent sign factor, which is indicated in the centers of the supercells
as ``$+$'' or ``$-$''.
The important fact is that, despite the sign change of  
$\Psi_a$,  the Josephson coupling between $\Psi_a$ and the order parameter of  
the conventional superconductor  is not averaged to zero along each 
of the two interfaces shown in Fig.~\ref{fig-corner}. 
Now I assume that, at these interfaces,
the order parameter of  the conventional superconductor preferentially
couples to $\Psi_a$ (as opposed to $\Psi_b$). In such a case, 
the opposite signs  of  $\Psi_a$ along the two interfaces
imply the required phase shift of $\pi$.

The interpretation of the tricrystal experiments\cite{Tsuei-etal-94,Tsuei-etal-04}, 
in which vortices carrying half of the 
flux quantum  were observed,  is not
practical at this time, because it should
depend critically on the unknown boundary conditions.  
Here, I can only mention, that, 
the design geometry of the tricrystal experiments 
is such that,
if the postulated 2D stripe superstructure 
exists in each of the three adjacent crystals, then
the three interfaces between the intersecting stripe superstructures are 
nearly equivalent to each other.
Therefore, the explanation of these experiments will likely amount to showing that 
each of the junctions generates a $\pi$ phase shift.

Finally, I discuss  the observation 
of the c-axis Josephson pair tunneling between 
YBCO single crystals  and the films of Pb\cite{Sun-etal-94}. 
(Pb is a conventional $s$-wave superconductor.) 
This experimental fact 
is difficult to explain on the basis of the ``d-wave'' picture\cite{AGL}.
At the same time, in the framework of the present proposal,
it can be easily interpreted as follows:
When a conventional superconductor is placed on the top
of  the 2D striped system shown in Fig.~\ref{fig-corner}, 
the contribution of the 
$\Psi_a$-component to the Josephson coupling 
changes sign and thus averages to zero,
but the contribution from the $\Psi_b$-component, 
which does not change sign,
can lead to a finite Josephson current.

\section{Conclusions}
\label{conclusions}

In the present work, I have analysed several consequences of the hypothesis that
holes doped into high-$T_c$ cuprates organize themselves in two-dimensional arrays
of deep stripes.  In particular, on the basis of this hypothesis, I have
formulated and solved a model of superconductivity. From that model,
I have obtained
the tunneling spectrum and the superfluid density, which show
good agreement with experiments. The symmetry
of the SC order parameter derived from the model is different from that of
$d_{x^2-y^2}$ BCS order parameter. The order parameter obtained in this work has
two components, at least one of which changes sign as a function of the
absolute position of the pair on the spatial scale of the stripe superstructure.
A number of other features of this proposal such as the geometry of the pseudogap
and the effect of the pseudogap on the superconducting transition temperature appear
to be in qualitative agreement with the phenomenology of high-$T_c$ cuprates.
The checkerboard pattern of LDOS observed by STM has been interpreted as coming
from the centers of stripe elements in the 2D arrangement of diagonal stripes.
This work also indicates the possibility, that, in  underdoped cuprates, there
may exist two different kinds of SC states, and that, at the
critical doping concentration, the pseudogap may change sign.

Even if a future work demonstrates the inadequacy of the
theoretical assumptions of the present one, 
the systematics of the asymmetry in the tunneling spectra discussed 
in Section~\ref{tunneling} and the scaling
of the superfluid density data shown in Fig.~\ref{fig-sfexpCombined}
should retain the status of useful empirical facts.

\acknowledgments

The author is grateful to M.~Turlakov, A.~J.~Leggett, R.~A.~Klemm, 
P.~Fulde, K.~Maki, B.~Altshuler, P.~McHalle, V.~Krasnov, 
J.~C.~Davis, C.~Panagopoulos, L. P. Gor'kov and J. Zaanen 
for helpful discussions.

\appendix

\section{Approximate solution via a non-canonical transformation}
\label{non-canonical}

In this Appendix, I present an approximate scheme of finding the variational
ground state of Hamiltonian~(\ref{H}) 
using the following Bogoliubov-like {\it non-canonical} transformation
of a-states {\it in real space}
\begin{equation}
a_i = u A_i + v \eta_i \sum_{j(i)} A_j^+;
\label{aA}
\end{equation}
together with a regular Bogoliubov transformation of b-states
\begin{eqnarray}
b_{ij,+} &=& s B_{ij,+} + w B_{ij,-}^+;
\label{bB+}
\\
b_{ij,-} &=& s B_{ij,-} - w B_{ij,+}^+,
\label{bB-}
\end{eqnarray}
where $A_i$ and $B_{ij,\sigma}$ are the annihilation operators of
the Bogoliubov quasiparticles; $u$, $v$, $s$ and $w$ are the 
transformation coefficients. 
These coefficients 
can be chosen real. They must
then obey the following normalization constraints:
\begin{equation}
u^2 + 4 v^2 =1,
\label{uv}
\end{equation}
\begin{equation}
s^2 + w^2 = 1.
\label{sw1}
\end{equation}
(Variables $u$ and $v$ of this Appendix should not be confused
with functions $u({\mathbf{k}})$ and $v({\mathbf{k}})$
defined by Eqs.(\ref{aeA},\ref{aoA}).

The  second term
on the right-hand-side of Eq.(\ref{aA})  changes sign from supercell
to supercell following the sign of 
$\eta_i$. This sign change is necessary to ensure that
the canonical fermionic anticommutation relations between operators 
$A_i$ and $A_j$ corresponding to neighboring AF
domains are not violated 
in the first order of $v$. This transformation is still non-canonical,
because it violates the anticommutation relations 
in the second and higher orders
of $v$. In order to see this,
one can assume that A-operators represent true fermions
and then check the anticommutation relation between operators
$a_i$ and $a_p^+$ corresponding to a pair of next nearest neighbors.

Despite the fact that transformation (\ref{aA}) is not canonical, I will
substitute it (together with transformations (\ref{bB+},\ref{bB-}) 
into  the Hamiltonian (\ref{H})
and then  handle A-operators as if they were true fermionic operators. 

The justification for such a scheme is three-fold: 
(i) The non-canonical transformation (\ref{aA})
is very natural for the structure of Hamiltonian (\ref{H}). 
(ii) {\it A priori}, this scheme represents
a controllable approximation in the case of small $v$ (large $\varepsilon_a$). 
(iii) For  arbitrary values of
parameters $\varepsilon_a$,  $\varepsilon_b$ and  $g$, the ground state
energy and the quasiparticle excitation energies 
obtained in the present framework
turn out to be very close to those obtained with the help of the
fully canonical transformation of Section~\ref{canonical}.

Transformation (\ref{aA}-\ref{bB-}) minimizes the energy of the system when
\begin{eqnarray}
u &=& \left( {1 \over 2} + {\hbox{sign}(X) \over 2} \sqrt{{1 + {Y^2 \over Z^2} \over 1 + {Z^2 \over X^2}}} \right)^{1/2};
\label{u}
\\
v &=&  {1 \over 2} 
\left( {1 \over 2} - {\hbox{sign}(X) \over 2} \sqrt{{1 + {Y^2 \over Z^2} \over 1 + {Z^2 \over X^2}}} \right)^{1/2};
\label{v}
\\
s &=& \left( {1 \over 2} + {\hbox{sign}(Y) \over 2} \sqrt{{1 + {X^2 \over Z^2} \over 1 + {Z^2 \over Y^2}}} \right)^{1/2};
\label{s}
\\
w &=& - \left( {1 \over 2} - {\hbox{sign}(Y) \over 2} \sqrt{{1 + {X^2 \over Z^2} \over 1 + {Z^2 \over Y^2}}} \right)^{1/2},
\label{w}
\end{eqnarray}
where
\begin{eqnarray}
X &=& \varepsilon_a (1 - 2 n_A);
\label{X}
\\
Y &=& 4 \varepsilon_b (1 - 2 n_B);
\label{Y}
\\
Z &=&  g (1 - 2 n_A) (1 - 2 n_B).
\label{Z}
\end{eqnarray}
Here $n_A$ and $n_B$ are the occupation numbers of the
quasiparticle states described by operators 
$A_i$ and $B_{ij,\sigma}$ respectively, i.e.
\begin{eqnarray}
n_A &=& {1 \over \hbox{exp}\left({\varepsilon_A \over T}\right) +1};
\label{nA}
\\
n_B &=& {1 \over \hbox{exp}\left({\varepsilon_B \over T}\right) +1}.
\label{nB1}
\end{eqnarray}
In Eqs.(\ref{nA},{\ref{nB1}), 
variables $\varepsilon_A$ and $\varepsilon_B$ are the energies of the
respective quasiparticle states.

At $T=0$, the constraints $|u| < 1$ and $|s| < 1$ impose
the following condition for the existence of the physical solution:
\begin{equation}
g \geq \sqrt{4 | \varepsilon_a \varepsilon_b |}
\label{gee}
\end{equation}
This condition is satisfied for any non-zero value
of $g$, when either $\varepsilon_a =0$ or $\varepsilon_b=0$.

In the rest of this Appendix, I limit the calculations
only to Cases IA and IIA (in the classification of Section~\ref{classification}).

{\bf Case IA:} $\varepsilon_b =0$, $\varepsilon_a \geq 0$.

Condition
$\varepsilon_b =0$   implies that, according to Eqs.((\ref{s},\ref{w})),
$s= 1/\sqrt{2}$ and $w = -1/\sqrt{2}$, both independent of temperature.
The coefficients $u$ and $v$ given by 
Eqs.(\ref{u},\ref{v}) with $Y=0$ have temperature-dependent
values. 

A natural sign convention for this case  is 
$\varepsilon_A >0$ and $\varepsilon_B <0$, i.e., at $T=0$,
$n_A = 0$ and $n_B = 1$. 

One can then obtain the energies 
\begin{equation}
\varepsilon_A = \sqrt{\varepsilon_a^2 + 
g^2 (1 - 2 n_B)^2}
\label{epsA}
\end{equation}
and
\begin{equation}
\varepsilon_B = {g^2 (1 - 2n_A) (1-2 n_B) \over 4 \varepsilon_A}
\label{epsB}
\end{equation}
of   A- and B-quasiparticles, respectively.

The zero temperature value of 
$\varepsilon_A$ following from Eq.(\ref{epsA}) coincides
with that of the  Van Hove singularity (\ref{epsA0I})
characterizing the spectrum of the ``canonical solution''.
The values of $\varepsilon_B$ given by Eqs.(\ref{epsB}) and (\ref{epsBI}) 
for the two solutions are also close to each other.

The ground state energy in the present case  can be evaluated as:
\begin{equation}
E_{\hbox{\sc \scriptsize gs}} 
= -{N \over 2} \left[ \sqrt{\varepsilon_a^2 + g^2} - \varepsilon_a \right].
\label{epsGS}
\end{equation}
If $E_{\hbox{\sc \scriptsize gs}}$ given by Eq.(\ref{epsGS}) 
is compared with the ground state energy of the canonical solution
(Eq.(\ref{EI})), then the difference
is never greater than five per cent.

At finite temperatures, in order to obtain
$n_A$, $n_B$, $\varepsilon_A$ and $\varepsilon_B$
one has to solve the system of equations 
(\ref{nA}, \ref{nB1}, \ref{epsA}, \ref{epsB})
numerically.
It is easy to find, however, that 
the above system of equations always has one trivial
solution: $\varepsilon_A = \varepsilon_a$, \mbox{$\varepsilon_B = 0$} with 
$n_A$ and $n_B$ given by Eqs.(\ref{nA}, \ref{nB}).
The condition for the
existence of the second, non-trivial, solution can be found analytically.
This condition is: $T<T_c$, where the critical temperature $T_c$ 
is the solution
of Eq.(\ref{Tceq}).
Thus the remarkable fact is that $T_c$ obtained in the framework
of the present non-canonical scheme reproduces the ``canonical'' result
of Section~\ref{canonical}.


\

{\bf Case IIA:} $\varepsilon_a =0$, $\varepsilon_b \geq 0$.

According to Eqs.(\ref{u}-\ref{w}), the 
condition $\varepsilon_a =0$   implies that
$u= 1/\sqrt{2}$ and $v = 1/\sqrt{8}$, while
$s$ and $w$ have  temperature-dependent values. 

The sign convention in this case is: 
$\varepsilon_A < 0$, $\varepsilon_B > 0$, i.e., at $T=0$,
$n_A = 1$ and $n_B = 0$.  
The calculation  now gives:
\begin{eqnarray}
\varepsilon_A &=&  {g^2 (1 - 2n_A) (1-2 n_B) \over 4 \varepsilon_B};
\label{epsAII}
\\
\varepsilon_B &=&  \sqrt{\varepsilon_b^2 + {1 \over 16} g^2 (1 - 2 n_A)^2};
\label{epsBIIapp}
\end{eqnarray}
\begin{equation}
{E_{\hbox{\sc \scriptsize gs}}} = 
- 2 N \left[ \sqrt{\varepsilon_b^2 + {1 \over 16} g^2} - \varepsilon_b \right].
\label{epsGSII}
\end{equation}
The critical temperature in this case is again the same as obtained 
from the ``canonical'' solution, i.e. it is given by Eq.(\ref{TceqII}).

\

In summary:  
The non-canonical variational scheme based on 
Eqs.(\ref{aA},\ref{bB+},\ref{bB-})  predicts 
the same critical temperature as the canonical
scheme of Section\ref{canonical}. Furthermore, 
the non-canonical scheme predicts the ground state energy
and the important tunneling characteristics 
within a few per cent from the canonical result. 
The only significant feature of the canonical solution missing in the non-canonical 
one is 
the absence of the gap in the spectrum of A-quasiparticles in Case II. 
A related conceptual detail is that
the non-canonical scheme fails to
predict the coherent dispersion $\varepsilon_A({\mathbf{k}})$
of A-quasiparticles. 

\

I conclude this Appendix with the following comment:
 
In the present variational scheme one can easily find that 
neither the variational
energy nor the excitation spectrum will change,
if the phases of transformation (\ref{aA}-\ref{bB-})
are modified in the following way:
\begin{equation}
a_i = u A_i + v \eta_i \sum_{j(i)} e^{\varphi_{ij}} A_j^+;
\label{aA1}
\end{equation}
\begin{eqnarray}
b_{ij,+} &=& s B_{ij,+} + w  e^{\varphi_{ij}} B_{ij,-}^+;
\label{bB+1}
\\
b_{ij,-} &=& s B_{ij,-} - w  e^{\varphi_{ij}} B_{ij,+}^+,
\label{bB-1}
\end{eqnarray}
where phases $\varphi_{ij}$ can be different for different pairs of indices
$i$ and $j$. 

The freedom to vary phases $\varphi_{ij}$ in 
Eqs.(\ref{aA1}-\ref{bB-1})
is, at least in part, due to the fact that the parameter space of non-canonical
transformations is larger than that of canonical ones.
Therefore, one may  try to choose phases in 
Eqs.(\ref{aA1}-\ref{bB-1}) such that 
transformation (\ref{aA1}) becomes canonical. I have found,  that, in this way
(with the selection of phases shown later in Fig.~\ref{fig-phases}),
the next-nearest-neighbor anticommutation test described earlier 
in this Appendix can, indeed, 
be satisfied.  
However, a similar test for the pairs of supercells separated by one common neighbor 
cannot be satisfied independently of the choice of  phases $\varphi_{ij}$.  
Transformation (\ref{aA1}) cannot be made rigorously canonical,
because it involves only the nearest neighbors.
It is, however, possible to conform with the canonical anticommutation relations,
if transformation (\ref{aA1}) is modified to include more remote neighbors. 
Such a canonical transformation is much easier 
to describe in $k$-space --- subject of Section~\ref{canonical}.
One can thus conclude that the coherent dispersion of A-quasiparticles in
$k$-space is protected by the Fermi statistics. 

From a different perspective, one can also observe that,
in the canonical scheme, 
the non-intuitive combination of phases $\varphi_{\alpha}$ given by 
Eq.(\ref{phases} minimizes the pairing amplitude between
more remote neighbors, which leads to the maximum energy gain from
the nearest-neighbors interacting {\it via} the Hamiltonian~(\ref{H}).

\section{Minimization of energies (\ref{EI}) and (\ref{EII})
with respect to phases $\phi_{\alpha}$}
\label{phasemin}

The minimization procedure presented in 
this Appendix is equally applicable to the total energy expressions
for both Case I (Eq.(\ref{EI})) and Case II  (Eq.(\ref{EII})). 
Below, in order to be specific, I will focus
on the expression (\ref{EI}).
This expression can be considered as an implicit function of phases
$\varphi_{\alpha}$ entering it through the dependence on
a single function $|V({\mathbf{k}})|^2$. 

From Eq.(\ref{V}),
\begin{eqnarray}
\nonumber
|V({\mathbf{k}})|^2 &=&  4 
\left\{
\hbox{cos}^2 \left[ {\mathbf{k}} {\mathbf{R}}_1 + 
{\varphi_1 - \varphi_3 \over 2} \right]
\right.
\\
\nonumber
&&
\ \ 
+
\hbox{cos}^2 \left[ {\mathbf{k}} {\mathbf{R}}_2 + 
{\varphi_2 - \varphi_4 \over 2} \right] 
\\
\nonumber
&&
\ \ 
 + 
2 \ 
\hbox{cos} \left[ {\mathbf{k}} {\mathbf{R}}_1 + 
{\varphi_1 - \varphi_3 \over 2} \right] 
\\
\nonumber
&&
\ \ \ \ \ \times
\hbox{cos} \left[ {\mathbf{k}} {\mathbf{R}}_2 + 
{\varphi_2 - \varphi_4 \over 2} \right]
\\
&&
\ \ \ \ \ \times
\left.
\hbox{cos} \left[  
{\varphi_2 + \varphi_4 -\varphi_1 - \varphi_3 \over 2} \right]
\right\}
\label{V2}
\end{eqnarray}

I first note that $|V({\mathbf{k}})|^2$ is the periodic function of 
${\mathbf{k}}$  in the directions of ${\mathbf{R}}_1$ 
and ${\mathbf{R}}_2$.
Replacing sum in Eq.(\ref{EI}) by the integral
according to the prescription (\ref{sum-integral}),
I further note that  the symmetry of the function $|V({\mathbf{k}})|^2$
is such, that any shift  of the integration region 
does not change the value of the integral.
Therefore, the result of the integration does not depend on 
the values of
${\varphi_1 - \varphi_3 \over 2}$ and 
${\varphi_2 - \varphi_4 \over 2}$. 
In the following, in order to be specific, I  choose 
${\varphi_1 - \varphi_3 \over 2} =0 $ and 
${\varphi_2 - \varphi_4 \over 2} =0$.

The only phase combination to be constrained by the minimization
of energy (\ref{EI})
is ${\varphi_2 + \varphi_4 -\varphi_1 - \varphi_3 \over 2}$. 
Now I switch back to the language of summation and
note that the summation points in 
Eq.(\ref{EI}) can be divided in pairs 
$({\mathbf{k}}, {\mathbf{k}}^{\prime})$ as shown in Fig.~(\ref{fig-kpairs}).
Point ${\mathbf{k}}$ is chosen inside the white area surrounded by the dashed line, 
while ${\mathbf{k}}^{\prime}$ is the nearest mirror image of  ${\mathbf{k}}$
in the dark area outside of the dashed line.
For each such a pair, 
\begin{equation}
|V({\mathbf{k}})|^2 = 4 [ h({\mathbf{k}}) + p({\mathbf{k}}) \ \zeta ],
\label{Vk1}
\end{equation}
while
\begin{equation}
|V({\mathbf{k}}^{\prime})|^2 = 4 [ h({\mathbf{k}}) - p({\mathbf{k}}) \ \zeta ],
\label{Vk2}
\end{equation}
where
\begin{eqnarray}
h({\mathbf{k}}) = \hbox{cos}^2 \left[ {\mathbf{k}} {\mathbf{R}}_1  \right]  +
\hbox{cos}^2 \left[ {\mathbf{k}} {\mathbf{R}}_2  \right];
\label{h}
\\
p({\mathbf{k}}) = 2 \hbox{cos} \left[ {\mathbf{k}} {\mathbf{R}}_1  \right]  
\hbox{cos} \left[ {\mathbf{k}} {\mathbf{R}}_2  \right];
\label{p}
\\
\zeta = \hbox{cos} \left[  
{\varphi_2 + \varphi_4 -\varphi_1 - \varphi_3 \over 2} \right].
\label{zeta}
\end{eqnarray}
Thus the energy (\ref{EI}) can  be presented as
\begin{equation}
E = - \sum_{{\mathbf{k}}}^{\sim}  \left\{ \ F[h({\mathbf{k}}) + p({\mathbf{k}}) \zeta \ ] + 
F[h({\mathbf{k}}) - p({\mathbf{k}}) \zeta\ ] \ \right\},
\label{EIF}
\end{equation}
where symbol ``$\sim$'' in the sum superscript implies 
that the summation is limited to the area shown in Fig.~(\ref{fig-kpairs}) 
inside the dashed line. Function $F$ is implicitly defined by equation
\begin{equation}
F \left[ {1 \over 4} |V({\mathbf{k}})|^2 \right] = 
[1 - 2 n_A({\mathbf{k}})] \ \varepsilon_A({\mathbf{k}}) - \varepsilon_a, 
\label{F}
\end{equation}
where  $\varepsilon_A({\mathbf{k}})$
and \mbox{$n_A({\mathbf{k}}) \equiv n_A[\varepsilon_A({\mathbf{k}})]$}
are expressed as functions of $|V({\mathbf{k}})|^2$
with the help of Eqs. (\ref{nAk},\ref{epsAkI}).
Even without specifying function $F$ explicitly, 
one can take the derivative of $E$ with respect to 
$\zeta$ to find:
\begin{eqnarray}
\nonumber
{\partial E \over \partial \zeta} &=& 
- \sum_{{\mathbf{k}}}^{\sim} p({\mathbf{k}})  
 \left\{ 
F^{\prime}[h({\mathbf{k}}) + p({\mathbf{k}}) \zeta] 
\right. 
\\
&&
\ \ \ \ \ \ \ \ \ \ \ \ \ \ \ \ 
- \left.  F^{\prime}[h({\mathbf{k}}) - p({\mathbf{k}}) \zeta] 
\right\},
\label{dEdzeta}
\end{eqnarray}
where $F^{\prime}$ is the first derivative of function
$F$. Each term in  the sum (\ref{dEdzeta}) is equal to zero,
when $\zeta = 0$, which implies an extremum of $E$.
I have examined a large number examples numerically
and have found that, in all cases considered,  
the above extremum corresponds to the global maximum.
This result is also easy to derive analytically in the critical case
by showing that all pairs of terms 
\mbox{$F[h({\mathbf{k}}) + p({\mathbf{k}}) \zeta] + 
F[h({\mathbf{k}}) - p({\mathbf{k}}) \zeta]$}
in Eq.(\ref{EIF}) simultaneously
reach their maximal values, when $\zeta = 0$.
 
Condition $\zeta = 0 $ substituted into Eq.(\ref{zeta}),
then gives:
\begin{equation}
\hbox{cos} \left[  
{\varphi_2 + \varphi_4 -\varphi_1 - \varphi_3 \over 2} \right] =0,
\label{cosphi}
\end{equation}
from which Eq.(\ref{phases}) follows.


\begin{figure} \setlength{\unitlength}{0.1cm}

\begin{picture}(50, 70) 
{ 
\put(-9, 0)
{ \epsfxsize= 2.7in
\epsfbox{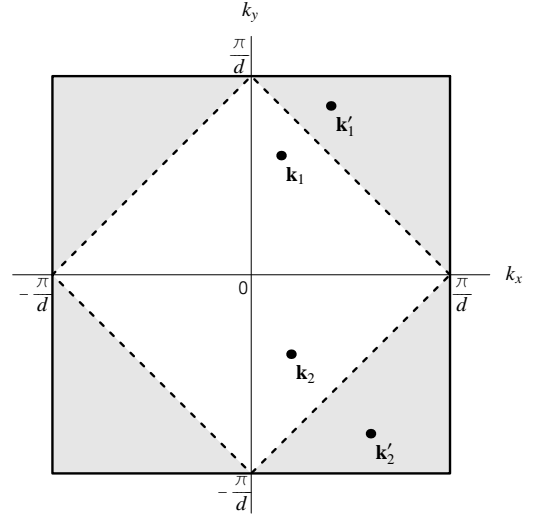} 
}}
\end{picture} 
\caption{ 
Two examples of the pairs  k-points referred to in the text:  
$({\mathbf{k}}_1,{\mathbf{k}}_1^{\prime})$ and $({\mathbf{k}}_2,{\mathbf{k}}_2^{\prime})$.
Thick solid lines represents the boundary of 
the first Brillouin zone of the stripe superstructure.
The first k-point of each pair should belong to the light region inside the dashed
boundary. The second point should be obtained from the first one
by reflection with respect to the nearest dashed line.
} 
\label{fig-kpairs} 
\end{figure}


\section{Derivation of the equation for the critical temperature}

\label{TcEq}

In order to obtain the critical temperature in Case I of Section~\ref{canonical},
I substitute into Eq.(\ref{epsBI}) the limiting values 
of all quantities as $T - T_c \rightarrow 0_-$.
In this limit, 
\begin{equation}
\varepsilon_A({\mathbf{k}})  \rightarrow  \varepsilon_a,
\label{epsATc}
\end{equation} 
\begin{equation}
n_A({\mathbf{k}})  \rightarrow  
{ 1 \over \hbox{exp} \left( {\varepsilon_a \over T_c} \right) + 1}
\label{nATc}
\end{equation}
$\varepsilon_B \rightarrow 0$, and
$n_B  \rightarrow 1/2$, i.e. $2 n_B -1  \rightarrow 0$.
In order to resolve the uncertainty associated
with substituting the limiting values of
$\varepsilon_B$ and $2 n_B -1$, it is
necessary to keep the next order of $\varepsilon_B$ 
in the expression for $2 n_B -1$, i.e.
\begin{equation}
2 n_B -1  \rightarrow - {\varepsilon_B \over 2 T_c}.
\label{nBTc}
\end{equation}
The substitution of Eqs.(\ref{epsATc}, \ref{nATc}, \ref{nBTc}) 
into Eq.(\ref{epsBI})
leads to the following equation:
\begin{equation}
T_c = { g^2 \left[\hbox{exp}\left({\varepsilon_a \over T_c}\right) - 1\right] \over 
16 \varepsilon_a \left[\hbox{exp}\left({\varepsilon_a \over T_c}\right) + 1\right]}
\ {1 \over N} \ \sum_{{\mathbf{k}}} |V({\mathbf{k}})|^2.
\label{TcI}
\end{equation}
Now I note, that, independently of the choice of phases
in Eq.(\ref{V2}),
\begin{equation}
\sum_{{\mathbf{k}}} |V({\mathbf{k}})|^2 = 2N.
\label{SV2}
\end{equation}
The substitution of Eq.(\ref{SV2}) into Eq.(\ref{TcI}) then gives
Eq.(\ref{Tceq}).

\bibliography{hitc1}

\begin{thebibliography}{98}
\bibitem[{Kol()}]{Koltun-etal-91-comment}
\bibinfo{note}{Figure 8 of Ref.\cite{Koltun-etal-91} contains a strongly
  asymmetric point-contact tunneling spectrum characterized by $\Delta /T_c =
  6.8$, where $\Delta$ is obtained as the half of the peak-to-peak separation.
  Citing Ref.\cite{Koltun-etal-91}, I imply the above ratio for that spectrum.
  However, the authors of Ref.\cite{Koltun-etal-91} have assumed that they
  accidentally measured an SIS contact and, therefore, listed $\Delta$ equal to
  the quater of the peak-to-peak separation.}
\end{thebibliography}

\end{document}